%% file: WZTagger.tex
\RequirePackage[l2tabu, orthodox]{nag}
\newcommand*{\ATLASLATEXPATH}{}
\documentclass[UKenglish,texlive=2011,cernpreprint,PAPER]{\ATLASLATEXPATH atlasdoc}
\usepackage[full]{\ATLASLATEXPATH atlaspackage}
\usepackage{\ATLASLATEXPATH atlasbiblatex}
\usepackage{\ATLASLATEXPATH atlascontribute}
\usepackage{\ATLASLATEXPATH atlasphysics}
\usepackage{float}
\usepackage{a4wide}

\graphicspath{{}}

\hypersetup{pdftitle={ATLAS draft},pdfauthor={The ATLAS Collaboration}}

\usepackage{overpic}
\usepackage{subfig}

\usepackage{comment}

\input{WZTagger-meta-data}

\begin{document}

\maketitle

\section{Introduction}
\label{sec:intro}

Processes involving the production and decay of $W$ and $Z$ bosons provide benchmarks for testing the Standard Model (SM), as well as probes of physics beyond the SM (BSM).  Since the cross section for the direct strong production of events with multiple jets (QCD multijets) at the Large Hadron Collider (LHC) is much larger than for $W$ and $Z$ boson production, it is usually the case that the leptonic decays of bosons must be used to reduce the overwhelming background.   However, when the momentum $p^V$ of a boson $V$ is comparable with its mass, $m_V$,  the spatial proximity of the decay products provides a new set of tools that can be used to distinguish between jets from hadronic boson decays and jets originating from QCD multijet backgrounds.  In particular, since the angle between the decay products of a boson $V$ scales with $2m_V/p^V$, for large $p^V$, {\it jet substructure} techniques become powerful tools.  This leads to a tradeoff between using relatively pure leptonic decays and high-branching-ratio hadronic decays.  In some BSM theories, new particles similar to $W/Z$ bosons do not couple directly to leptons, so searching for hadronic decays of heavy particles is essential.

Jet substructure techniques developed to distinguish hadronically decaying $W$ and $Z$ bosons from QCD multijet background processes have become increasingly sophisticated.  A recent review is given in Ref.~\cite{Altheimer:2013yza}.  Both ATLAS~\cite{ATL-PHYS-PUB-2014-004} and CMS~\cite{Khachatryan:2014vla} have performed detailed comparisons of the various tagging variables and jet-grooming techniques with the overall conclusion that large QCD multijet suppression factors\footnote{$\mathcal{O}(1\%)$ QCD multijet efficiency at ~50\% signal efficiency.} are possible while maintaining acceptable levels of boson tagging efficiency.   Given a $W/Z$-boson tagger, a natural next step is to distinguish boson types.

There are several important possible applications of a boson-type tagger at the LHC.  First, a type tagger could enhance the SM physics program with $W$ and $Z$ bosons in the final state.  Measurements of this kind include the determination of the cross sections for $V$+jets, $VV$, and $t\bar{t}+V$.  Another important use of a boson-type tagger is in searches for flavour-changing neutral currents (FCNC).  Due to the Glashow--Iliopoulos--Maiani (GIM) mechanism~\cite{PhysRevD.2.1285}, FCNC processes in the SM are highly suppressed.   Many models of new physics predict large enhancements to such processes.  Both ATLAS and CMS have performed searches for FCNC~\cite{Aad:2012ij,Chatrchyan:2013nwa} of the form $t\rightarrow Zq$ in the leptonic channels, but these could be extended by utilizing the hadronic $Z$ decays as well.  FCNC processes mediated by a leptophobic $Z'$ such as $t\rightarrow Z'q$ may be detected only via hadronic type-tagging methods.   A third use of a boson-type tagger is to categorize the properties of new physics, if discovered at the LHC.  For instance, if a new boson were discovered as a hadronic resonance, a boson-type tagger could potentially distinguish a $W'(\rightarrow qq)$ from a $Z'(\rightarrow qq)$ (where mass alone may not be useful).  This is especially relevant for leptophobic new bosons, which could not be distinguished using leptonic decays.

Labelling jets as originating from a $W$ or $Z$ boson is less ambiguous than quark/gluon labelling.   A $W$ boson can radiate a $Z$ boson, just like a quark can radiate a gluon, but this is heavily suppressed for the former and not for the latter.  The radiation pattern of jets from $W$- and $Z$-bosons is less topology dependent because it is largely independent of the other radiation in the event as $W$ and $Z$ bosons are colour singlets.  Aside from the production cross section and subtle differences in differential decay distributions, the only features that distinguish between $W$ and $Z$ bosons are their mass, charge, and branching ratios.  Experimentally, this means that the only variables that are useful in discriminating between hadronic decays of $W$ and $Z$ bosons are those which are sensitive to these properties.  The three variables used in the analysis presented here are {\it jet mass}, sensitive to the boson mass, {\it jet charge}, sensitive to the boson charge, and a {\it b-tagging} discriminant which is sensitive to the heavy-flavour decay branching fractions of the bosons.  The application of a boson-type tagger in practice will be accompanied by the prior use of a boson tagger (to reject QCD multijet processes).  The type-tagger variables are largely independent of typical boson-tagger discriminants like $n$-subjettiness~\cite{Thaler:2010tr}, which rely on the two-prong hard structure of both the $W$ and $Z$ decays\footnote{See Sect.~\ref{sec:tau21} for details.}.  

This paper introduces a jet tagging method to distinguish between hadronically decay $W$ and $Z$ bosons at the LHC, and documents its performance with the ATLAS detector at $\sqrt{s}=8$ TeV.  The paper is organized as follows.  Section~\ref{sec:datasets} describes the simulated datasets used in constructing and evaluating the boson-type tagger.  Following a discussion of the differences between the properties of $W$ and $Z$ bosons in Sect.~\ref{sec:distinguish}, Sect.~\ref{sec:objs} defines the three discriminating variables.  The construction and performance of the tagger are detailed in Sect.~\ref{sec:perfm} and the sensitivity to systematic uncertainties is described in Sect.~\ref{sec:systs}.  The input variables are studied in a dataset enriched in boosted $W$ bosons in Sect.~\ref{sec:data}.   The paper ends with a discussion of possible uses of the tagger in Sect.~\ref{sec:outlook} and conclusions in Sect.~\ref{sec:conc}.

\section{Datasets}
\label{sec:datasets}

Two sets of Monte Carlo (MC) simulations are generated, one to study the tagger's $W$ versus $Z$ performance and the other to compare the tagger inputs for $W$ bosons with the data.  Simulations of hypothetical $W'~\rightarrow~WZ$ production and decay provide a copious source of boosted $W$ and $Z$ bosons whose $p_\text{T}$ scale is set by the mass of the $W'$ boson.  Such events are used to construct a tagger to separate hadronically decaying boosted $W$ and $Z$ bosons, as well as to evaluate its performance.   It is not possible to measure the performance directly in the data due to the lack of a pure sample of boosted, hadronically decaying $Z$ bosons, but the modelling of the tagger inputs can be studied using hadronically decaying $W$ bosons from $t\bar{t}$ events in the data.

A simulated sample of $W'$ bosons is generated with {\tt PYTHIA}~8.160~\cite{Sjostrand2008852} using the leading-order parton distribution function set (PDF) {\tt MSTW2008}~\cite{Lai:2010vv,Gao:2013xoa} and the {\tt AU2}~\cite{ATL-PHYS-PUB-2012-003} set of tunable parameters (tune) for the underlying event.  The baseline samples use {\tt PYTHIA} for the $2\rightarrow 2$ matrix element calculation, as well as $p_\text{T}$-ordered parton showers~\cite{Sjostrand:2004ef} and the Lund string model~\cite{string} for hadronization.  Additional samples are produced with {\tt HERWIG++}~\cite{Bahr:2008pv}, which uses angular ordering of the parton showers~\cite{Gieseke:2003rz}, a cluster model for hadronization~\cite{Webber:1983if}, as well as the {\tt EE3}~\cite{Gieseke:2012ft} underlying-event tune.   The $W$' differs from the SM $W$ boson only in its mass and the branching ratio $W'\rightarrow WZ$ is set to 100\%.  The $W$ and $Z$ bosons are produced with a mixture of polarizations, but the longitudinal polarization state dominates because $m_{W},m_{Z}\ll m_{W'}$.  In order to remove artifacts in the $p_\text{T}$ distributions of the $W$ and $Z$ bosons due to the generation of $W'$ particles with discrete masses, the $p_\text{T}^V$ spectra are re-weighted to be uniform in the range $200$~GeV$ <p_\text{T}^V<400$~GeV.  As is discussed in Sect.~\ref{sec:intro}, for $p_\text{T}>200$~GeV, a jet with large radius is expected to capture most of the $W$ or $Z$ boson decay products.  The range is truncated to $p_\text{T}<400$~GeV because hadronically decaying $W$ bosons can be probed with data in this $p_\text{T}$ range; there are too few events in the 8 TeV dataset for $p_\text{T}>400$ GeV.

Top-quark pair production is simulated using the next-to-leading-order (NLO) generator {\tt POWHEG-BOX} \cite{pow1,pow2,pow3} with the NLO PDF set {\tt CT10}~\cite{Gao:2013xoa} and parton showering from {\tt PYTHIA} {\tt 6}~\cite{pythia6}.   The single-top ($s$-, $t$-, and $Wt$-channel) backgrounds are modelled with {\tt POWHEG-BOX} and {\tt PYTHIA} {\tt 6}, as for the nominal $t\bar{t}$ simulation.  The PDF set {\tt CT10f4}~\cite{Lai:2010vv} is used for the $t$-channel and {\tt CT10} is used for the $s$- and $Wt$-channels.  For the $Wt-$channel, the `inclusive Diagram Removal' (DR) scheme is used for overlap with $t\bar{t}$~\cite{Frixione:2008yi}.   The $W$+jets and $Z$+jets backgrounds are modelled with {\tt ALPGEN} {\tt 2.1.4}~\cite{Mangano:2002ea}, {\tt PYTHIA} {\tt 6} and the {\tt CTEQ6L1} PDF set~\cite{Pumplin:2002vw}.  Dibosons are generated with {\tt HERWIG} {\tt 6.520.2}~\cite{Corcella:2002jc} using the {\tt CTEQ6L1} PDF set and the AUET2 tune~\cite{ATL-PHYS-PUB-2011-008}. Version {\tt 6.426} is used everywhere for {\tt PYTHIA} {\tt 6}, with the {\tt Perugia2011C} tune~\cite{Skands:2010ak}.  

Events are processed with a full simulation of the ATLAS detector and trigger~\cite{ATLAS2010} based on the $\mathrm{\tt Geant4}$~\cite{Geant} toolkit, and reconstructed using the same software as for the experimental data.  The average number of additional $pp$ collisions per bunch crossing (pileup interactions) was $20.7$ over the full 2012 run.  The effects of pileup are modelled by adding multiple minimum-bias events, which are simulated with {\tt PYTHIA} 8.{\tt 160}, to the generated hard-scatter events. The distribution of the number of interactions is then weighted to reflect the pileup distribution in the 2012 data.  A sample of $W$ bosons is selected
from data taken in 2012 at centre-of-mass energy of $\sqrt{s}=8\,{\rm TeV}$ from $t\bar{t}$ candidates as described in Sect.~\ref{sec:data}.

\section{Distinguishing a $Z$ boson from a $W$ boson}
\label{sec:distinguish}

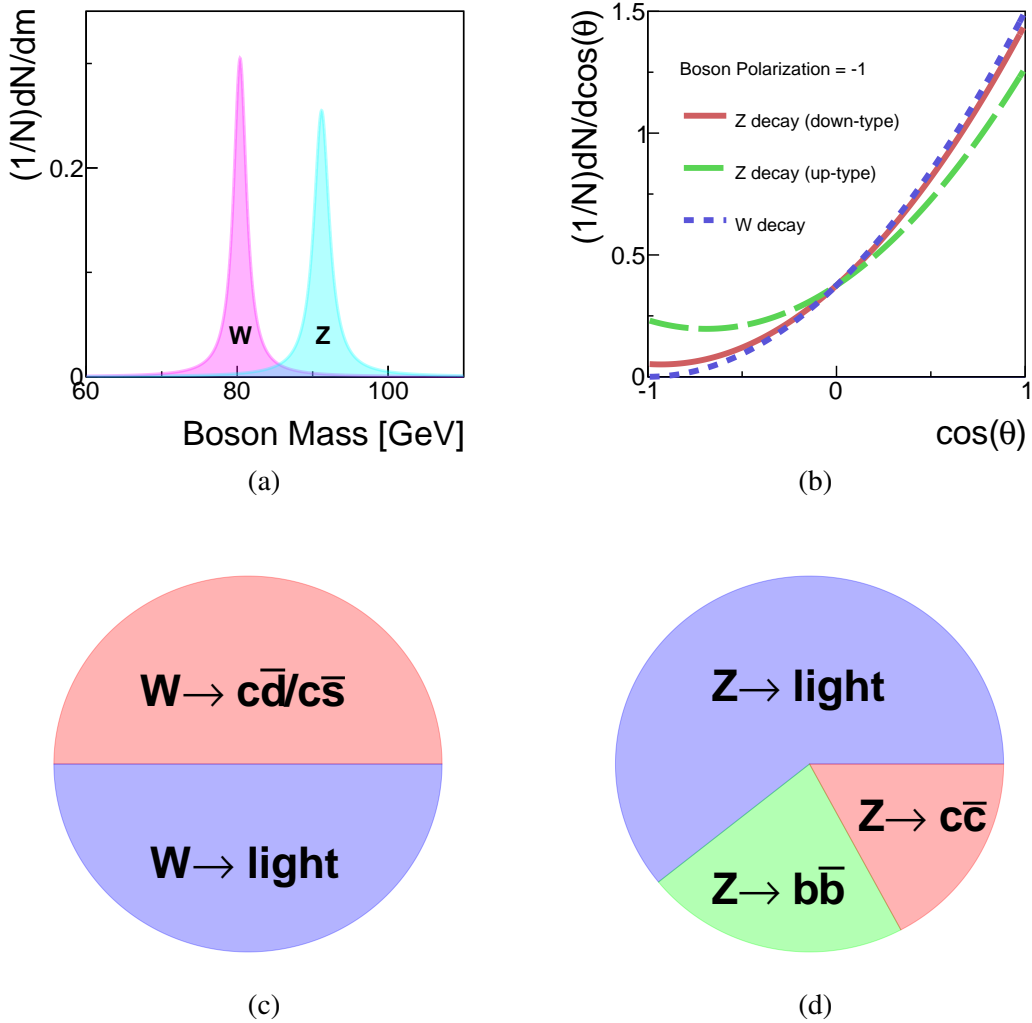
\begin{figure}[h!]
\begin{overpic}[width=0.95\textwidth]{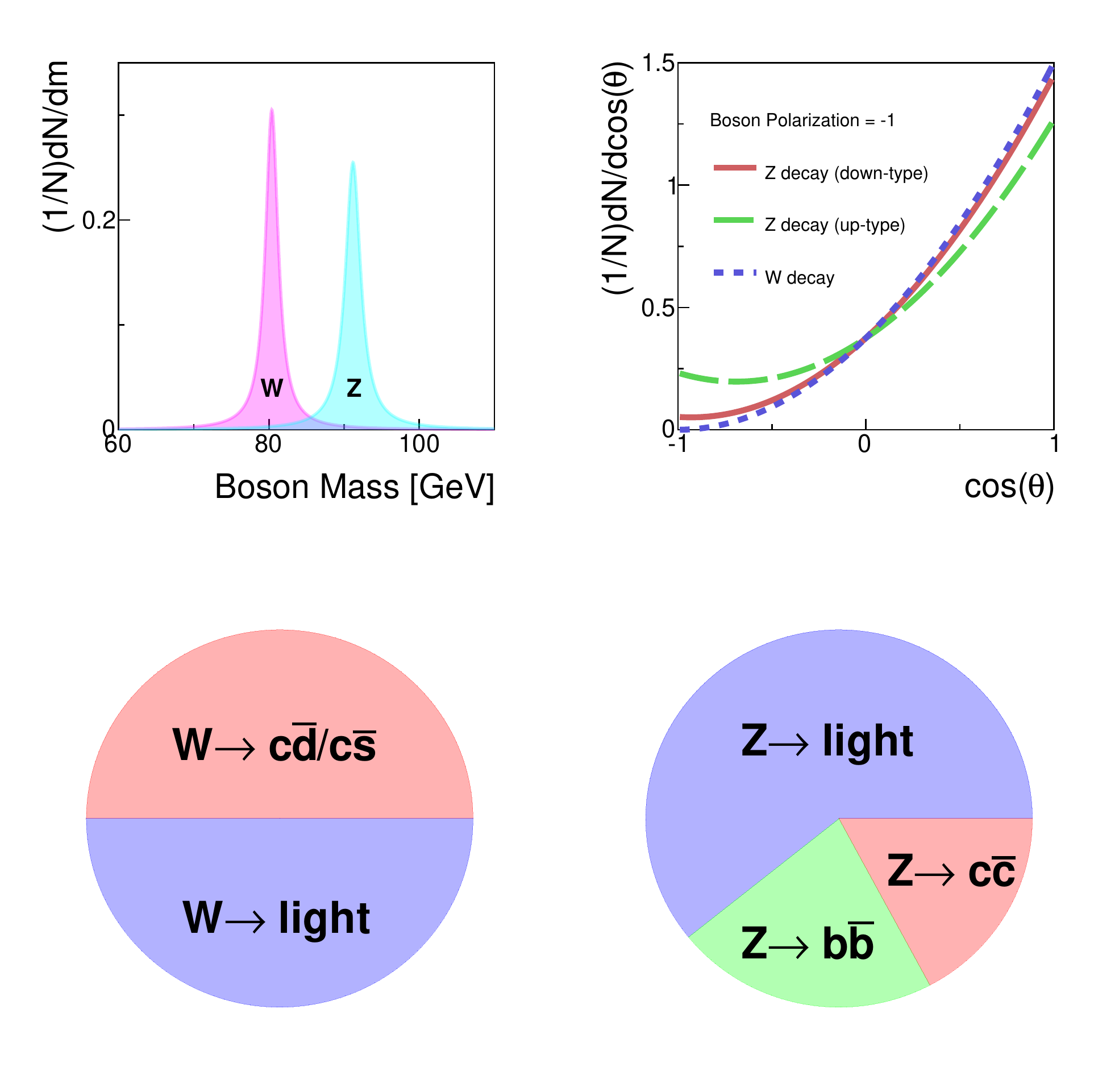}
\put(25,49){(a)}
\put(74,49){(b)}
\put(25,2){(c)}
\put(74,2){(d)}
\end{overpic}
\vspace{3mm}
\caption{(a) Breit--Wigner resonances for the $W$ (red) and $Z$ (blue) bosons, (b) angular distribution of the decay products of transversely polarized $W/Z$ bosons with respect to the spin direction in the boson rest frame, (c) hadronic branching fractions of the $W^+$ boson, and (d) of the $Z$ boson.  In (c) and (d), {\it light} stands for decay modes not involving $c$ and $b$ quarks.}
\label{fig:mass}
\end{figure}

Decays of $W$ or $Z$ bosons are characterized by the boson's mass and coupling to fermions.   The mass difference between the $W$ and $Z$ boson is about $10$~GeV and if produced from a hard scatter or the decay of a heavy enough resonance, both bosons are produced nearly on-shell since the width $\Gamma_V=2.1$ ($2.5$)~GeV is much less than the mass $m_V=80.4$ ($91.2$)~GeV for $W$ ($Z$) bosons~\cite{pdg}.  The Breit--Wigner resonance curves for $W$ and $Z$ bosons are shown in Fig.~\ref{fig:mass}(a).  The separation between the curves is a theoretical limit on how well mass-sensitive variables can distinguish between $W$ and $Z$ bosons.  For hadronic boson decays, the mass peaks measured with jets are broader. This is because the jet-clustering algorithm for final-state hadrons loses particles at large angles to the jet axis and includes extra particles from the underlying event and pileup.

The generic coupling of a boson $V$ to fermions is given by $g_\text{V} \gamma_\mu[c_\text{V}-\nobreak c_\text{A}\gamma_5]$, where $g_\text{V}$ is a boson-dependent overall coupling strength, and $c_\text{V}$ and $c_\text{A}$ are the vector and axial-vector couplings, respectively.   The $W$ boson couples only to left-handed fermions so $c_\text{V}=c_\text{A}=1$ with $g_W\propto kN_\text{C}G_\text{F}m_W^3|V_{ij}|^2$, where $G_\text{F}$ is the Fermi coupling constant, $V_{ij}$ is a Cabibbo--Kobayashi--Maskawa (CKM) matrix element~\cite{PhysRevLett.10.531,Kobayashi:1973fv}, $k$ represents higher-order corrections, and $N_\text{C}=3$ for the three colours of quarks and $N_\text{C}=1$ for leptons.  The CKM matrix is nearly diagonal so $W^+\rightarrow u\bar{d}$ and $W^+\rightarrow c\bar{s}$ are the dominant decay modes.  Small off-diagonal elements contribute to the other possible decay modes, and the overall hadronic branching ratios are approximately $50\%$ for $W\rightarrow cX$ and $50\%$ for $W\rightarrow \text{light-quark pairs}$.  The $W$ boson has electric charge $\pm 1$ in units of the electron charge, so by conservation of charge, its decay products have the same net charge.   The scalar sum of the charge of all the final-state hadrons originating from a $W$ boson decay is not infrared safe (directly sensitive to the non-zero detection threshold), so there are limits to the performance of charge tagging dictated by the energy threshold placed on charged particles in the event reconstruction.

In contrast to $W$ boson decays, $Z$ bosons decay to both the left- and right-handed fermions.  The partial width for $Z\rightarrow f\bar{f}$ is proportional to $kN_\text{C}G_\text{F}m_Z^3[c_\text{V}^2+c_\text{A}^2]$.  The factors $c_\text{V}$ and $c_\text{A}$ are slightly different for up- and down-type fermions.  The $b\bar{b}$ branching ratio is 22\%, the $c\bar{c}$ branching ratio is $17\%$ and the sum of the remaining branching ratios is $61\%$.  $W$ boson decays to $b$-quarks are highly suppressed by the small CKM matrix elements $V_{cb}$ and $V_{ub}$, so that identifying $b$-hadron decays associated with a hadronically decaying boson is a powerful discriminating tool.  Branching ratios are plotted in Fig.~\ref{fig:mass}(d) for $Z$ decays to light quarks, $c$-quarks, and $b$- quarks, and in Fig.~\ref{fig:mass}(c) for the $W$ boson decays to light quarks and $c$-quarks.

Since the coupling structure is not identical for $W$ and $Z$ bosons, the total decay rates differ, and the angular distributions of the decay products also differ slightly.  However, even at parton level without any combinatoric noise, the differences in the angular distributions are subtle.   There is no difference for the two bosons with longitudinal polarization because the distributions for right- and left-handed fermions are the same.  The distributions are different for right- and left-handed fermions for transversely polarized $W$ and $Z$ bosons, as shown in Fig.~\ref{fig:mass}(b).  The relative contribution of left- and right-handed components for the $Z$ decays depends on the quark flavour; for up-type quarks the relative contribution from right-handed fermions is $15\%$ while it is only $3\%$ for down-type quarks.  In $t\bar{t}$ decays, the fraction of longitudinally polarized $W$ bosons (ignoring the $b$-quark mass) is $m_t^2/(m_t^2+2m_W^2)\sim 0.7$.  In contrast, the boson is mostly transversely polarized in inclusive $V$+jets events.  Any discrimination shown in Fig.~\ref{fig:mass}(b) is diluted by the longitudinal polarization, combinatorics, non-perturbative effects, and detector reconstruction, so angular distributions are not considered further in this paper\footnote{The impact of polarization on distinguishing boosted $W$ boson jets from QCD multijets has been studied in Ref.~\cite{Khachatryan:2014vla}.  There are small differences in performance between transversely and longitudinally polarized bosons, but any differences are less relevant for $W$ versus $Z$ tagging where the angular distributions are identical for longitudinally polarized bosons and only slightly differ for transversely polarized bosons.}.

\section{Definitions of reconstructed objects}
\label{sec:objs}

\begin{figure}[h!]
 \centering
\subfloat[]{\includegraphics[width=0.33\textwidth]{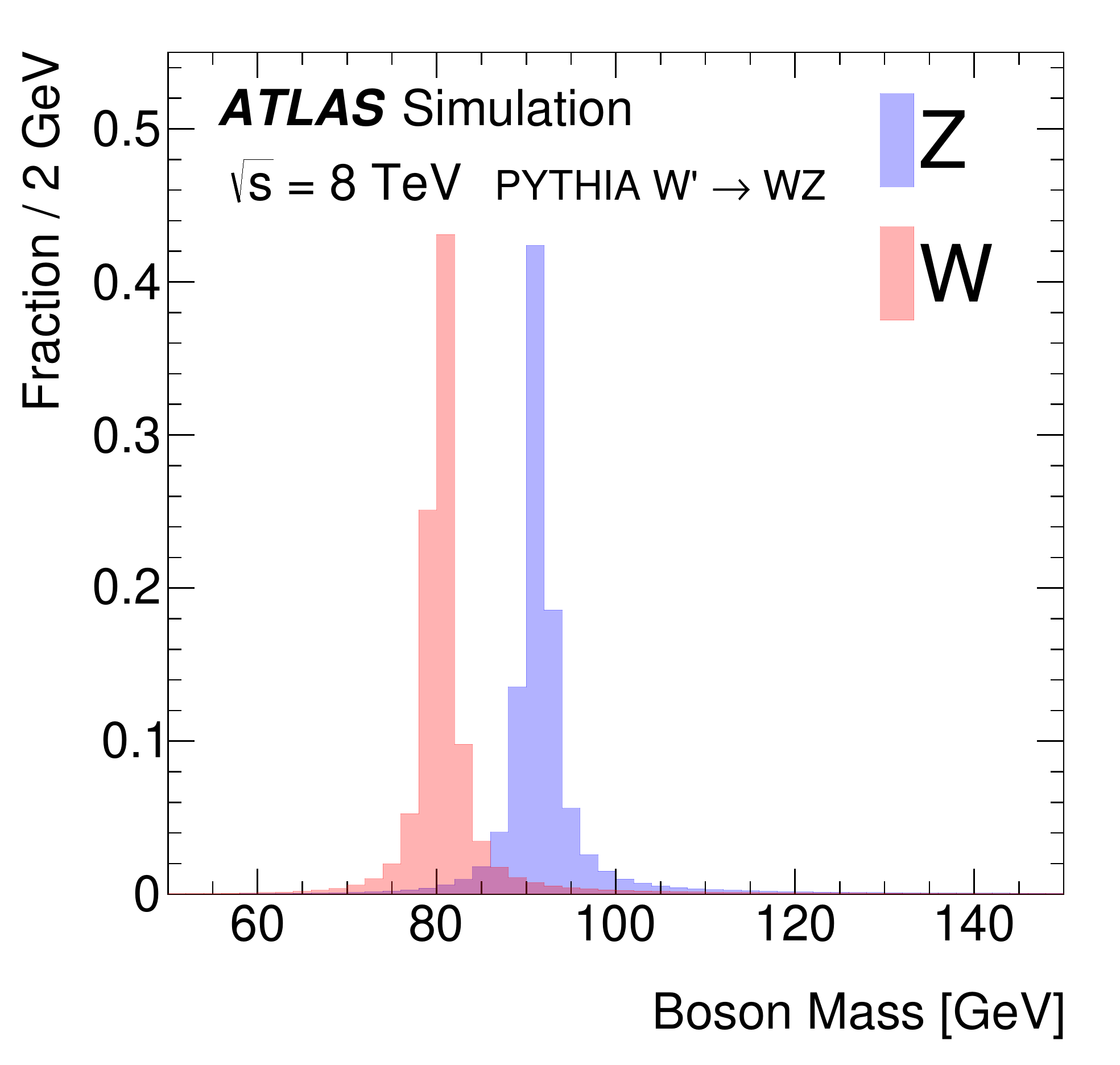}}
\subfloat[]{\includegraphics[width=0.33\textwidth]{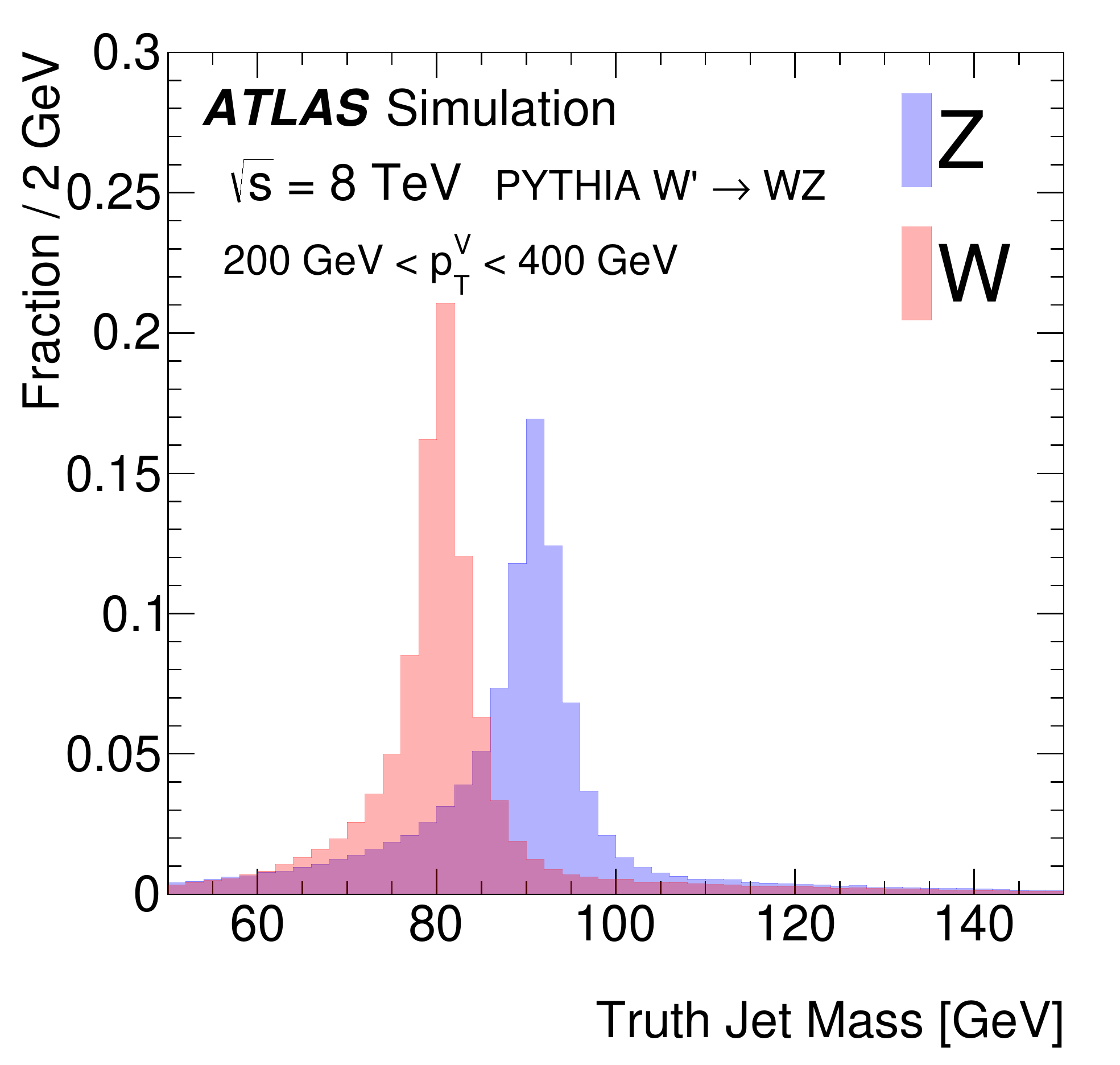}}
\subfloat[]{\includegraphics[width=0.33\textwidth]{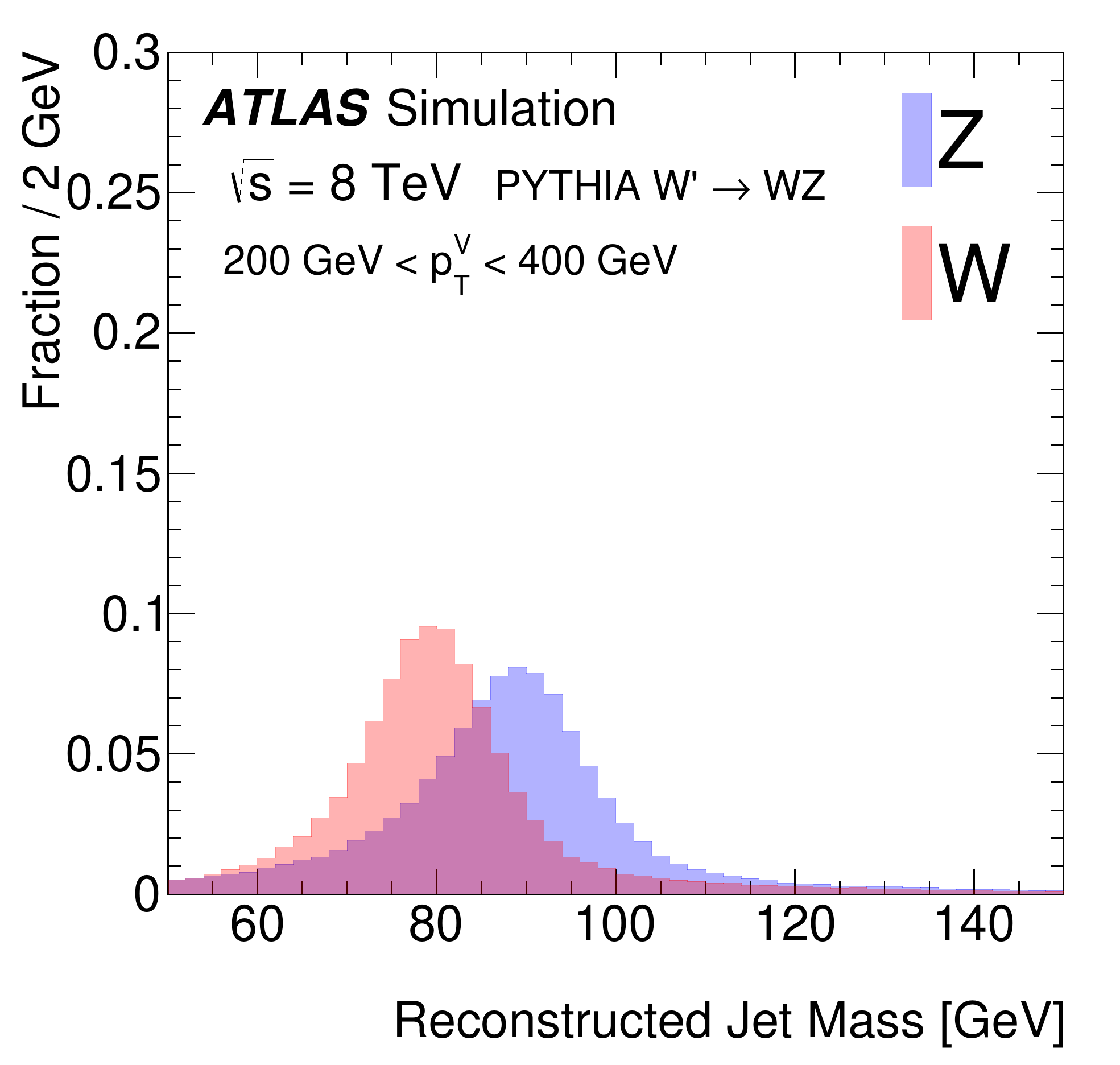}}
\caption{(a) The boson mass at generator level, (b) `truth' jet mass (at particle level) after parton fragmentation, and (c) reconstructed jet mass distributions.  The left plot has a different vertical scale than the right two plots and also has no $p_\text{T}$ requirement.}
\label{fig:fat_Jmv0}
\end{figure}

ATLAS is a multi-purpose particle detector~\cite{atlas-detector} with nearly $4\pi$ coverage in solid angle.\footnote{
ATLAS uses a right-handed coordinate system with its origin at the nominal interaction point (IP) in the centre of the detector and the $z$-axis along the beam pipe.  The $x$-axis points from the IP to the centre of the LHC ring, and the $y$-axis points upward.  Polar coordinates $(r,\phi)$ are used in the transverse plane, $\phi$ being the azimuthal angle around the beam pipe.  The pseudorapidity is defined in terms of the polar angle $\theta$ as $\eta=-\mathrm{ln}\tan(\theta/2)$.  Transverse momentum and energy are defined in the $x$--$y$ plane as $p_\text{T}=p\cdot \sin(\theta)$ and $E_\text{T}=E\cdot\sin(\theta)$.
}  The energy of the hadronic decay products of boosted bosons is measured by a system of calorimeters.  The electromagnetic calorimeter consists of a Pb/liquid-argon sampling calorimeter split into barrel ($|\eta|$ < 1.5) and endcap ( $1.5< |\eta| < 3.2$) sections.  The hadronic calorimetry is provided by a barrel steel/scintillating-tile calorimeter ($|\eta| < 1.7$) and two endcap Cu/liquid-argon sections ($1.5 < |\eta| < 3.2$).  Finally, the forward region ($3.1 < |\eta| < 4.9$) is covered by a liquid-argon calorimeter with Cu (W) absorber in the electromagnetic (hadronic) section.  Energy depositions are grouped into topological calorimeter-cell clusters~\cite{TopoClusters} and then calibrated using the local cluster weighting algorithm~\cite{EndcapTBelectronPion2002,Barillari:2009zza}.  Jets are formed from clusters using two different jet algorithms.  {\it Small-radius jets} are built with the anti-$k_t$ algorithm with jet radius parameter $R=0.4$~\cite{Cacciari:2008gp}.  {\it Large-radius jets} are formed using the anti-$k_t$ algorithm with $R=1.0$ and then trimmed~\cite{Krohn:2009th} by re-clustering the jet constituents with the $k_t$ algorithm using $R=0.3$ and removing the constituents with $p_\text{T}$ less than 5\% of the original jet $p_\text{T}$.  Both the small- and large-radius jets are further calibrated to account for the residual detector response effects.  For small-radius jets, this is a $p_\text{T}$- and $\eta$-dependent energy calibration, plus a correction to mitigate the contribution from additional $pp$ collisions and to suppress jets from these additional collisions~\cite{areasATLAS}.  In addition to $p_\text{T}$- and $\eta$-dependent energy corrections, large-radius jets $J$ have a calibrated {\it jet mass}:

\begin{align}
m_J^2=\Bigg(\sum_{j\in J} E_j\Bigg)^2-\Bigg(\sum_{j\in J} \vec{p}_j\Bigg)^2,
\end{align}

\noindent where $E_j$ is the energy of cluster $j$ and $\vec{p}_j$ is a vector with magnitude $E_j$ and direction $(\phi_j,\eta_j)$.  The jet mass calibration depends on the calibrated jet energy and on the jet $\eta$~\cite{Aad:2013gja}.  When a $W$ or $Z$ boson is produced with large enough momentum, its decay products are collimated.  When $2m_V/p^V\sim 1$, an $R=1.0$ trimmed jet captures a large fraction of the decay products and the jet-mass scale is set by $m_V$.  Since the $W$ and $Z$ boson masses differ by about $10$~GeV, the jet mass can be used to discriminate between these two particles.  The distributions of the boson masses and jet masses for hadronically decaying $W$ and $Z$ bosons are shown in Fig.~\ref{fig:fat_Jmv0}.  The particle-level (`truth') jet mass is constructed from stable particles in the MC simulation ($c\tau > 10$ mm), excluding neutrinos and muons, clustered with the same jet algorithm as for calorimeter-cell clusters.  The QCD processes that govern the formation of stable particles from the $W$ and $Z$ decay products create a broad distribution of jet masses even without taking into account detector resolution.  Constructing the jet mass from calorimeter-cell clusters further broadens the distribution.  The jet-mass resolution (physical $\oplus$ detector) is large compared to the natural width of the $W$ and $Z$ bosons and comparable to the difference in their masses.  For example, the standard deviation of the detector resolution $\sigma(m^\text{reco jet}/m^\text{truth jet})$ is approximately 10\%.  The jet-mass variable nevertheless has some discriminating power.

\begin{figure}[h!]
\begin{center}
\includegraphics[width=0.65\textwidth]{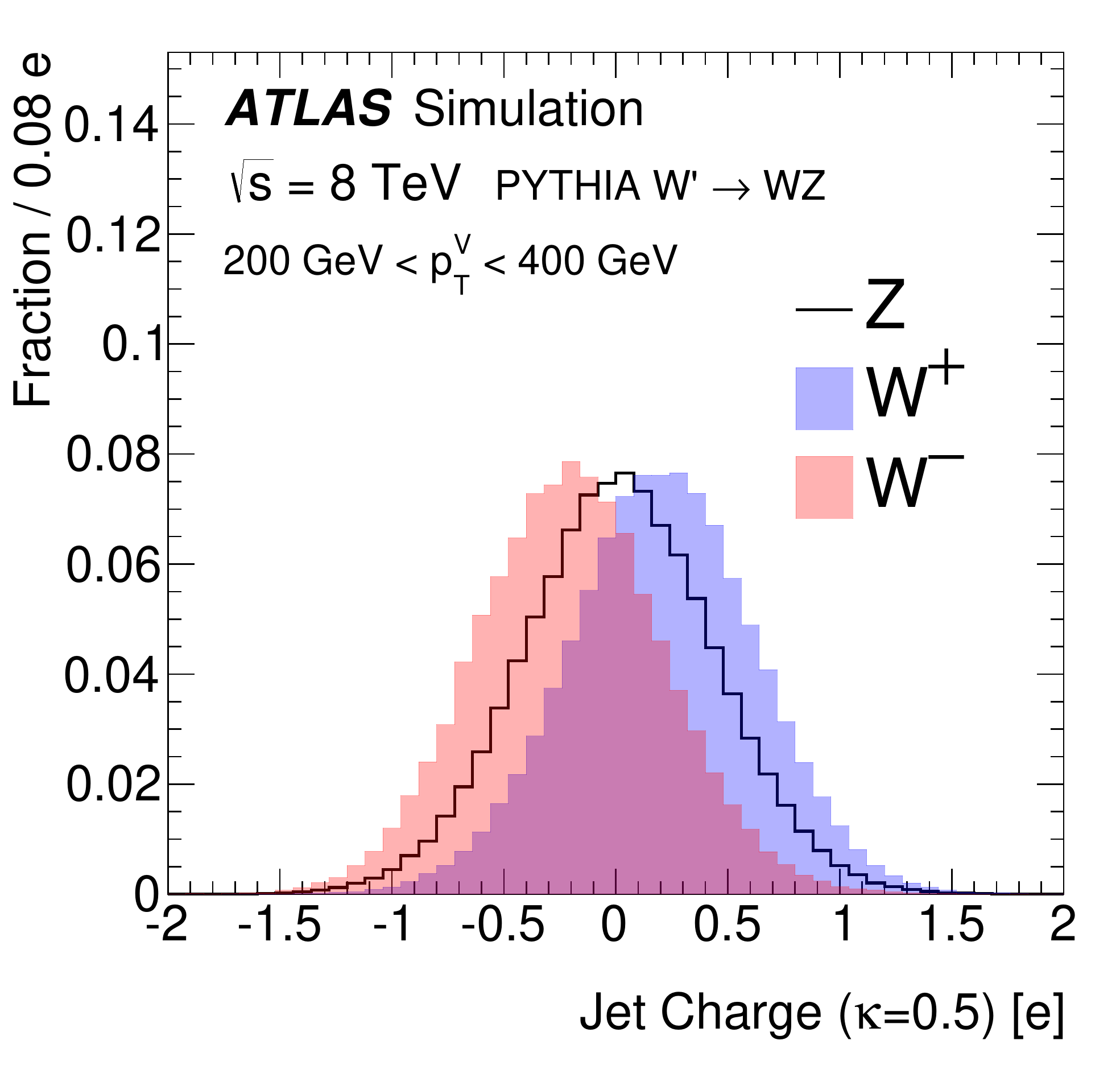}
\caption{The jet charge distribution for jets originating from $W^\pm$ and $Z$ bosons in simulated $W'$ decays.  Each distribution is normalized to unity.  The parameter $\kappa$ controls the $p_\text{T}$-weighting of the tracks in the jet charge sum.}
\label{fig:fat_Jqv0}
\end{center}
\end{figure}

The momentum and electric charge of particles traversing the detector contain information about the charge of their parent boson.  The tracks of charged particles are measured in a 2 T axial field generated by a solenoid magnet which surrounds the inner detector (ID) consisting of silicon pixels, silicon micro-strips, and a transition radiation tracking detector.  Charged-particle tracks are reconstructed from all three ID technologies with a full coverage in $\phi$,  $|\eta|<2.5$ and $p_\text{T}>400$ MeV.  The charge $q$ of a track is determined as part of the reconstruction procedure, which uses a fit with five parameters: the transverse and longitudinal impact parameters, $\phi$, $\theta,$ and $q/p$, where $p$ is the track  momentum.   To suppress the impact of pileup, tracks are required to originate from the primary collision vertex, which is defined as the vertex with the largest $\sum p_\text{T}^2$ computed from associated tracks.  Additionally, tracks must satisfy a very loose quality criterion for the track fit $\chi^2$ per degree of freedom, which must be less than three.  Tracks are associated with jets using ghost association~\cite{area}.  The charge of tracks associated with a jet is sensitive to the charge of the initiating parton.  In order to minimize the fluctuations due to low-$p_\text{T}$ particles, the {\it jet charge} is calculated using a $p_\text{T}$-weighting scheme~\cite{Feynman1978}:

\begin{align}
  \label{chargedef}
  Q_J = \frac{1}{({p_\text{T,J}})^\kappa}\sum_{i\in \text{\bf Tracks}} q_i\times (p_\text{T}^i)^\kappa,  
  \end{align} 

\noindent where $\text{\bf Tracks}$ is the set of tracks with $p_\text{T}>500$ MeV associated with jet $J$, $q_i$ is the charge (in units of the electron charge) determined from the curvature of track $i$ with associated $p_\text{T}^i$, $\kappa$ is a free parameter, and ${p_\text{T,J}}$ is the transverse momentum of the jet measured in the calorimeter.  The calorimeter energy is used in the denominator to determine $p_\text{T}$ instead of the sum of track momenta to account for the contribution from neutral particles.   Dedicated studies have shown that $\kappa=0.5$ is generally best for determining the charge of partons from the jets they produce~\cite{ATLAS-CONF-2013-086}.   The distributions of the jet charge for jets initiated by $W^+,W^-$ and $Z$ bosons are shown in Fig.~\ref{fig:fat_Jqv0}.  There is an observable separation between positive and negative $W$ bosons.  The expected charge composition of a $W$ sample is process dependent.  For example, there are more $W^+$ than $W^-$ bosons in inclusive $W'$ production because of the initial charge asymmetry of quarks in the proton resulting in more $W'{}^{+}(\rightarrow W^+Z)$ than $W'{}^{-}(\rightarrow W^-Z)$.  The discrimination between $Z$ bosons and a near even mixture of $W^\pm$ is greatly diminished with respect to e.g. $Z$ versus $W^+$.   In that case charge sensitive variables are not very useful for the tagger and so all results are shown also without such variables.  In a variety of physics processes, the charge of the hadronically decaying $W$ boson is known from other information in the event.  For example, in searches for FCNC effects in $t\bar{t}$ events with one leptonically decaying $W$ boson, the charge of the lepton is opposite to the charge of the hadronically decaying $W$ boson.  Henceforth, only $W^+$ bosons are used for constructing the boson-type tagger; the results are the same for $W^-$ bosons.

The tracks from charged particles can be used further to identify the decays of certain heavy-flavour quarks inside jets due to the long $b$-hadron lifetime.  This is useful for boson-type tagging because the $Z$ boson couples to $b\bar{b}$ while decays of the $W$ boson to $b$-quarks are highly suppressed and can be neglected.  ATLAS has commissioned a $b$-tagging algorithm called MV1 (defined in Refs.~\cite{btag1} and~\cite{btag2}) which combines information about track impact-parameter significance with the explicit reconstruction of displaced $b$- and $c$-hadron decay vertices.   The boson-type tagger presented here uses multiple bins of the MV1 distribution simultaneously.  Five bins of MV1 are defined by $b$-tag efficiencies (probability to tag a $b$-quark jet as such) of 0\%--50\%, 50\%--60\%, 60\%--70\%, 70\%--80\%, and 80\%--100\% as determined in simulated $t\bar{t}$ events.  A lower $b$-tag efficiency leads to higher light-quark jet rejection.   The five $b$-tagging efficiency bins are exclusive and MV1 is constructed as a likelihood with values mostly between zero and one (one means more like a $b$-jet).  For example, a $100\%$ $b$-tagging efficiency corresponds to a threshold of MV1 $>0$ and an $80\%$ $b$-tagging efficiency corresponds to a threshold value of MV1 $>z$ for $z\ll1$.  The 80\%--100\% $b$-tag efficiency bin then corresponds to jets with an MV1 value between $0$ and $z$.  Constructed in this way, the fraction of true $b$-jets inside an efficiency bin $x$\%--$y$\% should be $(y-x)\%$.  

\begin{figure}[h!]
\begin{center}
\includegraphics[width=0.45\textwidth]{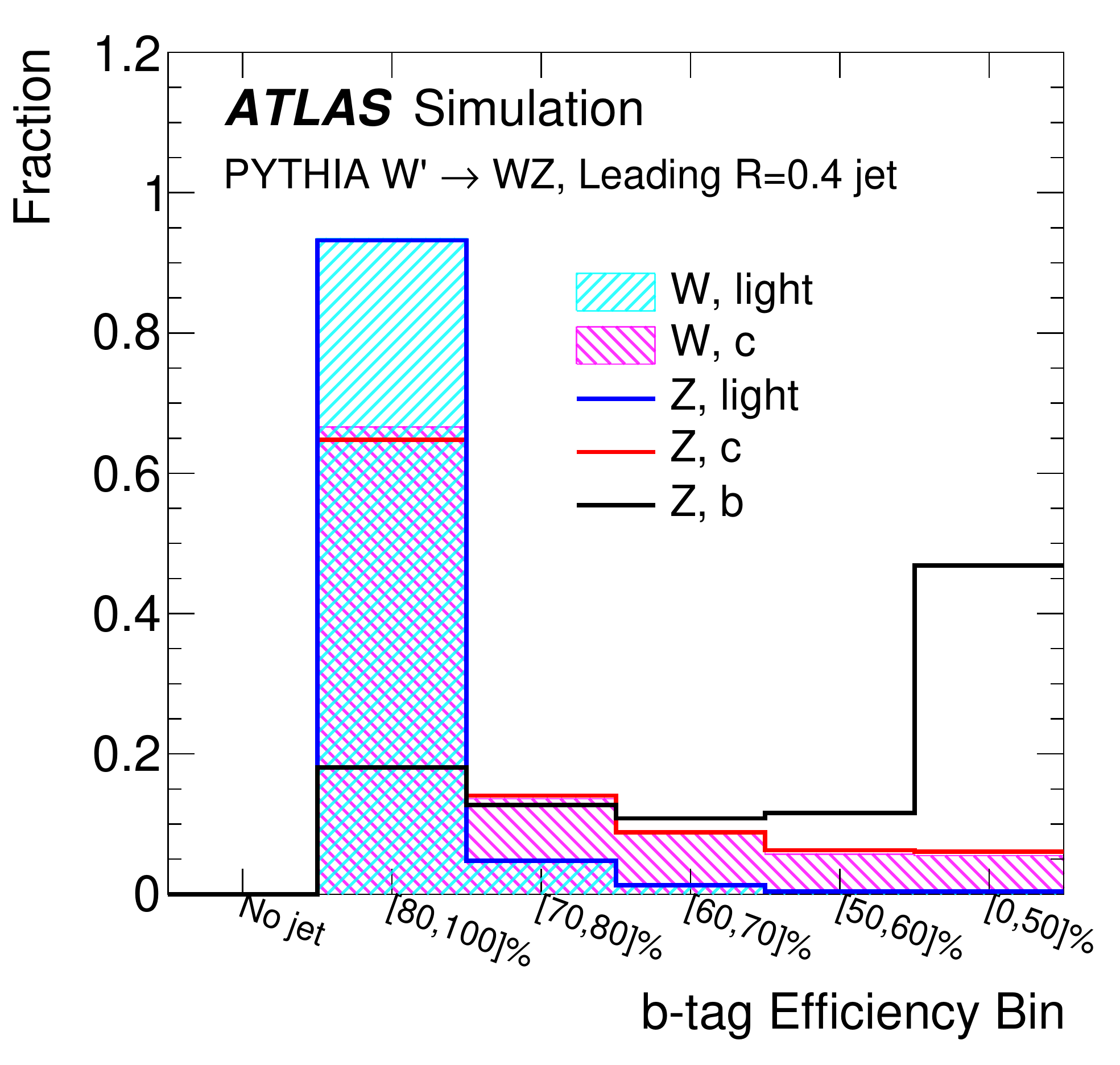}\includegraphics[width=0.45\textwidth]{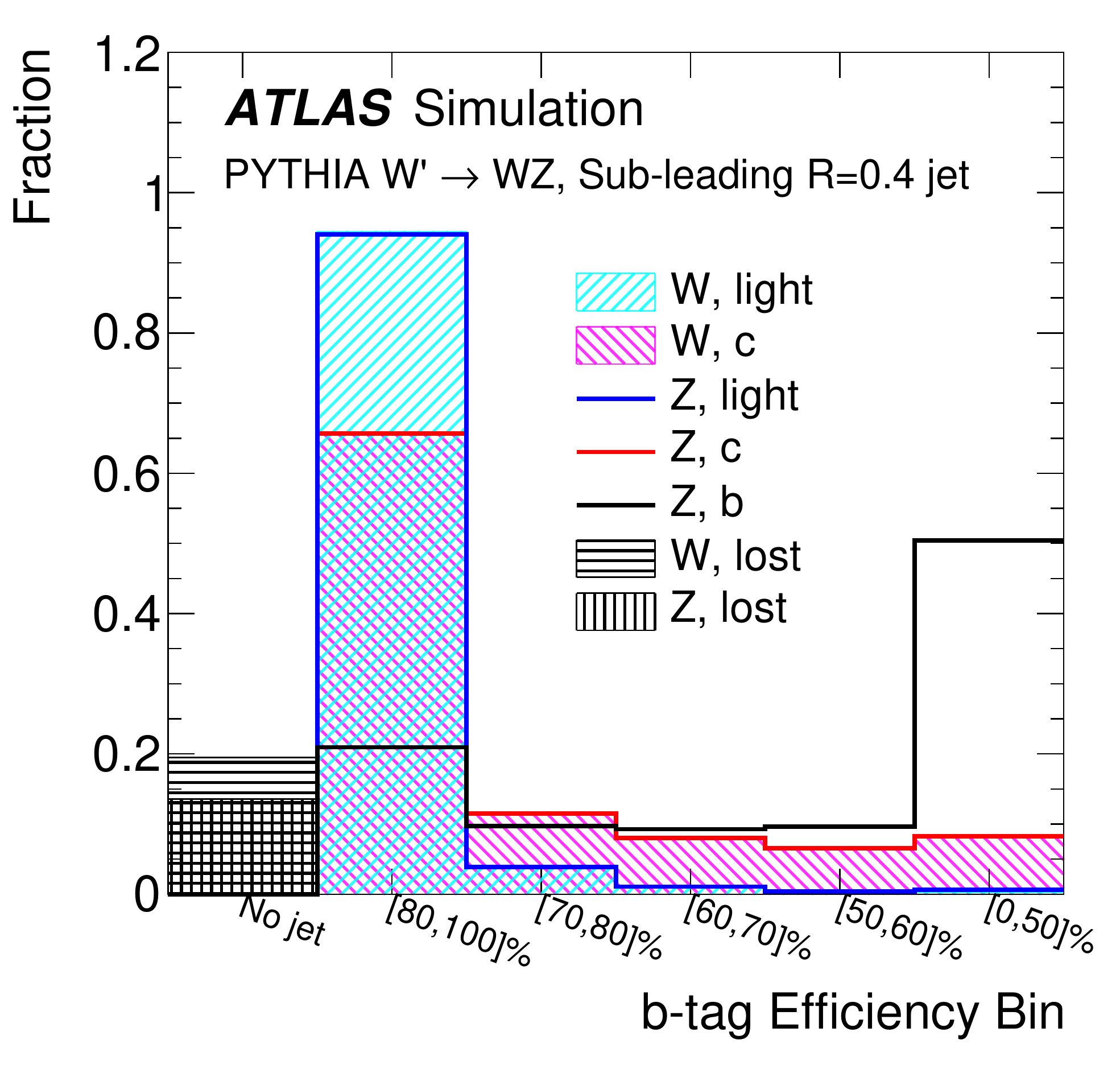}
\caption{The efficiency-binned MV1 distribution for small-radius jets associated with large-radius jets resulting from $W$ and $Z$ boson decays.  The left (right) plot shows the leading (sub-leading) small-radius jet MV1 distribution.  The bins correspond to exclusive regions of $b$-jet efficiency.  As such, the bin content of the black line ($b$-tagging for $b$-jets) should be proportional to the size of the efficiency window: about 50\% for the rightmost bin, 10\% for the three middle bins and 20\% for the second bin.}
\label{fig:fat_Jbv0}
\end{center}
\end{figure}

Small-radius jets are matched to a large-radius jet by geometric matching\footnote{In the definition of jets, $R$ is the characteristic size in $(y,\phi)$ and the rapidity $y$ is used in the jet clustering procedure, whereas geometrical matching between reconstructed objects is performed using $(\Delta R)^2=(\Delta \phi)^2+(\Delta\eta)^2$, where $\eta$ is the pseudorapidity.} ($\Delta{R}<1.0$).   Of all such small-radius jets, the two leading ones are considered.  There are thus 30 possible bins of combined MV1 when considering the leading and sub-leading matched small-radius jet.  The number of bins is $25$ from the $5\times 5$ efficiency-binned MV1 distributions in addition to five more for the case in which there is no second small-radius jet matched to the large-radius jet.  The distribution for the efficiency-binned MV1 variable for the leading and sub-leading matched small-radius jets is shown for  $W$ and $Z$ bosons in Fig.~\ref{fig:fat_Jbv0}.  The flavour of a small-radius jet is defined as the type of the highest energy parton from the parton shower record within $\Delta R < 0.4$.   As expected, a clear factorization is seen in Fig.~\ref{fig:fat_Jbv0} -- the MV1 value depends on the flavour of the small-radius jet and not the process that created it.  This means that $c$-jets from $W$ decays have the same MV1 distribution as $c$-jets from $Z$ decays; the same is true for light jets.  Small-radius jets originating from $b$-hadron decays tend to have a larger value of MV1, which means they fall in a lower efficiency bin.  Small-radius jets not originating from $b$- or $c$-decays are called light jets and are strongly peaked in the most efficient bin of MV1.  There is always one small-radius jet matched to the large-radius jet, but about $20\%$ of the time there is no sub-leading small-radius jet with $p_\text{T}>25$~GeV matched to the large-radius jets.  These cases are all predicted to originate from light-quark decays of the $W$ and $Z$ bosons.

\section{Tagger performance}
\label{sec:perfm}

\begin{figure}[h!]
\begin{center}
 \centering
\subfloat[]{\includegraphics[width=0.45\textwidth]{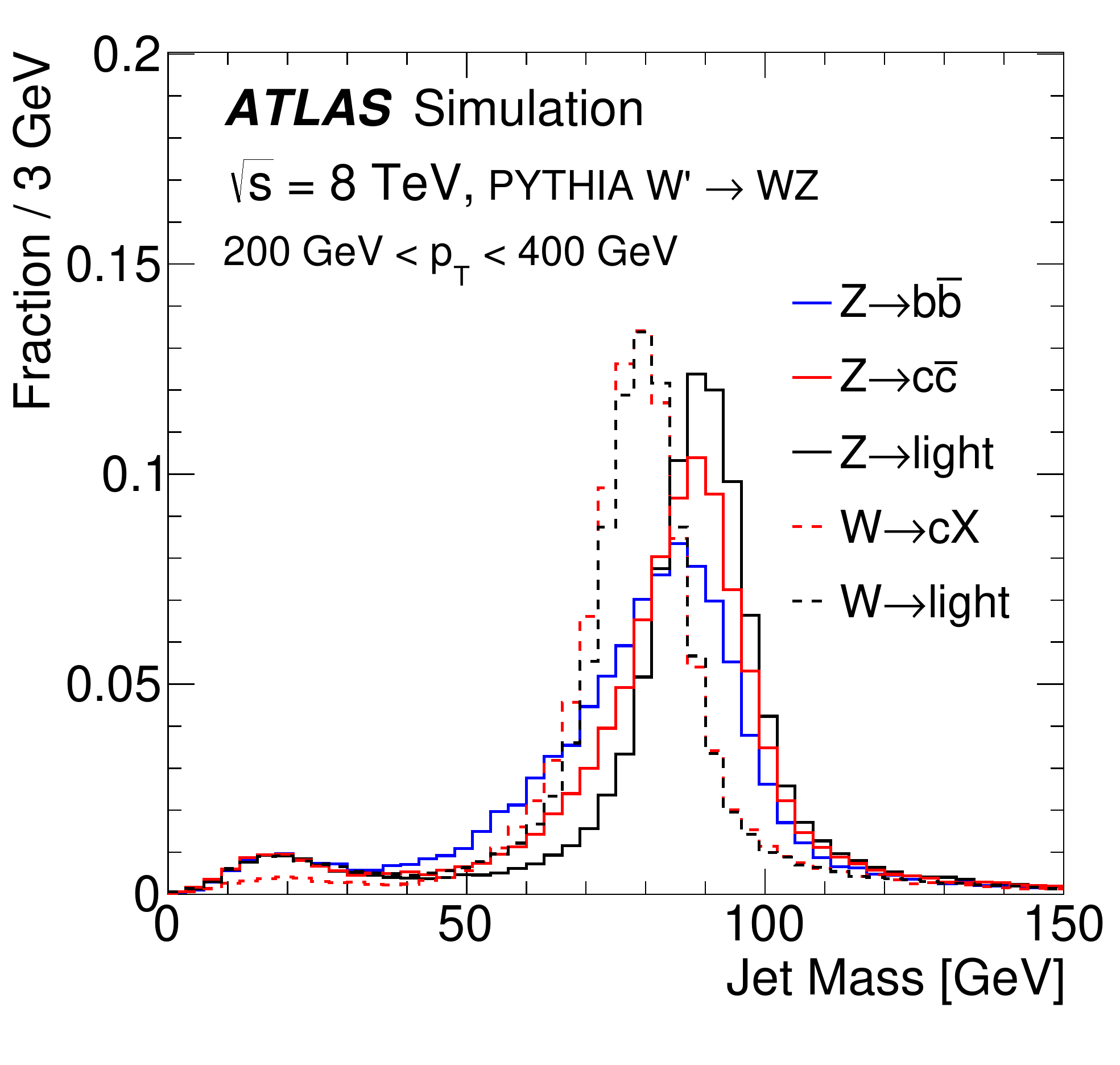}}
\subfloat[]{\includegraphics[width=0.45\textwidth]{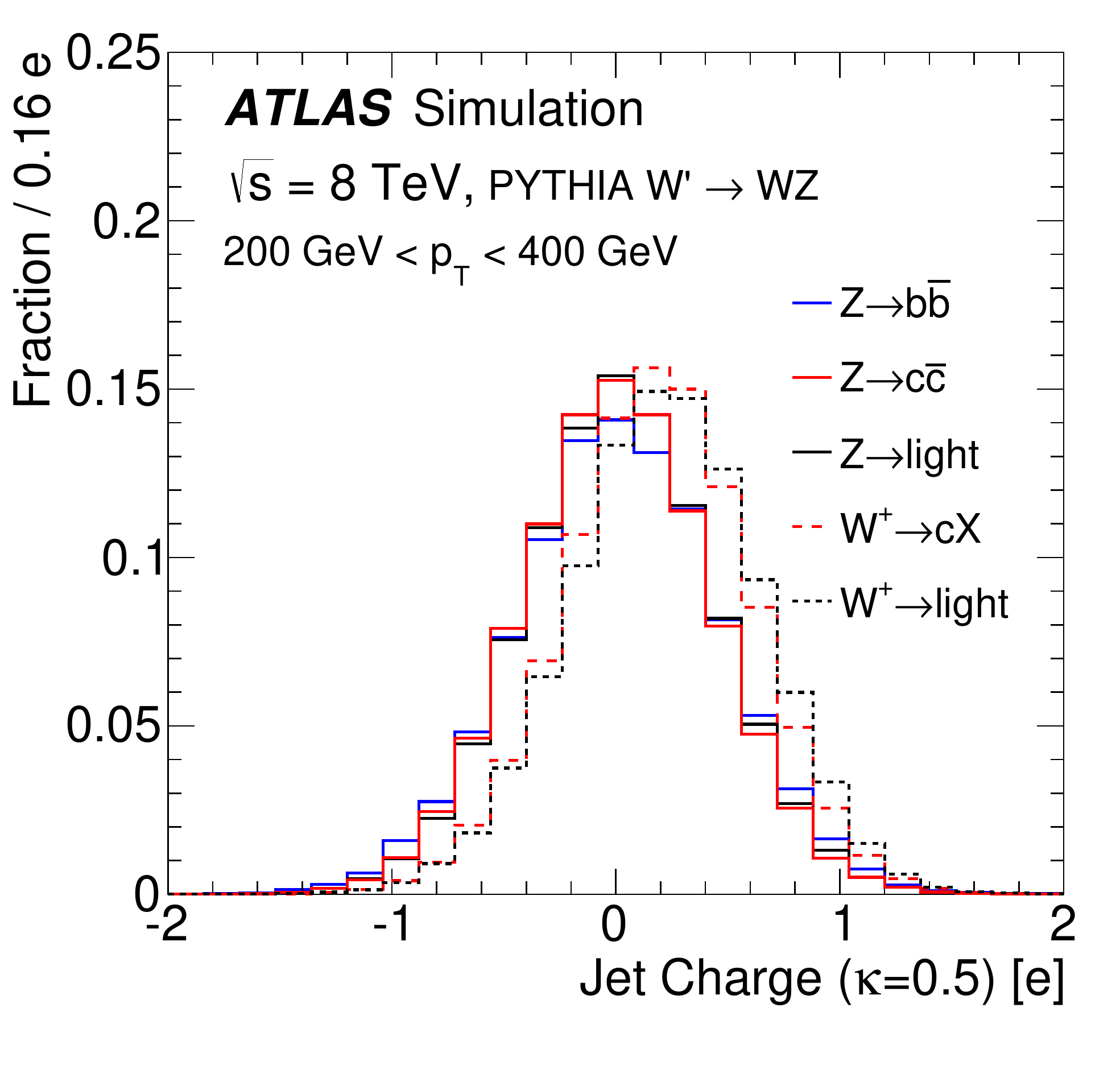}}
\caption{(a) The jet mass $p(M|\mathcal{F},V)$ and (b) jet charge $p(Q|\mathcal{F},V)$ templates conditioned on the flavour $\mathcal{F}$ of the boson $V$ decay for jets with $200$ GeV $<p_\text{T}<400$ GeV.   The solid lines are for $Z$ boson decays and the dashed lines are for $W$ boson decays.}
\label{fig:templates}
\end{center}
\end{figure}

The optimal multivariate tagger combining jet mass, jet charge, and the MV1 of matched small-radius jets is constructed from a three-dimensional (3D) likelihood ratio.  For $N$ bins each of jet mass and jet charge, as well as $30$ combined MV1 bins, the 3D likelihood ratio would have $30\times N^2$ total bins.  Populating all of these bins with sufficient MC events to produce templates for the likelihood ratio requires an unreasonable amount of computing resources, especially for the high-efficiency bins of combined MV1. Estimating the 3D likelihood as the product of the 1D marginal distributions, where all variables but the one under consideration are integrated out, is a poor approximation for jet mass and combined MV1 due to the correlation induced by the presence of semileptonic $b$-decays, which shift the jet mass to lower values due to the presence of unmeasured neutrinos.\footnote{The muons from semileptonic decays are added back to the jet using a four-momentum sum.  Muons are measured by the combination of a dedicated muon spectrometer with its own toroidal magnetic field outside the calorimeters, and the inner detector.  Adding back the muon has a negligible impact on the inclusive mass distribution due to the semileptonic branching ratios and lepton identification requirements.  For details about the muon reconstruction and selection, see Sect.~\ref{sec:data} (the only difference here is that the isolation is not applied).}  It is still possible to use a simple product by noting that all three tagger inputs are independent when the flavour of the decaying boson has been determined.  Thus, for each possible boson decay channel, templates are built for the jet mass, the jet charge, and the efficiency-binned MV1 distributions.  For a particular decay flavour, the joint distribution is then the product of the individual distributions.  Summing over all hadronic decay channels then gives the full distribution.  To ease notation, the efficiency-binned MV1 is denoted $B=(B_\text{lead},B_\text{sub-lead})$.  The distribution for $B_\text{lead}$ ($B_\text{sub-lead}$) is shown in the left (right) plot in Fig.~\ref{fig:fat_Jbv0}. Symbolically, for decay flavour channel $\mathcal{F}$, mass $M$, charge $Q$, and efficiency-binned MV1 $B$, the likelihood is given by:

\begin{align}
\label{eq:like}
p(M,Q,B|V)=\sum_\mathcal{F} \Pr(\mathcal{F}|V)p(M|\mathcal{F},V)p(Q|\mathcal{F},V)\Pr(B|\mathcal{F},V),
\end{align}

\noindent where\footnote{The symbol $p$ denotes a probability density whereas $\Pr$ denotes a discrete probability distribution.} $V\in\{W,Z\}$ and the sum is over $\mathcal{F}=bb,cc,cs,cd$ and light-quark pairs.  The distribution of $B$ is well approximated as the product of the distributions for $B_\text{lead}$ and $B_\text{sub-lead}$ when the flavours of the leading and sub-leading jets are known.  This is exploited for hadronically decaying $W$ bosons and for the light-quark flavour decays of $Z$ bosons to construct templates for $B$ that have a sufficient number of simulated events for large values of $B$, i.e. $\Pr(B|\mathcal{F},V)=\Pr(B_\text{lead}|\mathcal{F},V)\Pr(B_\text{sub-lead}|\mathcal{F},V)$.  The unit-normalized templates for $B$ are shown in Fig.~\ref{fig:fat_Jbv0} and the unit-normalized templates $p(M|\mathcal{F},V)$ and $p(Q|\mathcal{F},V)$ are shown in Fig.~\ref{fig:templates}.  For a given boson type, the jet-charge template is nearly independent of the flavour.  However, there is a dependence of the jet mass on the (heavy) flavour of the boson decay products.

\begin{figure}[h!]
\centering
\includegraphics[width=0.7\textwidth]{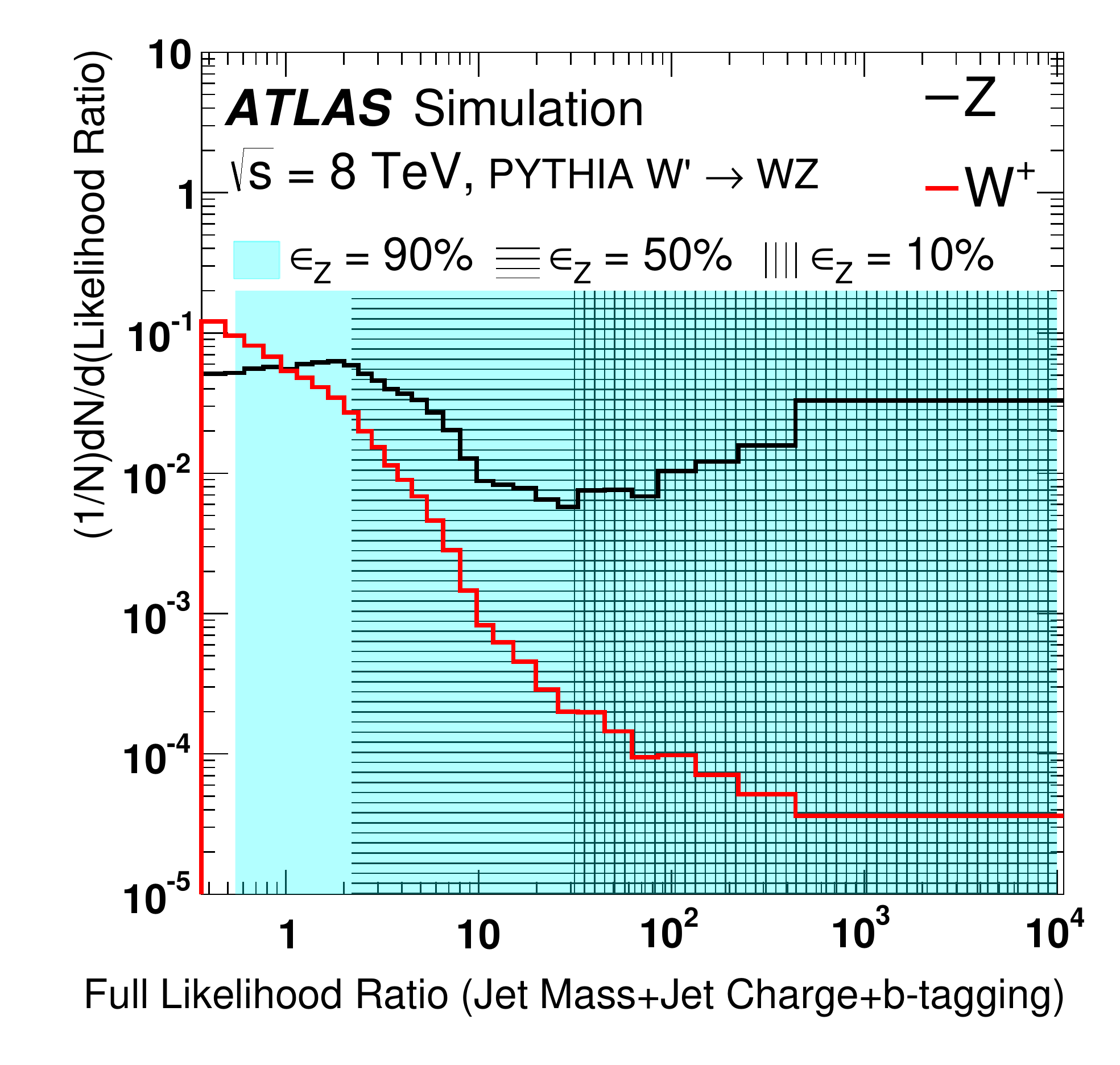}
\caption{The full likelihood ratio for the tagger formed from jet mass, jet charge, and a small-radius jet $b$-tagging discriminant.  The black histogram shows the likelihood ratio for $Z$ bosons and the red histogram is the likelihood ratio for $W^+$ bosons.  The shaded areas show the region of the likelihood ratio corresponding to $90\%$, $50\%,$ and $10\%$ working points of the $Z$-boson tagging efficiency.}
\label{fig:nomlike}
\end{figure}

The likelihood function is constructed by taking the ratio of the probability distribution functions $p(M,Q,B|V)$, for $V\in\{W,Z\}$, determined from the templates in Eq. (\ref{eq:like}).  Every bin $i$ of the 3D histogram that approximates $p(M,Q,B|V)$ is assigned a pair of numbers $(i,s_i/b_i)$ where $s_i$ is the overall fraction of the signal ($Z$ or $W$) in bin $i$ and $b_i$ is the fraction of the overall background (the other boson flavour) in bin $i$.  Bins are then sorted from largest to smallest $s_i/b_i$, with $f(i)$ defining a map from the old bin index to the new, sorted one. There are then two 1D histograms: for the signal, bin $j$ has bin content $s_{f^{-1}(j)}$ and for the background, bin $j$ has bin content $b_{f^{-1}(j)}$.   The optimal tagging procedure is then to set a threshold on the new 1D histograms.  The full likelihood ratio of the combined tagger is shown in Fig.~\ref{fig:nomlike} where the thresholds required for $90\%$, $50\%$, and $10\%$ $Z$-boson tagging efficiency are marked with shaded regions.

\begin{figure}[h!]
\centering
\subfloat[]{\includegraphics[width=0.5\textwidth]{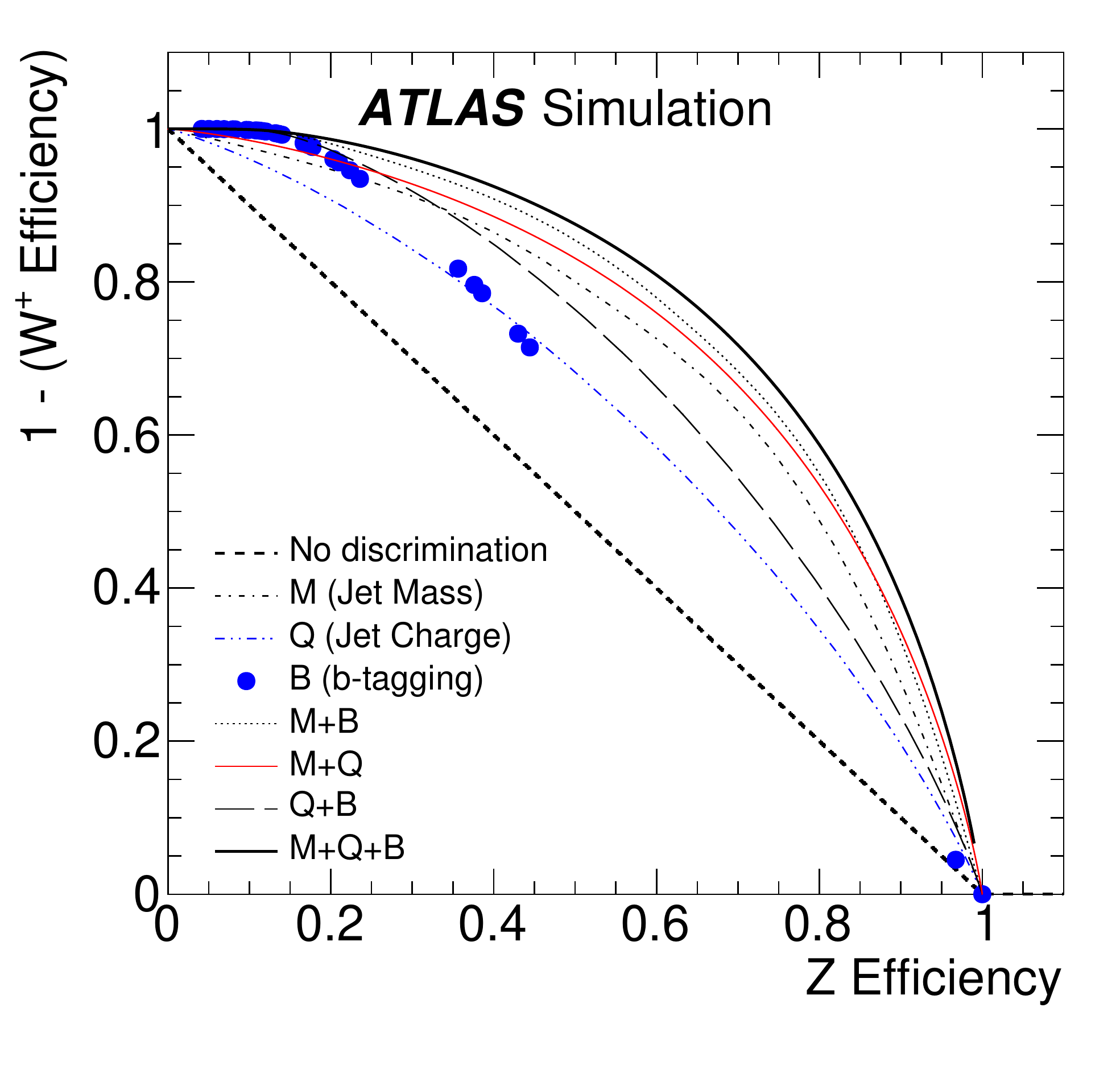}}
\subfloat[]{\includegraphics[width=0.5\textwidth]{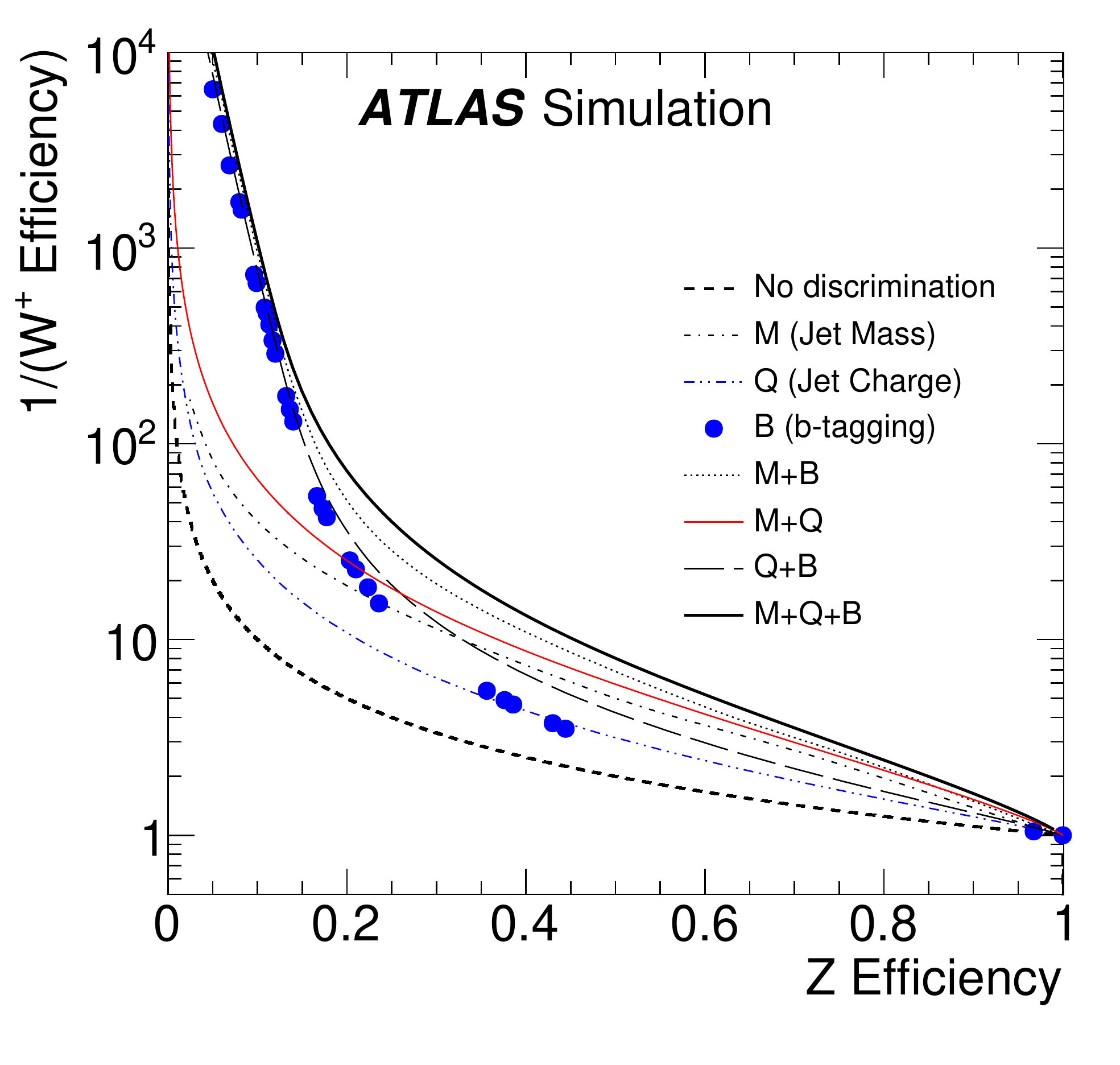}}
\caption{The tradeoff between $Z$ efficiency and (a) $1-$ ($W^+$ efficiency) (b) or $1/$($W^+$ efficiency) on (a) a linear scale and (b) a logarithmic scale.  Each curve is constructed by placing thresholds on the likelihood constructed from the inputs indicated in the legend.  Since the $b$-tagging discriminant is binned in efficiency, there are only discrete operating points for the tagger built only from $B$.}
\label{fig:nomROCs}
\end{figure}

\begin{figure}[h!]
\centering
\subfloat[]{\includegraphics[width=0.5\textwidth]{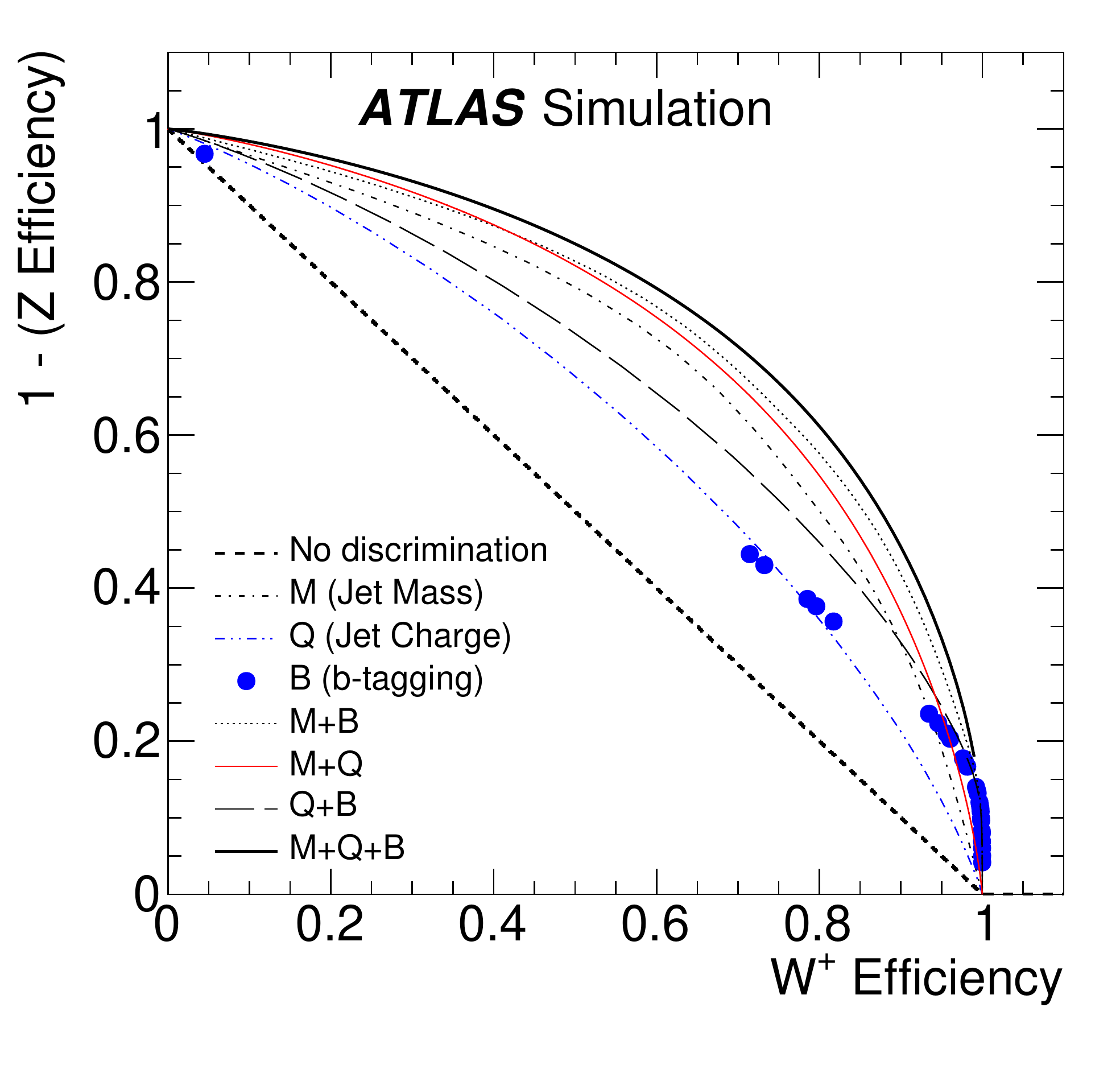}}
\subfloat[]{\includegraphics[width=0.5\textwidth]{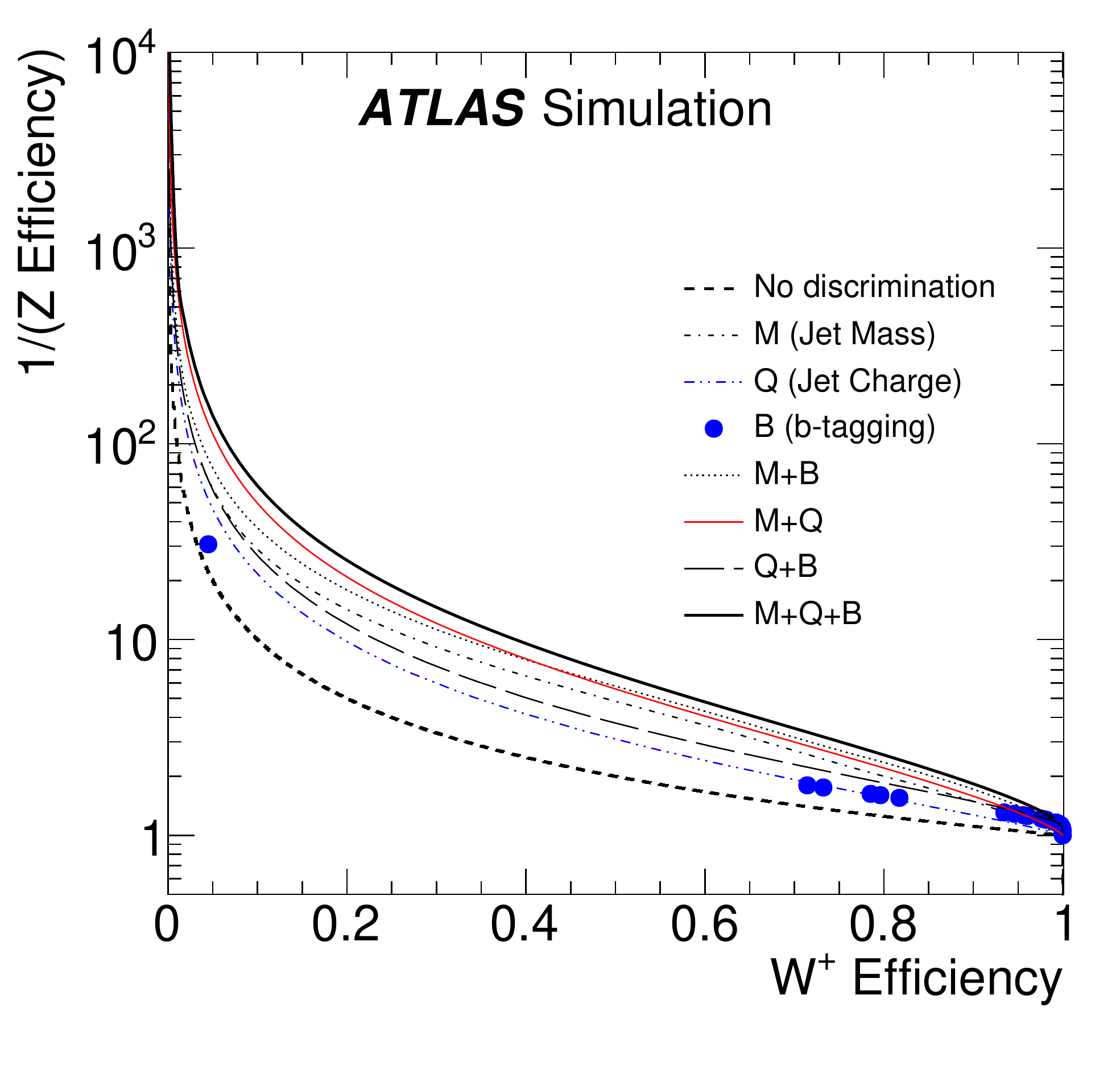}}
\caption{The tradeoff between $W^+$ efficiency and (a) $1-$ ($Z$ efficiency) or (b) $1/$($Z$ efficiency) on (a) a linear scale and (b) a logarithmic scale.  Each curve is constructed by placing thresholds on the likelihood constructed from the inputs indicated in the legend.  Since the $b$-tagging discriminant is binned in efficiency, there are only discrete operating points for the tagger built only from $B$.}
\label{fig:nomROCs_inverted}
\end{figure}

Curves displaying the tagging performance for all possible subsets of $\{M,Q,B\}$ are shown in Fig.~\ref{fig:nomROCs}.  There are $30$ possible values for $B$, which are therefore represented by discrete points.  The jet mass is the best performing single variable for medium to high $Z$-boson efficiencies, with visible improvement for $M$+$B$ and $M$+$Q$.  There is a significant gain from combining all three variables for $Z$-boson tagging efficiency above about $20\%$.  Below $20\%$, the combined tagger is dominated by $B$ where the $Z\rightarrow b\bar{b}$ branching fraction no longer limits $Z$-boson tagging efficiency.   For $Z$-boson efficiencies of about $50\%$, one can achieve $W^+$ rejection factors ($1/\epsilon_{W^+}$) of $3.3$ by using $Q$ or $B$ alone and about 5.0 using mass alone. For $Z$ efficiencies of $\epsilon_Z=90\%$, 50\%, and 10\%, $W^+$ rejection factors of $1.7$, $8.3$, and $1000$, respectively, can be achieved with the combined tagger.  Although most applications of boson-type tagging will target $Z$ bosons as the signal while rejecting $W$ bosons as background, the likelihood constructed in Fig.~\ref{fig:nomlike} can also be used to optimally distinguish $W^{+}$ bosons from $Z$ bosons.  The corresponding performance curves are shown in Fig.~\ref{fig:nomROCs_inverted}.  The locations of the $b$-tagging points are all now shifted to high efficiency with respect to Fig.~\ref{fig:nomROCs} because, for $W^+$ tagging, one wants to operate in the high-efficiency $b$-tagging bins (whereas the opposite is optimal for $Z$ tagging).  At an efficiency of $\epsilon_{W^+}=50\%$, a $Z$-boson rejection factor of $1/\epsilon_Z \approx 6.7$ can be achieved.  

\section{Systematic uncertainties}
\label{sec:systs}

The performance curves in Fig.~\ref{fig:nomROCs} are based on the nominal modelling parameters of the ATLAS simulation.  Additional studies show how the curves change due to the systematic uncertainties on the inputs to the likelihood function.   Sources of experimental uncertainty include the calibrations of the large- and small-radius jet four-momenta, the $b$-tagging (which incorporates e.g. impact parameter modelling), and the modelling of track reconstruction.  

\begin{figure}[h!]
\begin{center}
\subfloat[]{\includegraphics[width=0.5\textwidth]{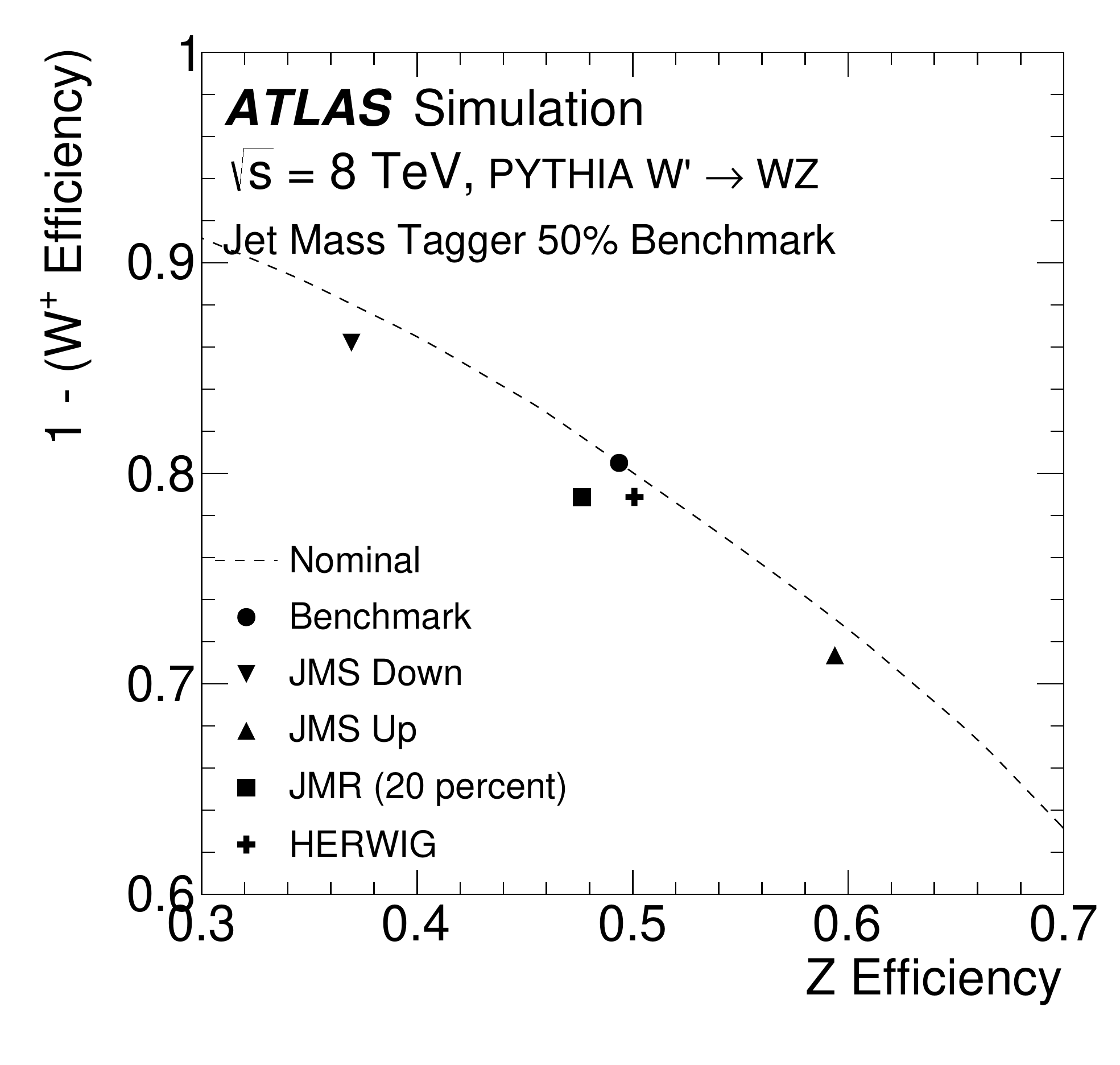}\includegraphics[width=0.5\textwidth]{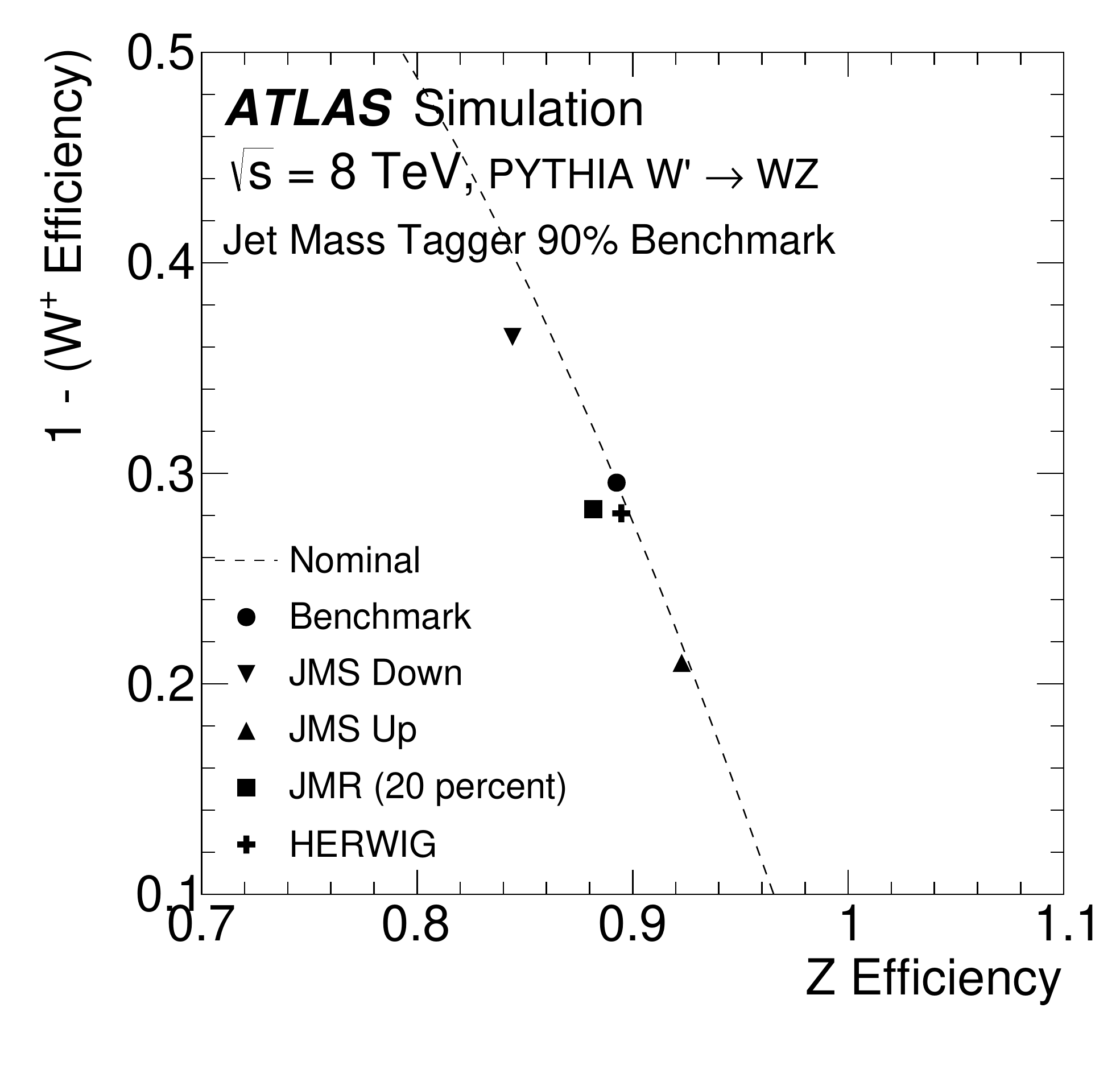}}\\
\vspace{-3mm}
\subfloat[]{\includegraphics[width=0.5\textwidth]{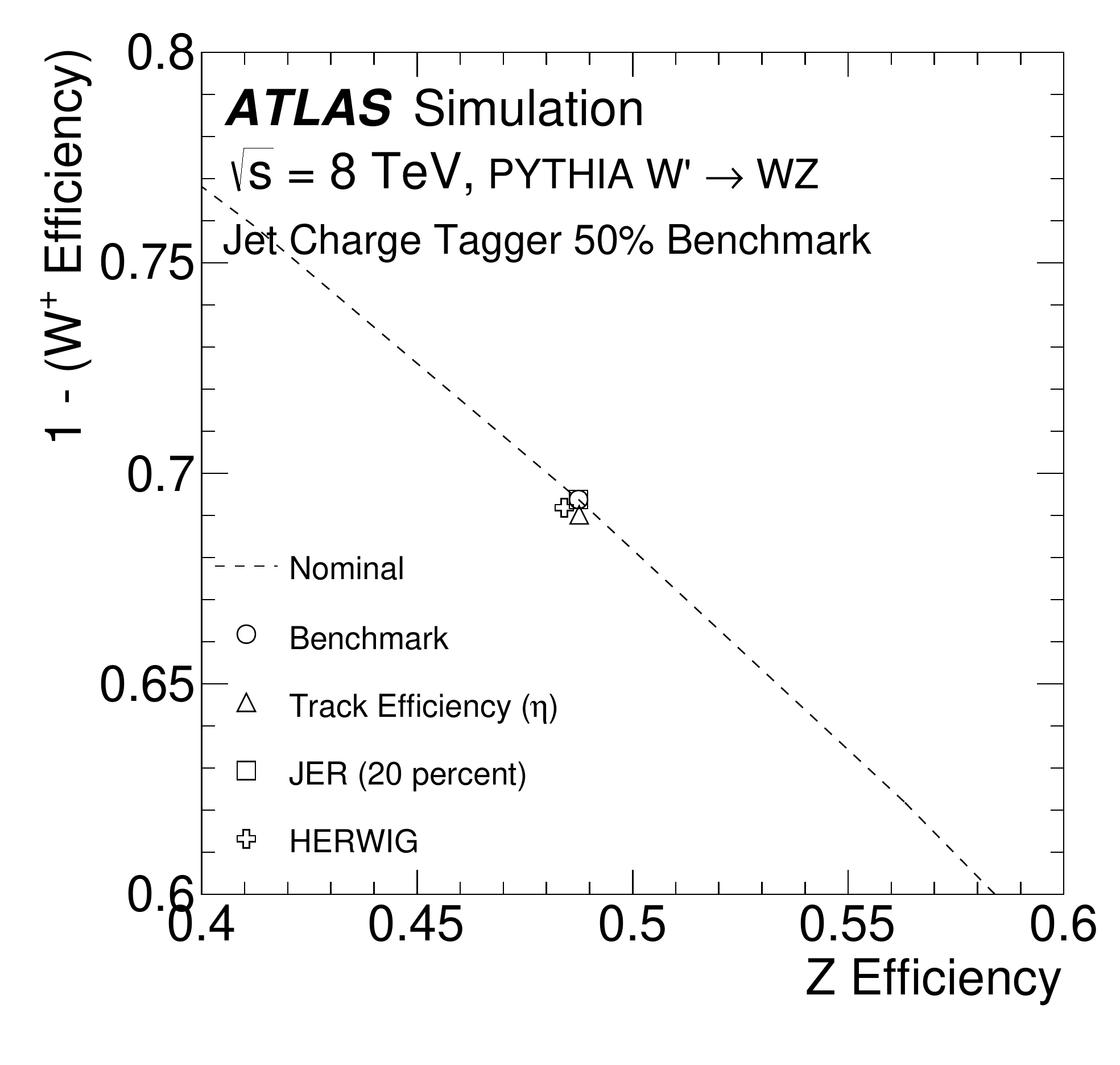}\includegraphics[width=0.5\textwidth]{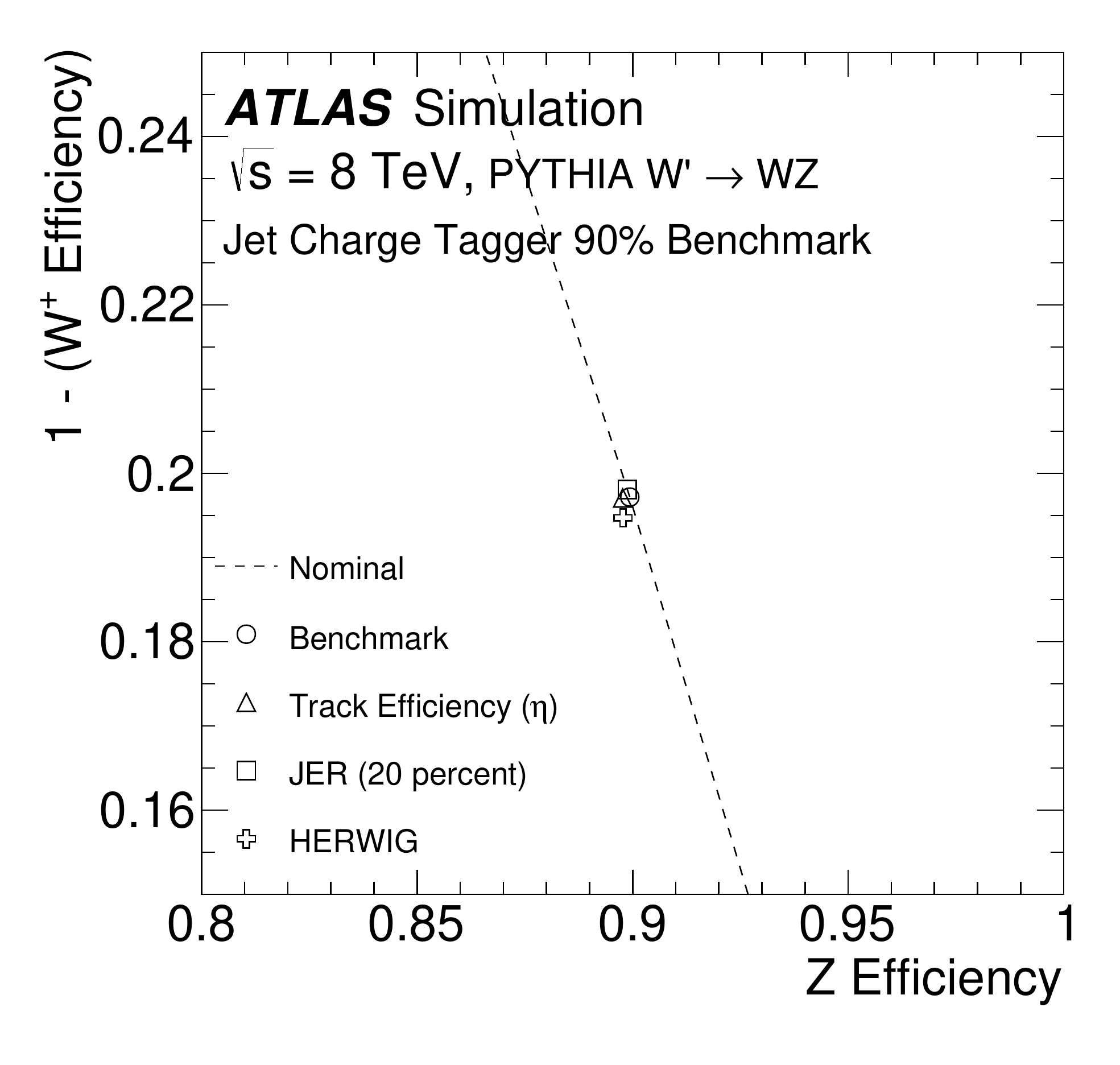}}\\
\vspace{-3mm}
\caption{The impact of selected systematic uncertainties on benchmark working points of the boson-type tagger.  (a) a jet-mass-only tagger, for 50\% (left) and 90\% $Z$ efficiency benchmarks.  (b) a jet-charge-only tagger, for 50\% (left) and 90\% $Z$ efficiency benchmarks. The point marked {\tt HERWIG} uses the alternative shower and hadronization model for the simulation, with the likelihood template from {\tt PYTHIA}. See the text for an explanation of the notation in the legend.}
\label{fig:nomROCs_syst}
\end{center}
\end{figure}

The uncertainty on the scale of the large-radius jet mass calibration is estimated using the double ratio in data and MC simulation of calorimeter jet mass to track jet mass~\cite{Aad:2013gja}.  Tracks associated with a jet are well measured and provide an independent observable correlated with the jet energy.  Uncertainties on the jet-mass resolution can have a non-negligible impact on the performance of the tagger.  The jet-mass resolution uncertainty is determined from the difference in the widths of the boosted $W$ boson jet-mass peak in semileptonic $t\bar{t}$ simulated and measured data events~\cite{Aad:2013gja} and also from varying the simulation according to its systematic uncertainties~\cite{ATLAS:2012am}.  The resolution is about 5~GeV in the Gaussian core of the mass spectrum and its uncertainty is about 20\%.  The impact of the jet-mass scale and resolution uncertainties on the boson-type tagger built using only the jet mass is shown in Fig.~\ref{fig:nomROCs_syst} for two nominal working points of $50\%$ and $90\%$  $Z$-boson tagging efficiency.   Both the likelihood map $f$ from Sect.~\ref{sec:perfm} and the threshold value are fixed.  Inputs to the tagger are shifted by their uncertainties and the 1D histograms described above are re-populated.  The efficiencies for $W$ and $Z$ bosons are recomputed and shown as markers in Fig.~\ref{fig:nomROCs_syst}(a).  Coherent shifts of the jet masses (JMS) for $W$ and $Z$ bosons result in movement along the nominal performance curve corresponding to $\pm10\%$ changes in the efficiency. However, there are also shifts away from the nominal curve because the optimal jet-mass cut is not a simple threshold.  Variation of the jet-mass resolution (JMR) preserves the scale and so the movement is nearly perpendicular to the original performance curve, at the $\lesssim 5\%$ level, because of the increased overlap in the $Z$ and $W$ mass distributions.\footnote{Although such shifts retain optimal use of the tagger (highest rejection for a fixed efficiency), they can degrade the quality of e.g. a cross-section measurement.}  Shifts along the nominal curve optimally use the input variables (albeit at different efficiencies), while shifts away from the nominal curve are a degradation in the performance.  The impact of the fragmentation is estimated by using input variables from {\tt HERWIG} but with the likelihood map from {\tt PYTHIA}.   {\tt PYTHIA} and {\tt HERWIG} have similar $W/Z$ efficiencies at both the $50\%$ and $90\%$ benchmark points.

The systematic uncertainty on the efficiency of the tracking reconstruction is estimated by removing tracks associated with jets using an $\eta$-dependent probability~\cite{Aad:2010ac}.  The probability in the region $2.3<|\eta|<2.5$ is $7\%$; it is 4\% for $1.9<|\eta|<2.3$, 3\% for $1.3<|\eta|<1.9$, and 2\% for $0<|\eta|<1.3$.  These probabilities are known to be conservative in the most central $\eta$ bins.  There is also an uncertainty on the modelling of track merging for high-$p_\text{T}$ jets, but the loss is expected to be negligible for jets with $p_\text{T}<400$~GeV.   Differences in the modelling of fragmentation can affect the expected performance for all the input variables, especially for the track-dependent observables.  The impact of various uncertainties on the boson-type tagger built using only the jet charge is shown in Fig.~\ref{fig:nomROCs_syst}(b).  Since $W$ and $Z$ boson decays produce on average many tracks (see Sect.~\ref{sec:data}), removing a small number of them does not have a big impact on the jet-charge tagger as a result of the $p_\text{T}$-weighting in the jet charge sum. 

\begin{figure}[h!]
\begin{center}
\includegraphics[width=0.65\textwidth]{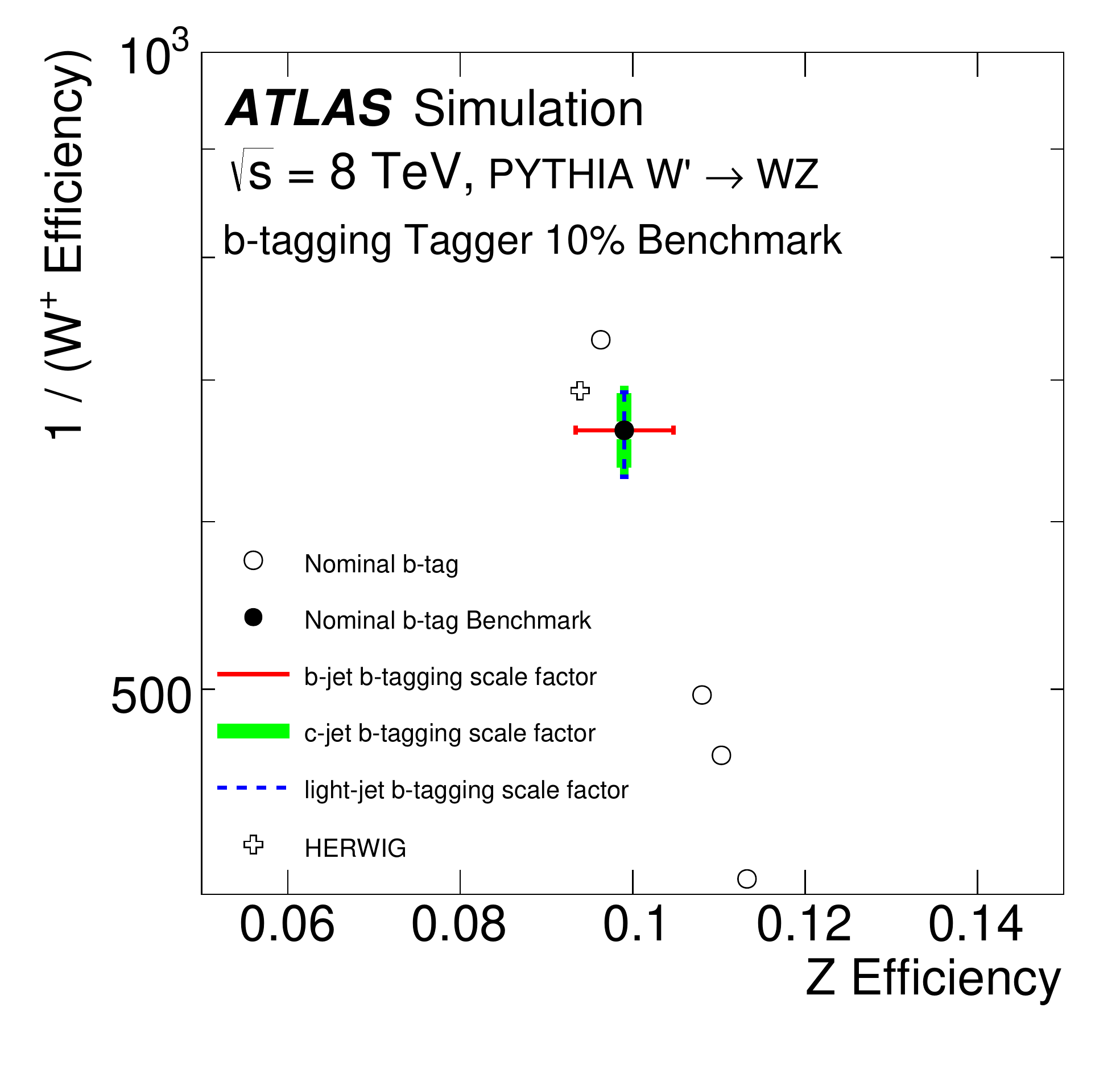}
\caption{The impact of selected systematic uncertainties on benchmark working points of a $b$-tagging-only tagger at a $10\%$ $Z$ efficiency benchmark.  The $b$-tagging discriminant is binned, so there are only discrete operating points.  The point marked {\tt HERWIG} uses the alternative shower and hadronization model for the simulation, with the likelihood template from {\tt PYTHIA}.  The $b$-tagging scale factor uncertainties are determined separately for $b$-, $c$-, and light-quark jets.  Variations are added in quadrature for each `truth' jet flavour.  There is no contribution from the $b$-jet scale factor uncertainties on the $W$ rejection because there are no `truth' $b$-jets.  Conversely, the $c$- and light-jet scale factor uncertainties do not impact the $Z$ bosons because at this low efficiency, all the selected $Z$ bosons decay into $b\bar{b}$.}
\label{fig:nomROCs_syst2}
\end{center}
\end{figure}

The efficiency to $b$-tag jets of various flavours ($b$, $c$, and light) is measured in data using $t\bar{t}$ events~\cite{btag1}, jets with identified charm hadrons, and multijet events~\cite{btag2}.  The differences between data and MC simulation are typically a few percent and are applied as independent correction factors on a per-jet basis.  The uncertainties on these scale factor measurements are used as estimates of the systematic uncertainty on the $b$-tagging.  The sources of uncertainty are decomposed into many uncorrelated components ($24$ for $b$-jets, $16$ for $c$-jets, and $48$ for light-flavour jets) and the impact on the rejection is added in quadrature for a fixed value of $\epsilon_\text{signal}$.  The $b$-tagging of matched small-radius jets is also affected by uncertainties on the jet-energy scale and resolution.  These quantities are varied within their uncertainties and if the shifted jet has $p_\text{T}<25$~GeV, its MV1 value is not considered.   The impact of various uncertainties on the boson-type tagger built using only the $b$-tagging discriminant for a 10\% nominal $Z$ efficiency is shown in Fig.~\ref{fig:nomROCs_syst2}.  At this efficiency, the full boson-type tagger is dominated by the $b$-tagging inputs, as seen in Fig.~\ref{fig:nomROCs}.   The scale factor uncertainty for $b$-jets has no impact on the $W$ efficiency (no real $b$-jets), but there is approximately a 10\% uncertainty on the $Z$ efficiency.  The uncertainties on the jet-energy scale for small-radius jets are relevant only because of the 25~GeV $p_\text{T}$ threshold.  Since all of the large-radius jets are required to have $p_\text{T}>200$~GeV, the threshold is relevant only in the rare case that one of the $W$ daughters is nearly anti-parallel in the $W$ rest frame to the direction of the $W$ boost vector.  

\begin{figure}[h!]
\begin{center}
\subfloat[]{\includegraphics[width=0.45\textwidth]{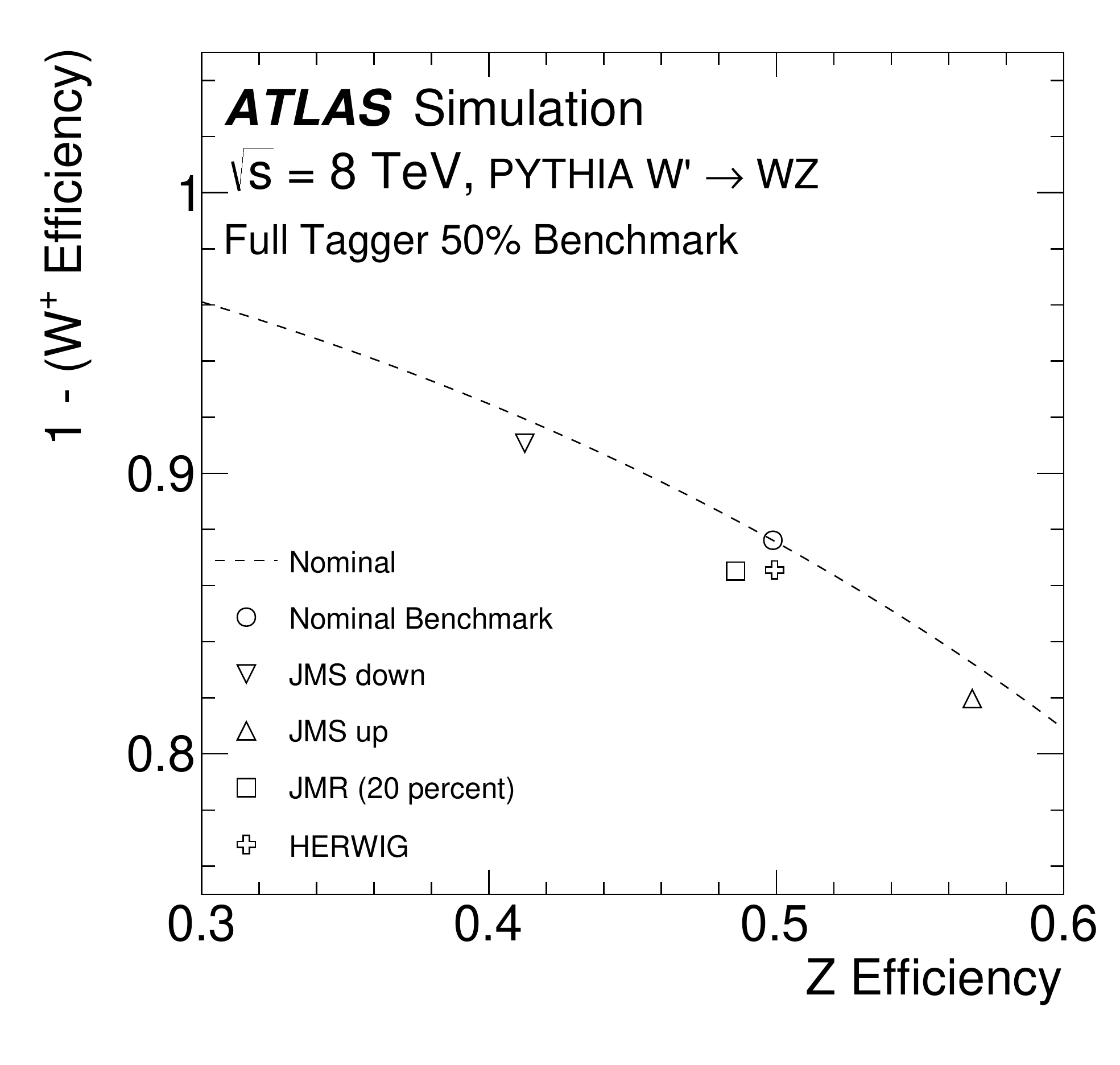}}\subfloat[]{\includegraphics[width=0.45\textwidth]{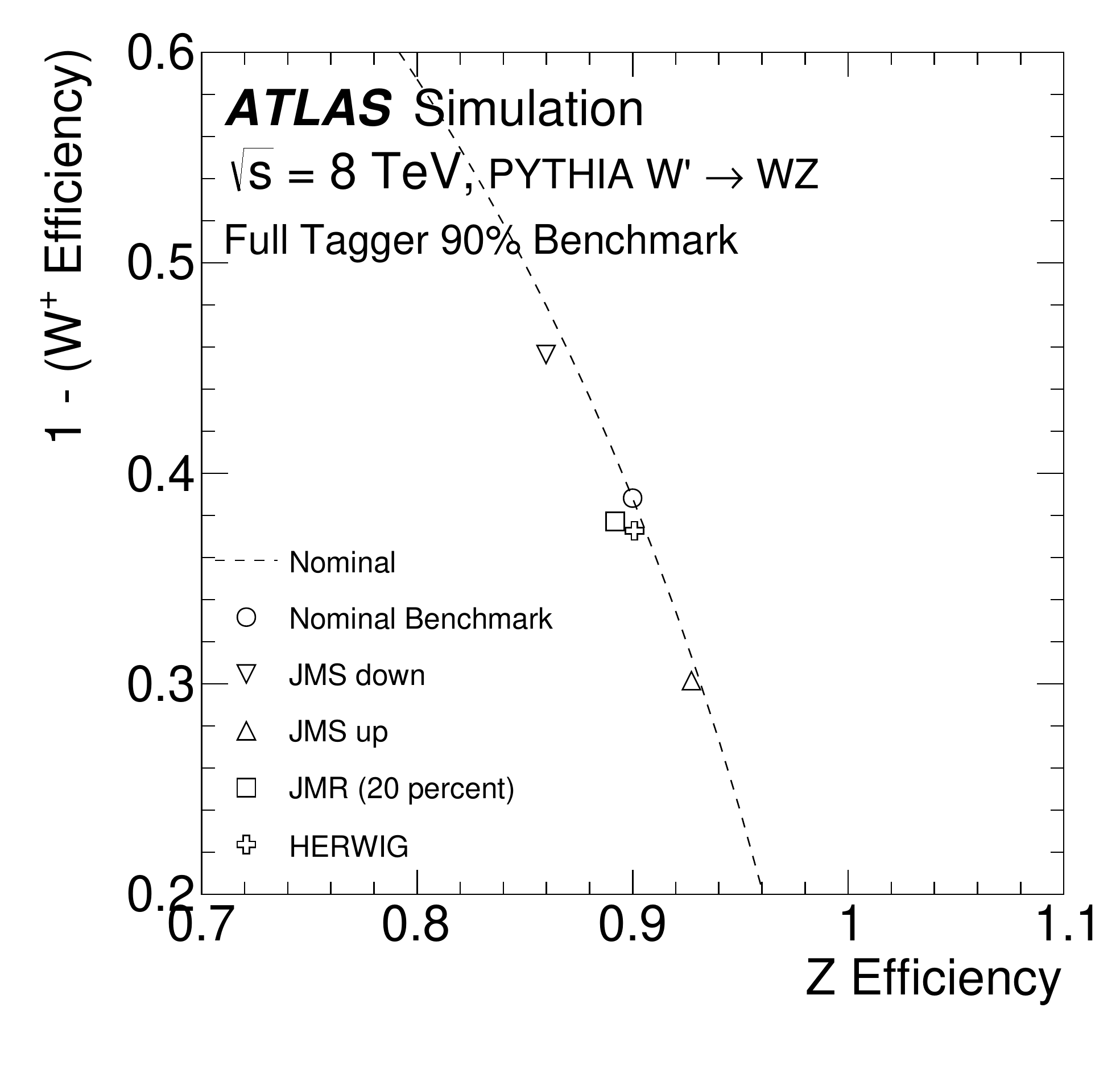}}\\
\caption{The impact of uncertainties on the jet-mass scale and resolution for 50\% (a) and 90\% (b) $Z$ efficiency working points of the full boson-type tagger. The point marked {\tt HERWIG} uses the alternative shower and hadronization model for the simulation, with the likelihood template from {\tt PYTHIA}.  }
\label{fig:nomROCs_sys3Dt}
\end{center}
\end{figure}

The impact of the uncertainties on the jet-mass scale and resolution on the boson-type tagger built using all of the inputs (jet mass, jet charge, and $b$-tagging) is shown in Fig.~\ref{fig:nomROCs_sys3Dt}(a).   At very low $Z$-boson tagging efficiency, the tagger is dominated by $b$-tagging, so Fig.~\ref{fig:nomROCs_syst2} is a good representation of the uncertainty on the full tagger's performance.  For higher efficiencies, the tagger is dominated by the jet mass, although the jet charge and $b$-tagging discriminant significantly improve the performance.  The uncertainty on the full tagger's performance at the $50\%$ and $90\%$ $Z$-boson tagging efficiency benchmark points is due mostly to the uncertainty on the jet mass, which is why these uncertainties are shown in Fig.~\ref{fig:nomROCs_sys3Dt}.

\section{Validation of tagging variables using data}
\label{sec:data}

The tagger cannot be fully tested with data because it is not possible to isolate a pure sample of hadronically decaying $Z$ bosons in $pp$ collisions.  However, the modelling of the variables used to design the tagger can be studied with a relatively pure and copious sample of hadronically decaying $W$ bosons in $t\bar{t}$ events which can be tagged by the leptonic decay of the other $W$ boson in the event (semileptonic $t\bar{t}$ events).  Single-lepton triggers are used to reject most of the events from QCD multijet background processes.  Candidate reconstructed $t\bar{t}$ events are chosen by requiring an electron or a muon with $p_\text{T} > 25$~\gev{} and $|\eta| < 2.5$, as well as a missing transverse momentum $\met > 20 \gev$.  The electrons and muons are required to satisfy a series of quality criteria, including isolation.\footnote{Leptons are considered isolated if they are well separated from jets ($\Delta R > 0.4$) and the track/calorimeter energy within a small cone, centred on the lepton direction but excluding the lepton itself, is below a fixed relative value. }  Events are rejected if there is not exactly one electron or muon.   In addition, the sum of the $\met$ and the transverse mass\footnote{The transverse mass, $m_\text{T}$, is defined as $m_{\text{T}}^2=2p_{\text{T}}^{\text{lep}}E_{\text{T}}^{\text{miss}}(1-\cos(\Delta\phi))$, where $\Delta\phi$ is the azimuthal angle between the lepton and the direction of the missing transverse momentum.} of the $W$ boson, reconstructed from the lepton and $\met$, is required to be greater than 60~\gev.   Events must have at least one $b$-tagged jet (at the 70\% efficiency working point) and have at least one large-radius trimmed jet with $p_\text{T}>200$~GeV and $|\eta|<2$.  Furthermore, there must be a small-radius jet with $p_\text{T}>25$ GeV, and $\Delta R<1.5$ to the selected lepton (targeting the decay chain $t\rightarrow bW(\rightarrow \ell\nu)$).  The other $W$ boson candidate is selected as the leading large-radius trimmed jet with $\Delta R>1.5$ from the small-radius jet that is matched to the lepton.   The $W$+jets and multijet backgrounds are estimated from the data using the charge asymmetry and matrix methods, respectively~\cite{Aad:2014zka}.  The other backgrounds are estimated directly from MC simulation.  Although the resulting event selection is expected to have a high $t\bar{t}$ purity (about 75\%), the events cannot be compared directly to the isolated $W$ bosons from the simulated $W'$ boson decays.  This is because there are several effects that make the typical large-radius jet in semileptonic $t\bar{t}$ events different from isolated $W$ and $Z$ boson jets in $W'$ boson events\footnote{When controlling for all differences, the distributions for isolated $W$ bosons from $t\bar{t}$ and from $W'$ are nearly identical.}:

\begin{enumerate}
\item The event selection is based on the reconstructed jet $p_\text{T}$ (earlier sections used $p_\text{T}^V$), so even if $p_\text{T}^\text{jet}\gtrsim 200$~GeV for an $R=1.0$ jet, the true hadronically decaying $W$ boson in the event may have $p_\text{T}^W<200$~GeV and thus the $W$ boson decay products might not be collimated within $\Delta R <1$.
\item There are more (close-by) jets in semileptonic $t\bar{t}$ events than in $W'$ boson events.  Jets not originating from the $W$ boson can form the leading large-radius jet, or the $b$-jet from the same top-quark as the hadronically decaying $W$ bosons can merge with the $W$ boson decay products to form a large-radius jet.
\end{enumerate}

\begin{figure}[h!]
\centering
\includegraphics[width=0.5\textwidth]{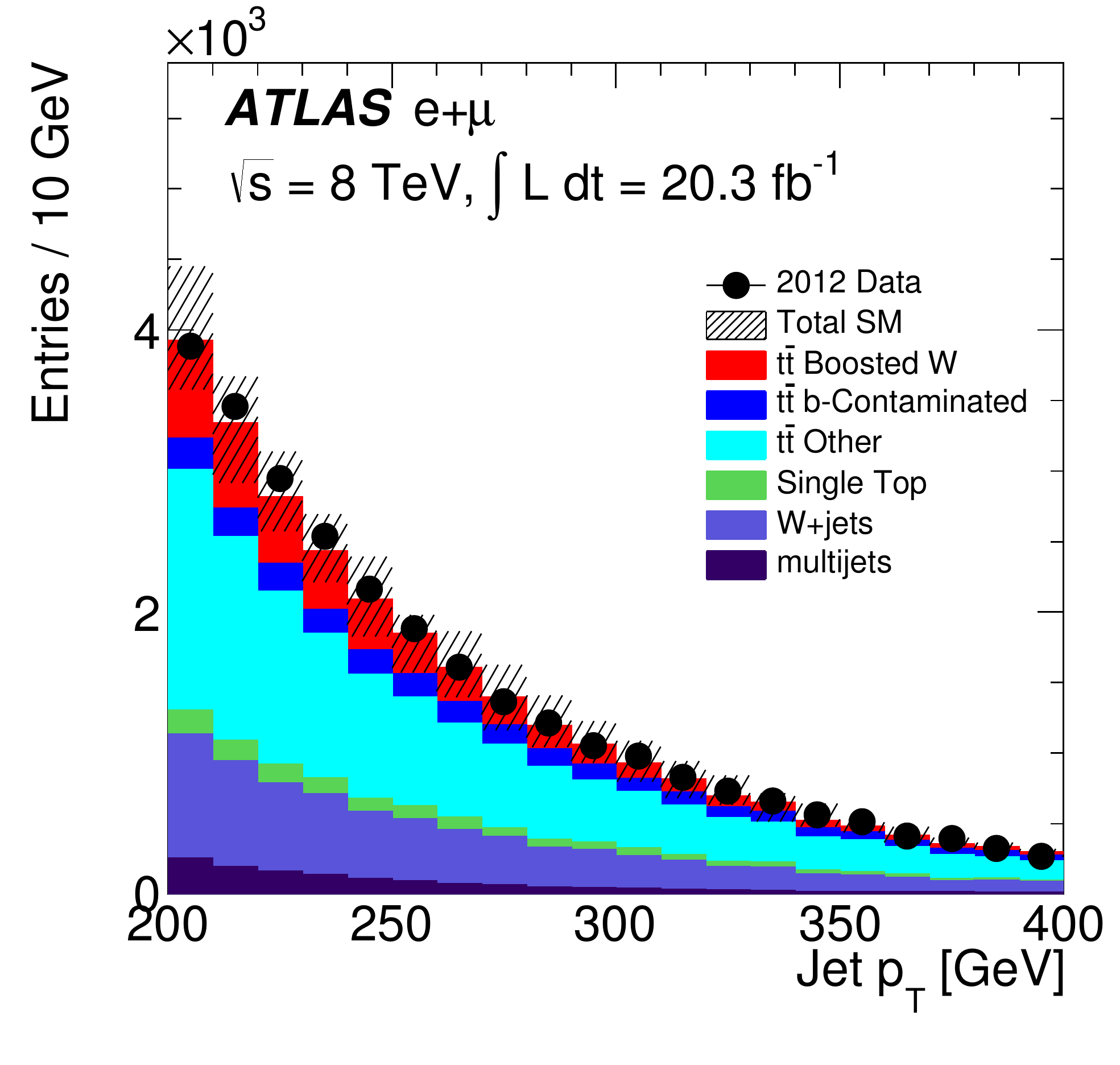}
\caption{The $p_\text{T}$ distribution of the selected large-radius jets.  The uncertainty band includes all the experimental uncertainties on the jet $p_\text{T}$ and jet mass described in Sect.~\ref{sec:systs}.}
\label{fig:data}
\end{figure}

\begin{figure}[h!]
\centering
\subfloat[]{\includegraphics[width=0.55\textwidth]{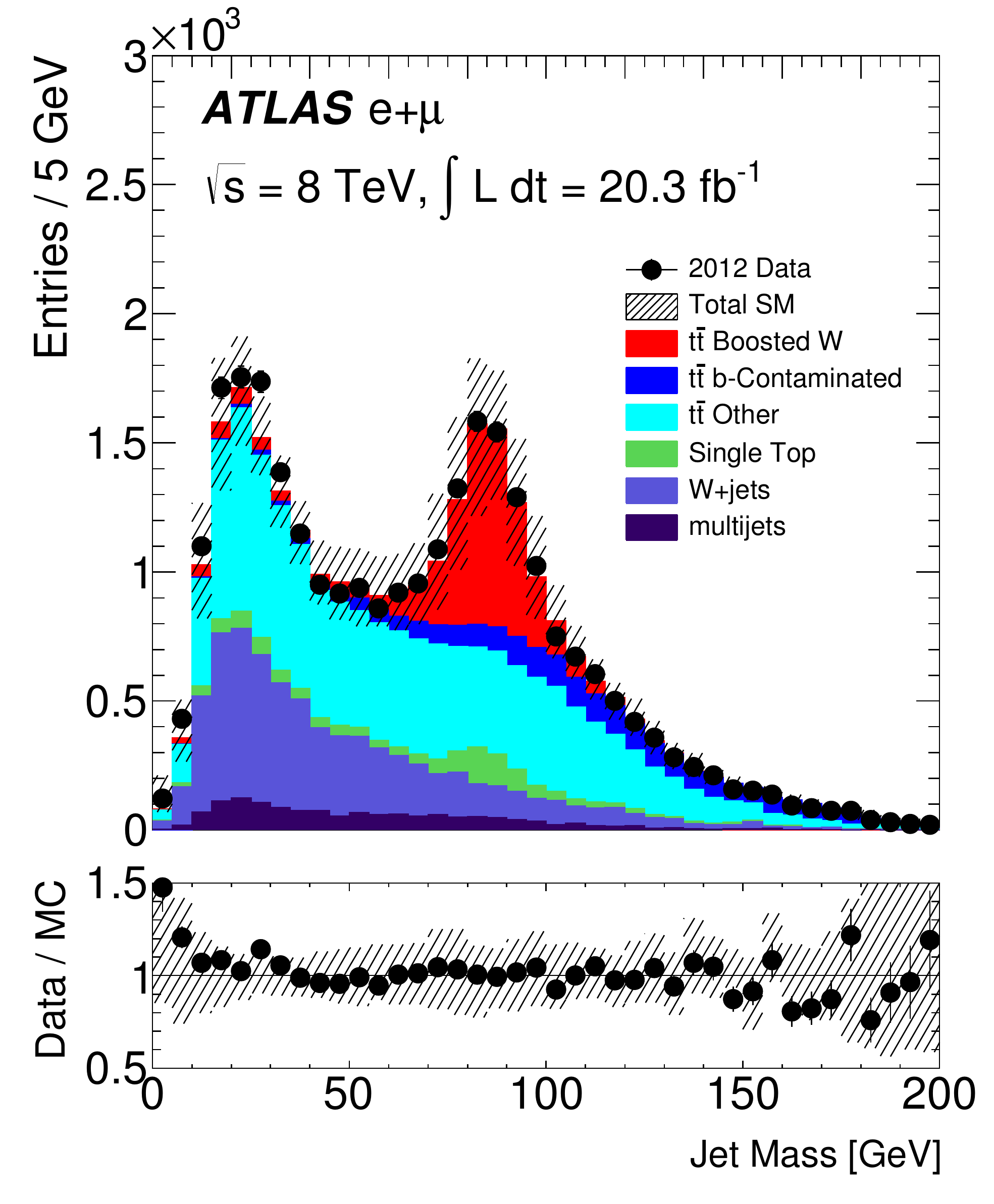}}\\\subfloat[]{\includegraphics[width=0.45\textwidth]{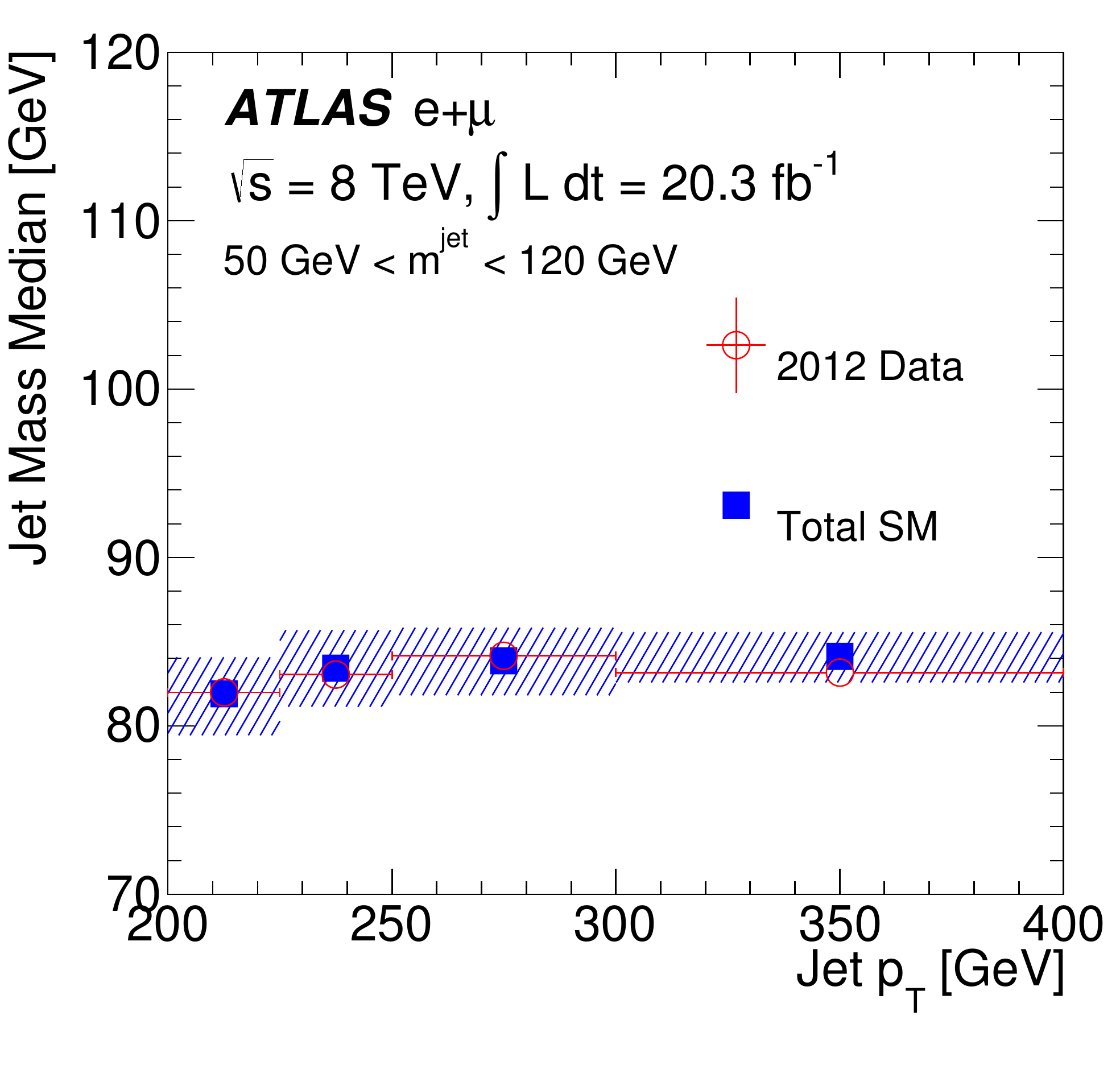}}\subfloat[]{\includegraphics[width=0.45\textwidth]{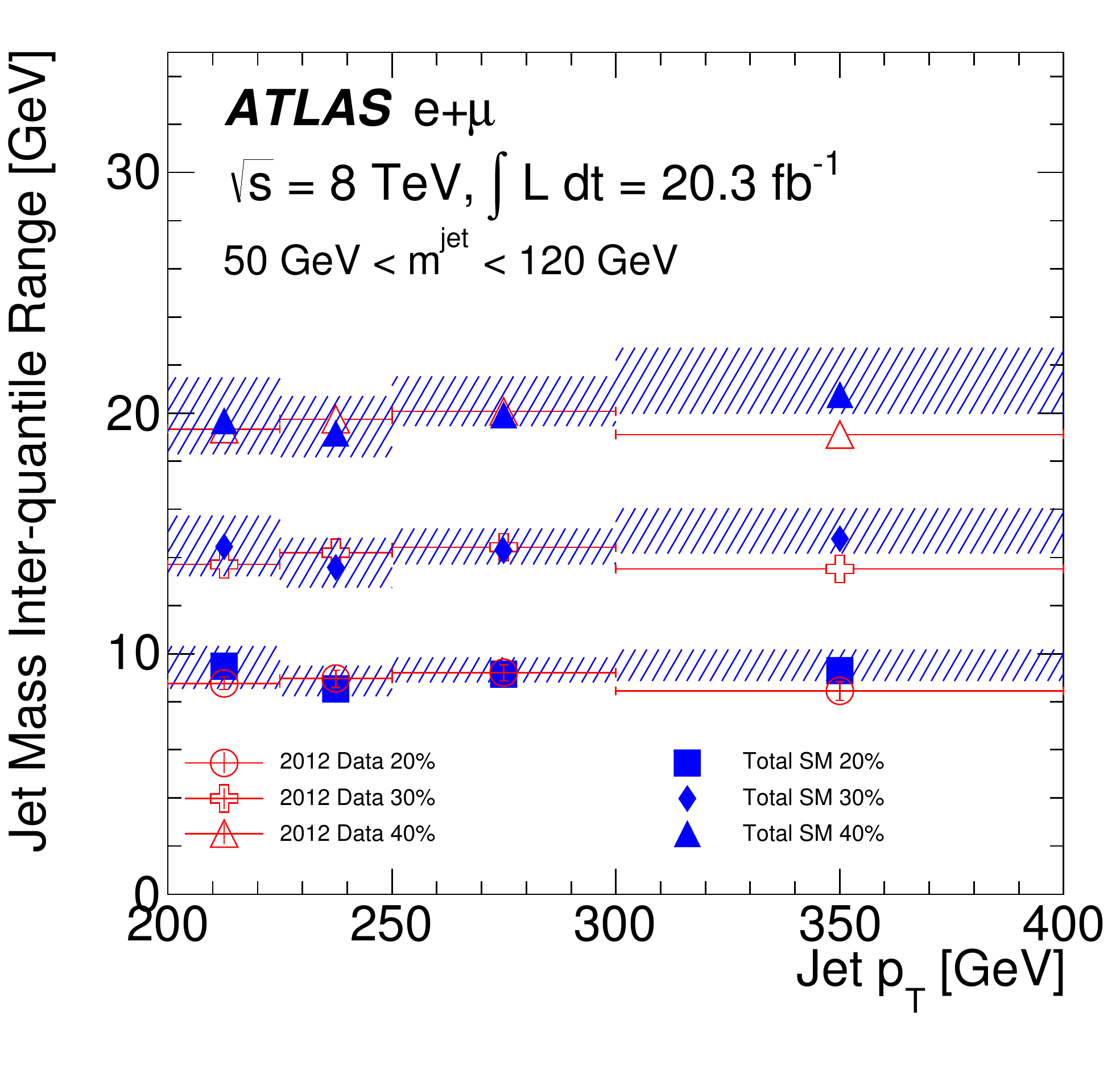}}
\caption{(a) The jet-mass distribution of the selected jets in semi-leptonic $t\bar{t}$ events.  (b) The median of the mass distribution as a function of the jet $p_\text{T}$ for events with the selected jet in the range $50$~GeV $<m^\text{jet}<$ $120$~GeV.  This includes the contributions from events which are not classified as Boosted $W$.  (c) For the same events as in (b), the inter-quantile range as a measure of spread.  The quantiles are centred at the median.   The uncertainty band includes all the experimental uncertainties on the jet $p_\text{T}$ and jet mass described in Sect.~\ref{sec:systs}.  The inter-quantile range of size $0\%<X<50\%$ is defined as the difference between the $50\%+X\%$ quantile and the $50\%-X\%$ quantile. Statistical uncertainty bars are included on the data points but are smaller than the markers in many bins.}
\label{fig:datamass}
\end{figure}

\noindent The variables $p_\text{T}^\text{jet}/p_\text{T}^W$ and $\Delta R(\text{jet},W)$, for the $W$ boson from the MC `truth' record and the selected large-radius jet, are used to classify the various $t\bar{t}$ event sub-topologies.  Events are labelled as having a {\bf Boosted $W$} if $|p_\text{T}^\text{jet}/p_\text{T}^W-1|<0.1$ and $\Delta R(\text{jet},W)<0.1$.  If the $b$-quark from the top-quark decay has an angular distance $\Delta R<1.0$ from the selected large-radius jet, this jet is labelled as {\bf $b$-contaminated}.  All other $t\bar{t}$ events, including events where both $W$ bosons decay into leptons, are labelled as {\bf Other}.  The $p_\text{T}$ spectrum of the jets from the classified events is shown in Fig.~\ref{fig:data}.  In Fig.~\ref{fig:data} and subsequent figures, systematic uncertainties on the simulation include the jet $p_\text{T}$ and jet mass uncertainties described in Sect.~\ref{sec:systs}, but exclude tracking uncertainties, which are sub-dominant.   Events are vetoed if the selected large-radius jet has $p_\text{T}>400$~GeV or if the $\Delta R$ between the selected large-radius jet and a tagged $b$-jet is less than $1.0$.  This suppresses the $b$-contaminated $t\bar{t}$ events.  The effectiveness of the $t\bar{t}$ event classification is most easily seen from the jet mass distribution, shown in Fig.~\ref{fig:datamass}(a).  The mass of the boosted $W$ bosons from $t\bar{t}$ events is peaked around $m_W$, as is a small contribution from the hadronically decaying $W$ bosons in single-top events in the $Wt$ channel.  There is no peak at $m_t$ in the $b$-contaminated spectrum because of the $b$-jet veto, but there is a small non-resonant contribution below the top-quark mass, due to events in which one $W$ daughter is matched with the $b$-jet.  This is akin to the $b$-jet+lepton invariant mass used in other circumstances to measure top-quark properties and naturally has a scale around $150$~GeV~\cite{Chatrchyan:2013boa}. The low-mass peak in $W$+jets and the `other' $t\bar{t}$ events is due to the Sudakov peak from QCD jets, the location of which scales with $R \times p_\text{T}$.  The dependence on $p_\text{T}$ of the $W$-peak position in Fig.~\ref{fig:datamass}(a) is shown in Fig.~\ref{fig:datamass}(b).  Events with the leading jet in a window around the $W$ mass, $50$~GeV $<m^\text{jet}<120$ GeV are selected and the median of the mass distribution is plotted in Fig.~\ref{fig:datamass}(b) as a function of the jet $p_\text{T}$.  The similar trend for the simulation and the data shows that the combination of the reconstructed jet-mass scale and `truth' jet-mass scale is well modelled.  To quantify the spread in the jet mass peak, various inter-quantile ranges are shown as a function of $p_\text{T}$ in Fig.~\ref{fig:datamass}(c).  The inter-quantile range of size $0\%<X<50\%$ is defined as the difference between the $50\%+X\%$ quantile and the $50\%-X\%$ quantile, and is a measure of the spread in the distribution.  The width of the boosted-$W$ mass peak is well modelled within the statistical precision of the 2012 data sample.

\begin{figure}[h!]
\centering
\subfloat[]{\includegraphics[width=0.55\textwidth]{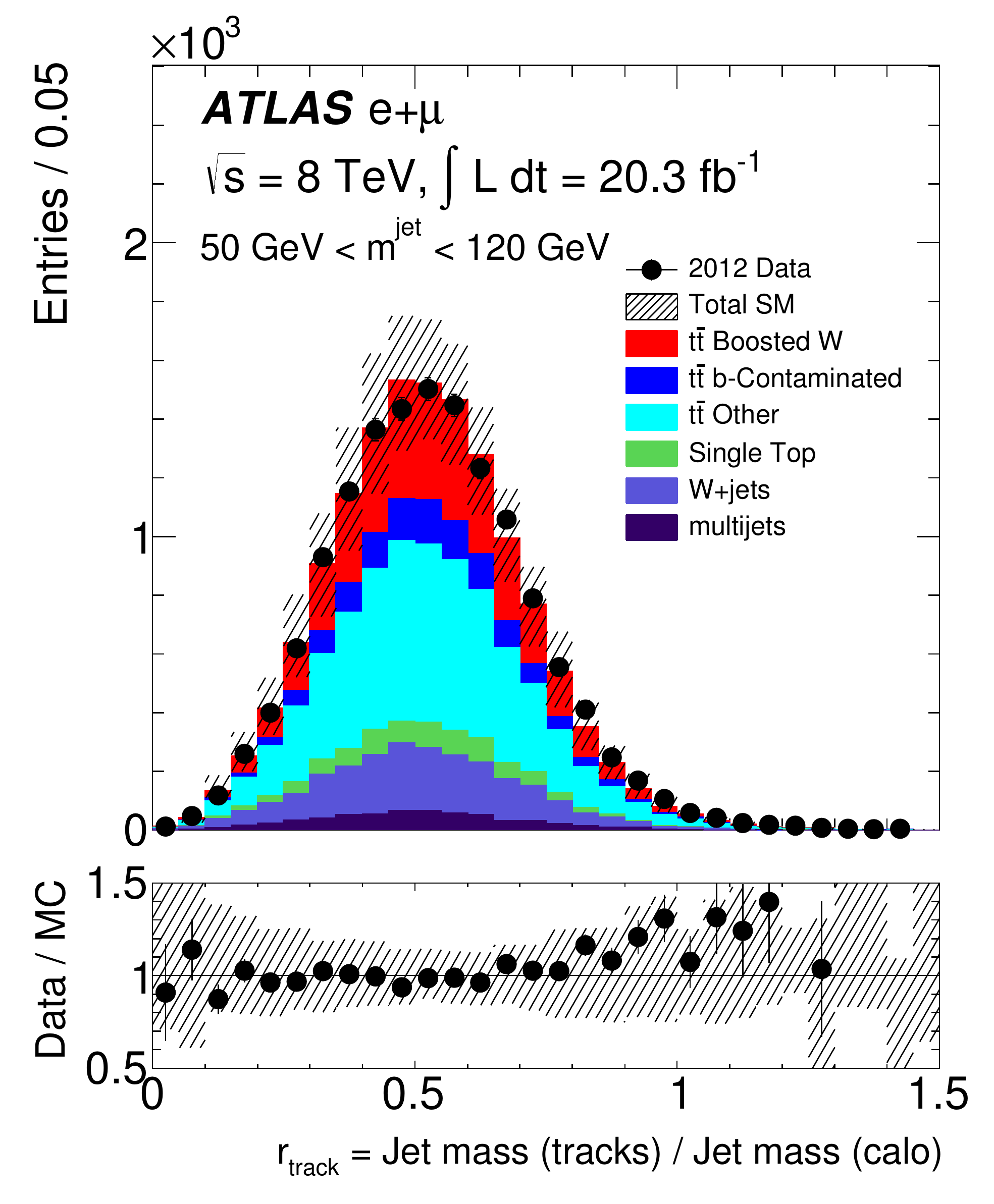}}\\\subfloat[]{\includegraphics[width=0.45\textwidth]{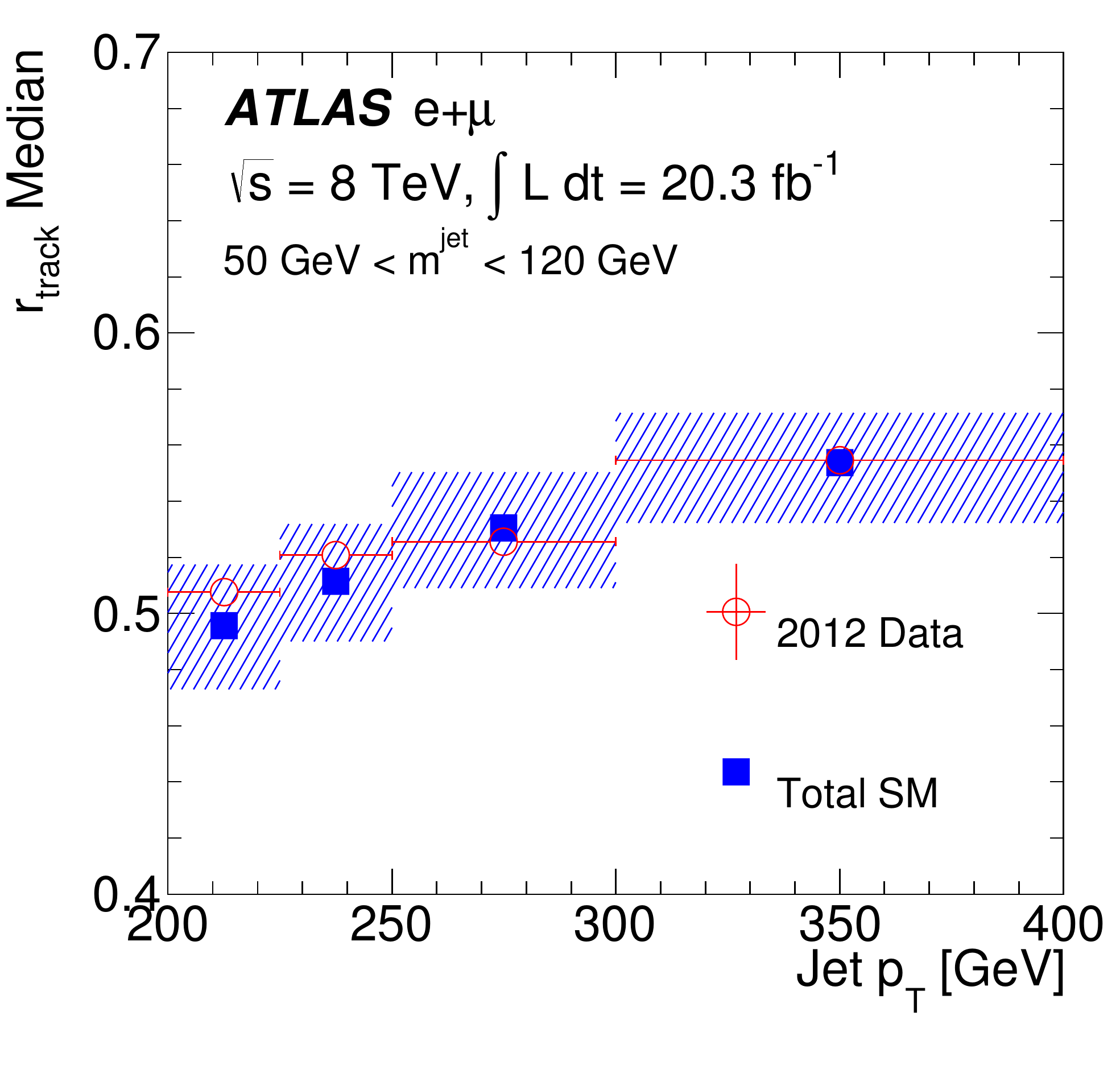}}\subfloat[]{\includegraphics[width=0.45\textwidth]{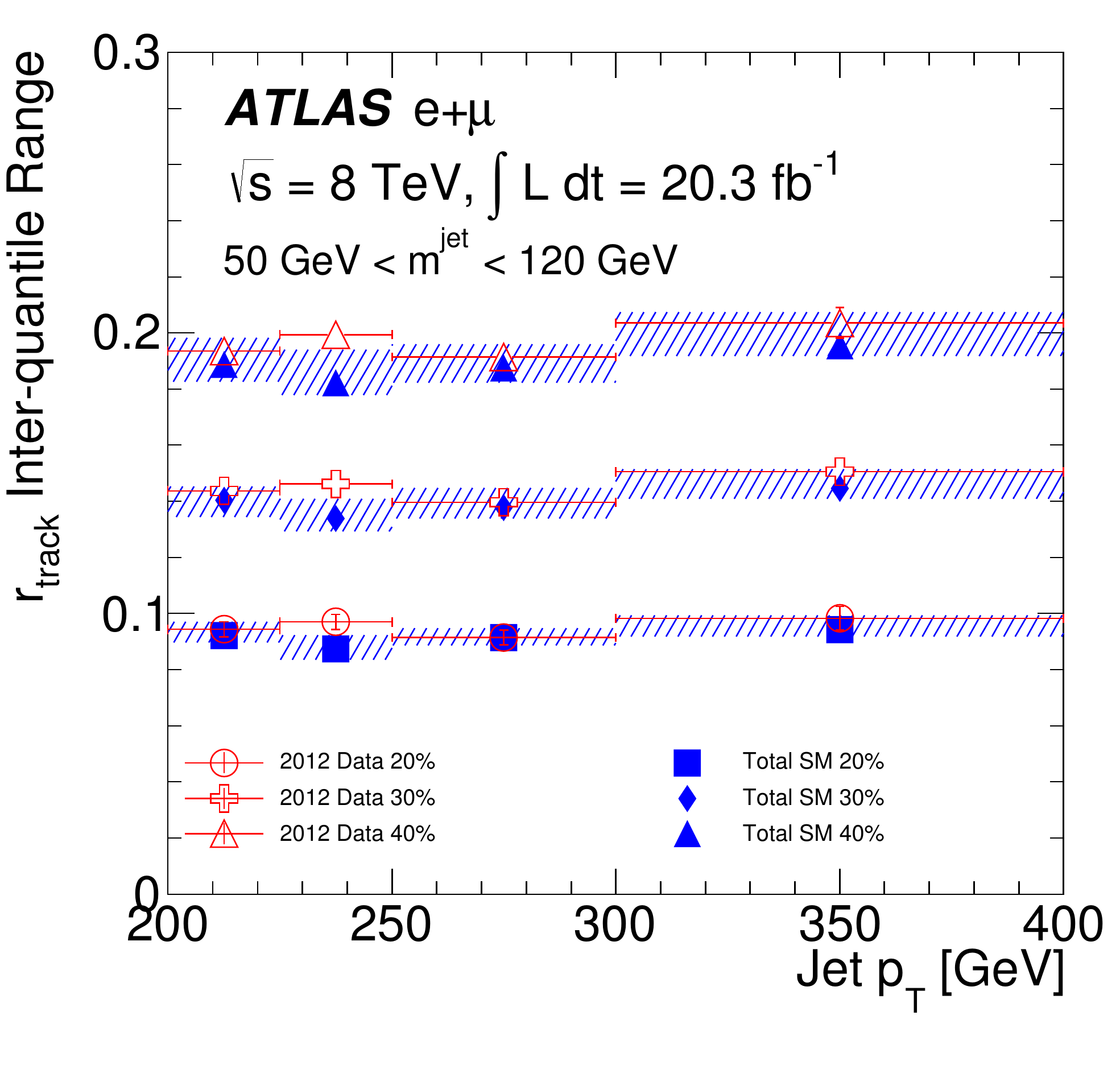}}
\caption{(a) The distribution of $r_\text{track}$ in the data for semi-leptonic $t\bar{t}$ events with the selected jet in the range $50$~GeV $<m^\text{jet}<$ $120$~GeV.  (b) The median of the $r_\text{track}$ distribution as a function of the jet $p_\text{T}$.  This includes the contributions from events that are not classified as Boosted $W$.  (c) The inter-quantile range as a measure of the width.  The quantiles are centred at the median.   The uncertainty band includes all the experimental uncertainties on the jet $p_\text{T}$ and jet mass described in Sect.~\ref{sec:systs}.  The inter-quantile range of size $0\%<X<50\%$ is defined as the difference between the $50\%+X\%$ quantile and the $50\%-X\%$ quantile. Statistical uncertainty bars are included on the data points but are smaller than the markers in many bins.}
\label{fig:datartrak}
\end{figure}

\begin{figure}[h!]
\centering
\subfloat[]{\includegraphics[width=0.55\textwidth]{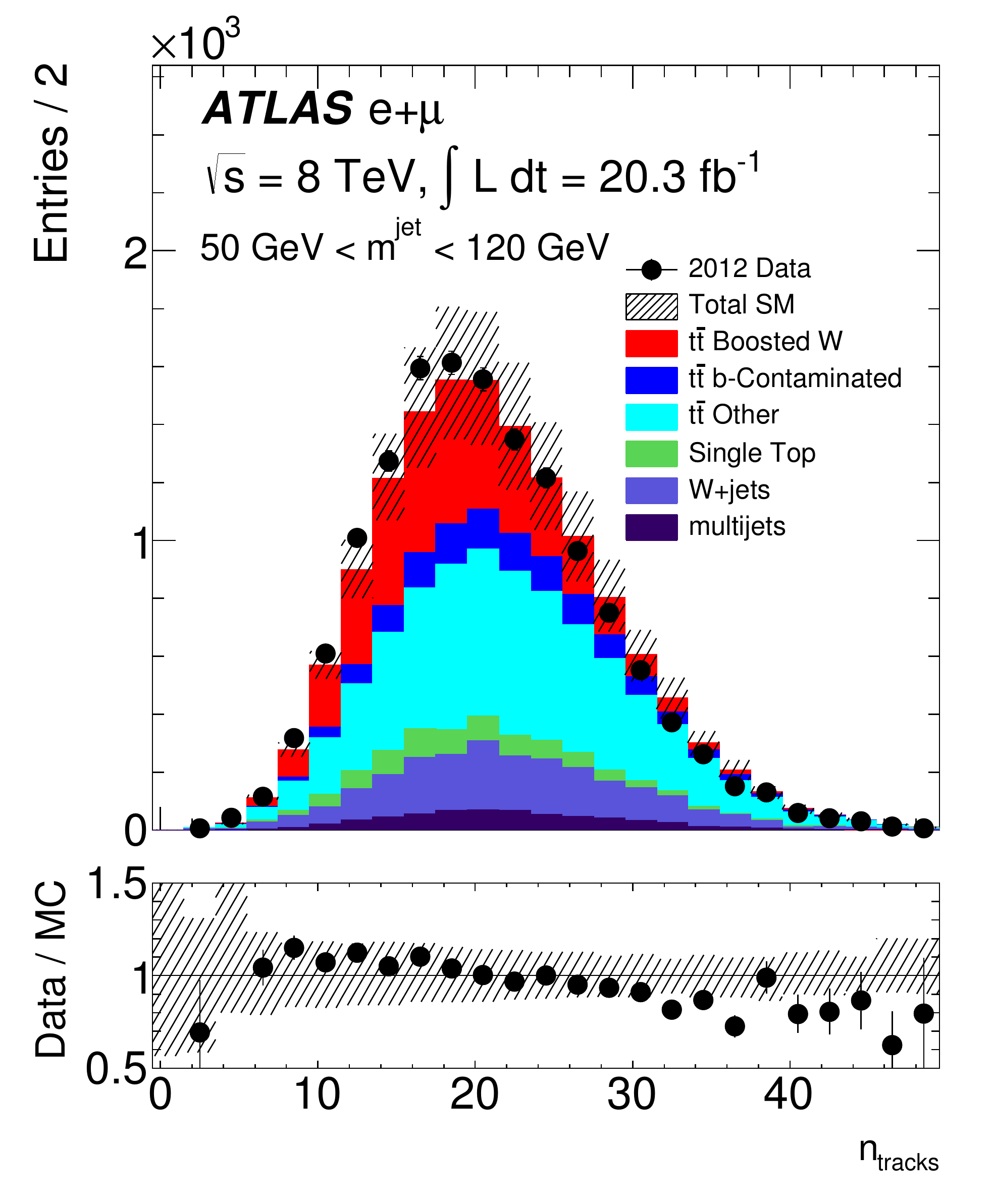}}\\\subfloat[]{\includegraphics[width=0.45\textwidth]{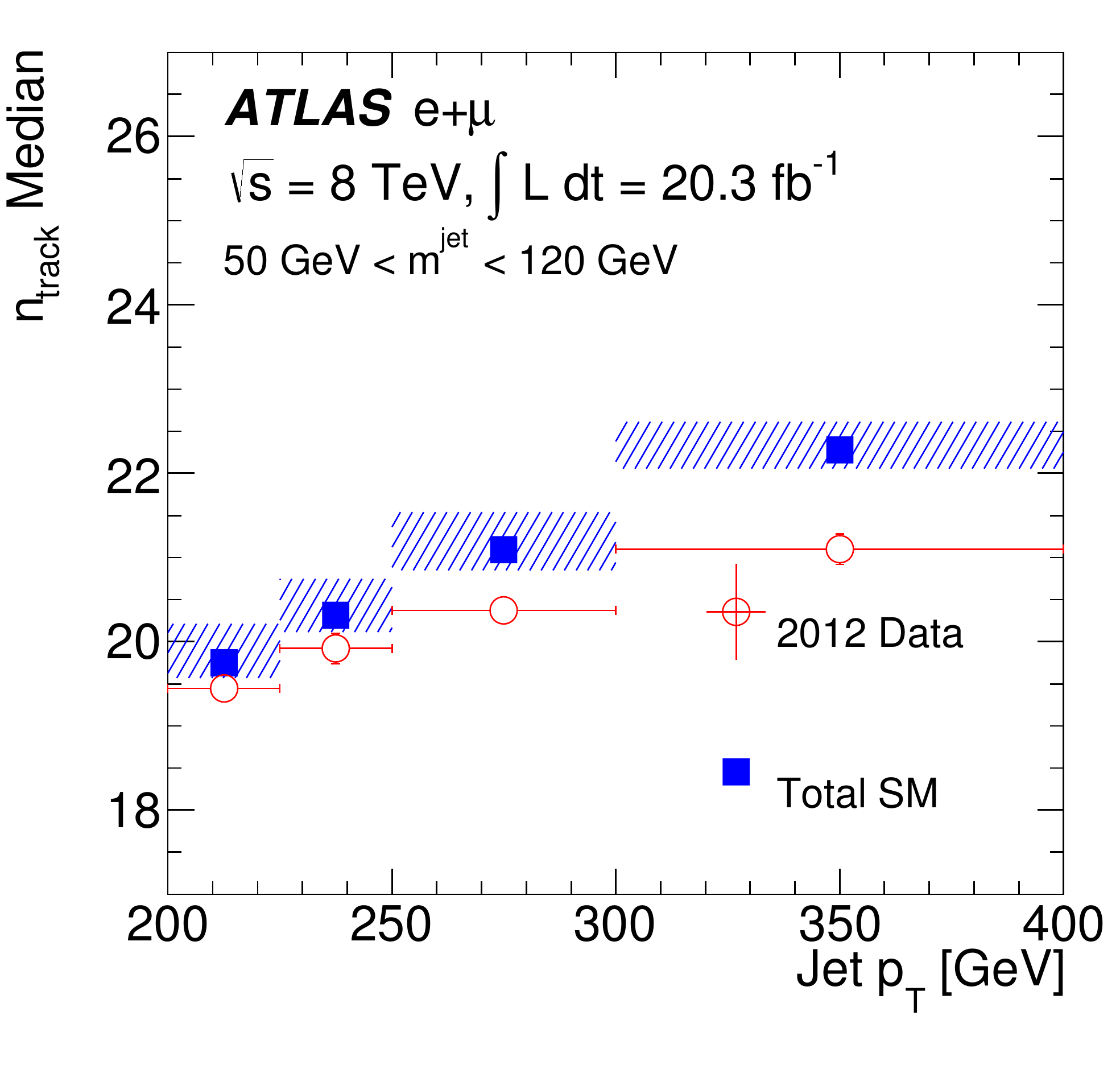}}\subfloat[]{\includegraphics[width=0.45\textwidth]{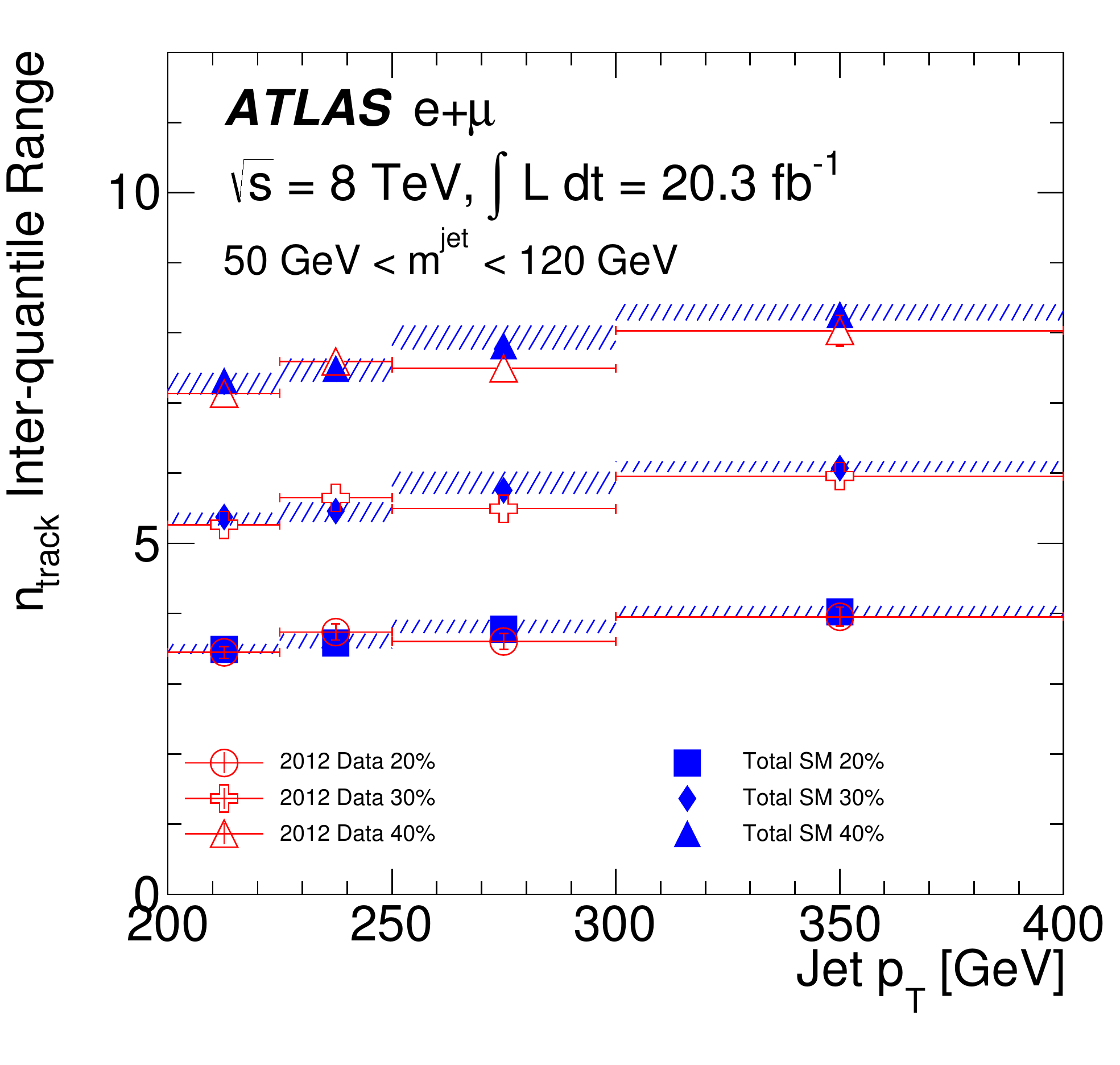}}
\caption{(a) The distribution of the number of tracks associated with the selected large-radius jet in the semi-leptonic $t\bar{t}$ data for events with the selected jet in the range $50$~GeV $<m^\text{jet}<$ $120$~GeV.  (b) The median of the distribution of the number of tracks as a function of the jet $p_\text{T}$.  This includes the contributions from events that are not classified as Boosted $W$.  (c) The inter-quantile range as a measure of the width.  The quantiles are centred at the median.  The uncertainty band includes all the experimental uncertainties on the jet $p_\text{T}$ and jet mass described in Sect.~\ref{sec:systs}.  The inter-quantile range of size $0\%<X<50\%$ is defined as the difference between the $50\%+X\%$ quantile and the $50\%-X\%$ quantile. Statistical uncertainty bars are included on the data points but are smaller than the markers in many bins.}
\label{fig:datantrack}
\end{figure}

The modelling of boosted $W$ bosons can also be studied using the jet-mass scale measured from tracks.  Defining the variable $r_{\mathrm{track}}$ as the ratio of the jet mass determined from tracks to the jet mass determined from the calorimeter, the jet mass scale uncertainty is related to the difference from unity of the ratio of $\langle r_\text{track}\rangle$ in data to $\langle r_\text{track}\rangle$ in MC simulation.  The mass scale uncertainty is calculated using the procedure described above, but with $r_{\mathrm{track}}^{-1}$.  If the jet consists only of pions, the natural scale for $r_\text{track}$ is 2/3, although there are significant physics and detector effects that introduce a large spread of values.  The distribution of $r_\text{track}$ in the $t\bar{t}$--enriched event sample with the same $p_\text{T}$ and $b$-jet veto requirements as in Fig.~\ref{fig:datamass} is shown in Fig.~\ref{fig:datartrak}(a).   Unlike the raw jet-mass distribution, the $r_\text{track}$ distribution is similar for all of the sub-processes, as expected.  The scale and spread of the $r_\text{track}$ distribution are quantified in Figures~\ref{fig:datartrak}(b) and ~\ref{fig:datartrak}(c) using the $p_\text{T}$ dependence of the median and inter-quantile ranges.   Previous studies have indicated that the track multiplicity, $n_\text{track}$, in quark and gluon jets is not well modelled, especially for gluon jets, where $n_\text{track}$ is lower in the data with respect to {\tt PYTHIA}~\cite{Aad:2014gea}.  The distribution of the track multiplicity for large-$R$ jets in the $t\bar{t}$-enriched event sample is shown in Fig.~\ref{fig:datantrack}.  The boosted $W$ events are peaked at slightly lower values of the number of associated tracks compared to the quark/gluon jets from the other processes.  The (charged) particle multiplicity increases for generic quark and gluon jets as a function of jet energy.  However, the mass-scale of the jets produced from $W$ boson decays is set by $m_W$ so that in the absence of detector reconstruction effects, the track multiplicity distribution should be largely $p_\text{T}$ independent.  The $p_\text{T}$ dependence of the track multiplicity is shown in Fig.~\ref{fig:datantrack}(b) and~\ref{fig:datantrack}(c) in the form of the median and the inter-quantile ranges.  The median does increase because of the large non-$W$ component as well as the finite detector acceptance for charged particles from the boosted $W$ boson decay.  The width is well modelled within the statistical precision of the data.  However, there is disagreement for the median.  Previous studies (including Rev.~\cite{Aad:2014gea}) suggest that this is due to fragmentation modelling and not the modelling of the detector response.

\begin{figure}[h!]
\centering
\subfloat[]{\includegraphics[width=0.55\textwidth]{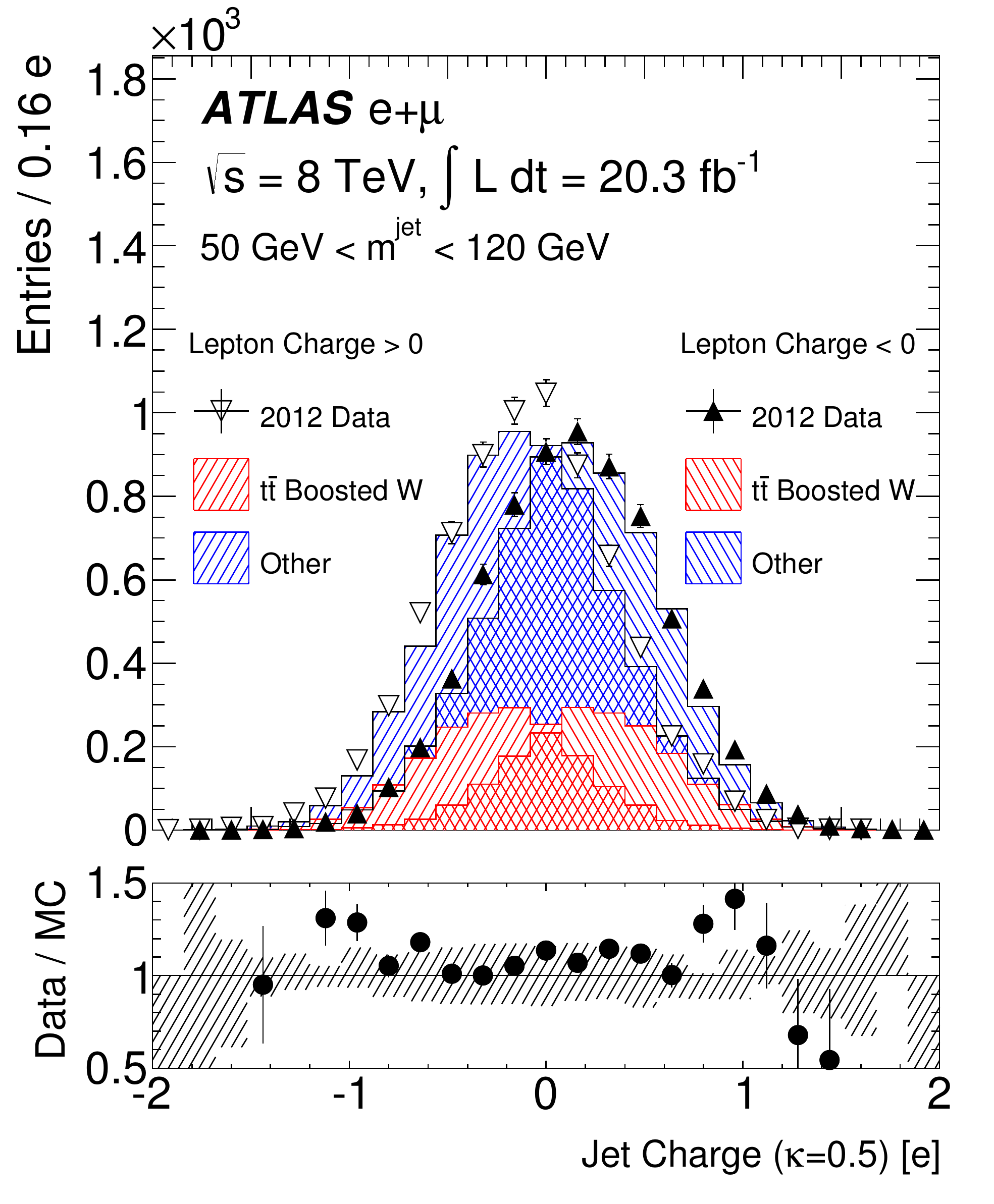}}\\\subfloat[]{\includegraphics[width=0.45\textwidth]{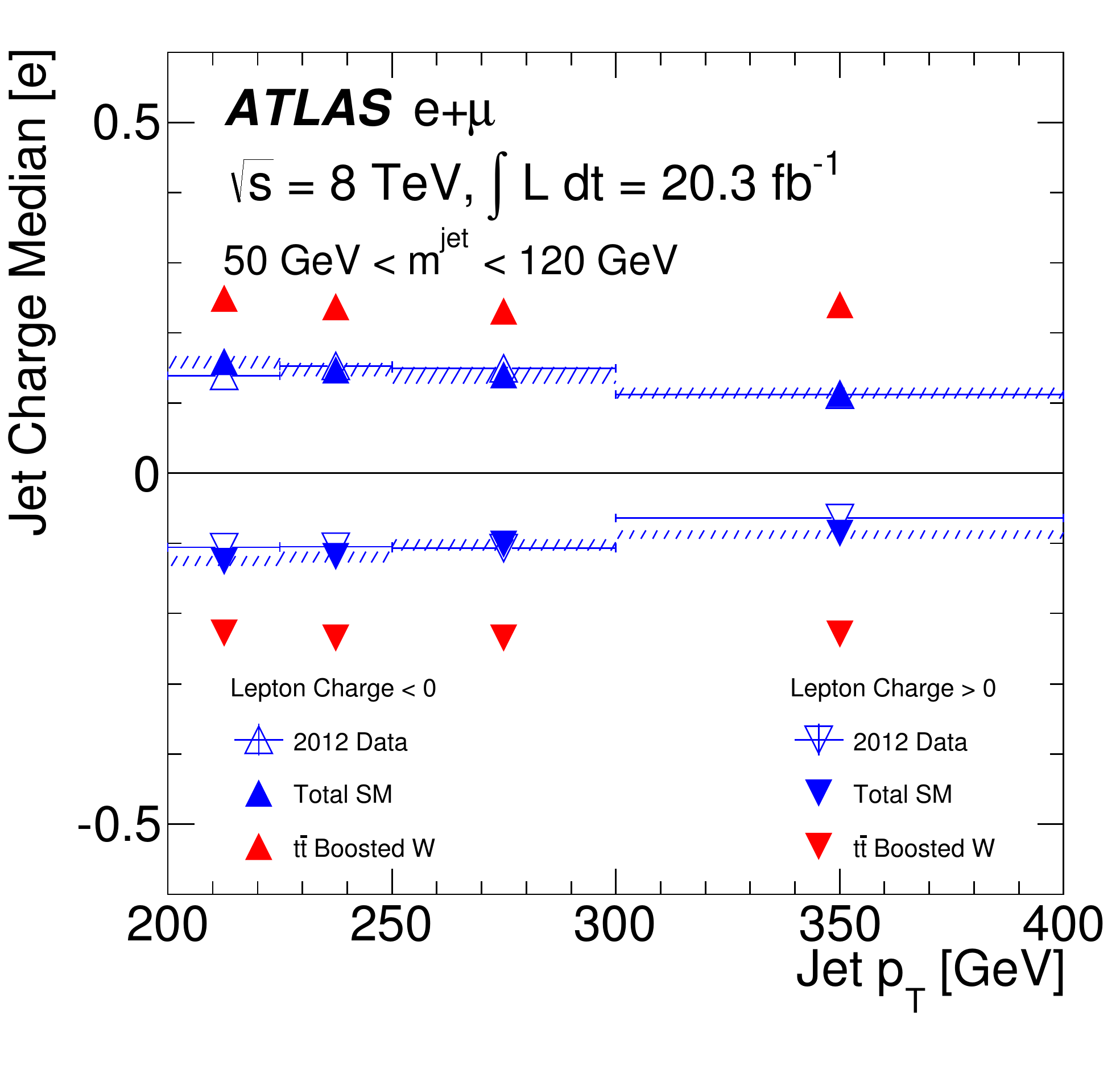}}\subfloat[]{\includegraphics[width=0.45\textwidth]{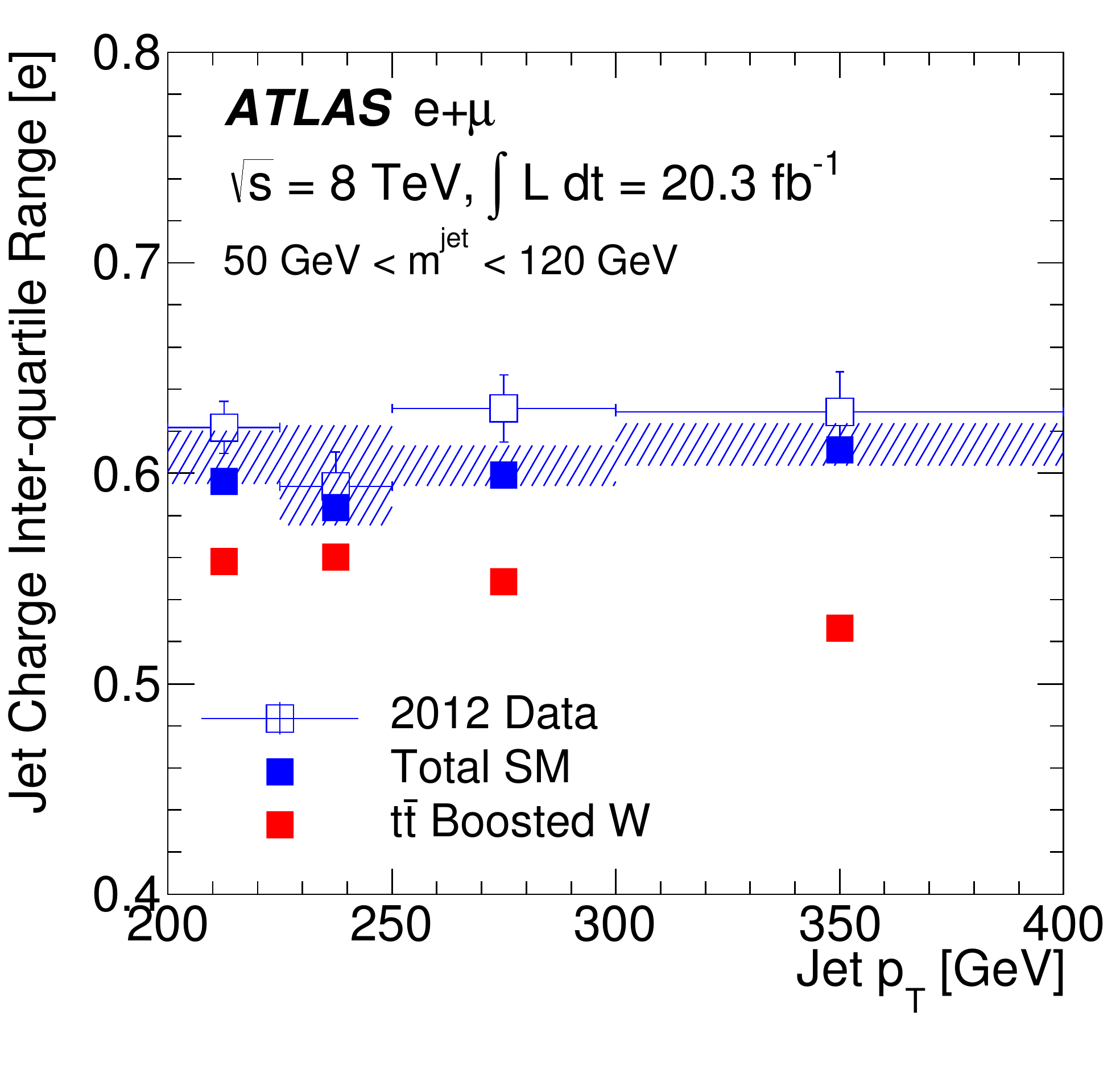}}
\caption{(a) The distribution of the jet charge in the data for semi-leptonic $t\bar{t}$ events with the selected jet in the range $50$~GeV $<m^\text{jet}<$ $120$~GeV.  The ratio uses the positive lepton charge.  (b) The median of the jet charge distribution as a function of the jet $p_\text{T}$.  This includes the contributions from events that are not classified as Boosted $W$ (except for the blue triangles, for which only the Boosted $W$ is included).  (c) The inter-quartile range as a measure of the width.  The quantiles are centred at the median.   The uncertainty band includes all the experimental uncertainties on the jet $p_\text{T}$ and jet mass described in Sect.~\ref{sec:systs}.  The inter-quantile range is defined as the difference between the $75\%$ quantile and the $25\%$ quantile.  Statistical uncertainty bars are included on the data points but are smaller than the markers in many bins.}
\label{fig:datacharge}
\end{figure}

The $p_\text{T}$-weighted distribution of the track charges defines the jet charge, which is shown in Fig.~\ref{fig:datacharge}(a).   The charge of the lepton from the leptonic $W$ boson decay determines the expected charge of the hadronically decaying $W$ boson candidate, allowing for a tag-and-probe study of the capability of charge tagging in hadronic $W$ boson decays~\cite{ATLAS-CONF-2013-086}.  The jet charge for boosted $W$ bosons for positively (negatively) charged leptons is clearly shifted to the left (right) of zero.  There is also some separation between positive and negative $W$ boson decays when the selected large-radius jet does not satisfy the criteria for being a boosted $W$ boson.  This is because the jet still contains some of the $W$ boson decay products, and the jet charge is correlated with the charge of the $W$ boson.  The difference between the inclusive and boosted $W$-boson jets is clearer in the $p_\text{T}$ dependence plot of the median jet charge shown in Fig.~\ref{fig:datacharge}(b).  The medians of the distributions for boosted $W$ jets are nearly twice as far apart as the medians for inclusive jets.  However, in both cases the spread is less than the width of the distribution, shown as the inter-quantile range (inter-quantile range with $X=25\%$) in Fig.~\ref{fig:datacharge}(c).  Even though there is some small disagreement for the median number of tracks, the $p_\text{T}$-weighted sum defining the jet charge is reasonably well modelled.

\begin{figure}[h!]
\centering
\subfloat[]{\includegraphics[width=0.5\textwidth]{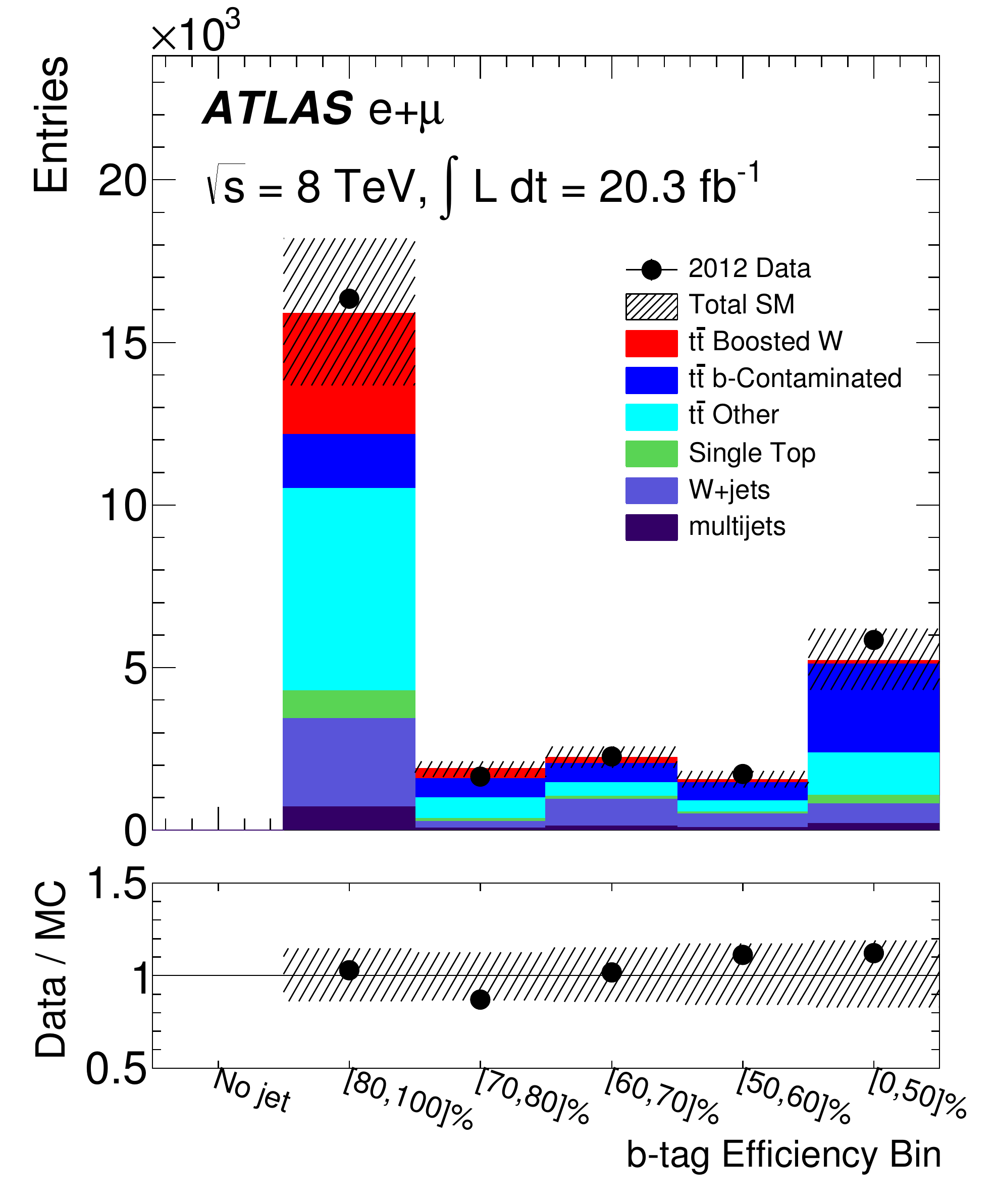}}\subfloat[]{\includegraphics[width=0.5\textwidth]{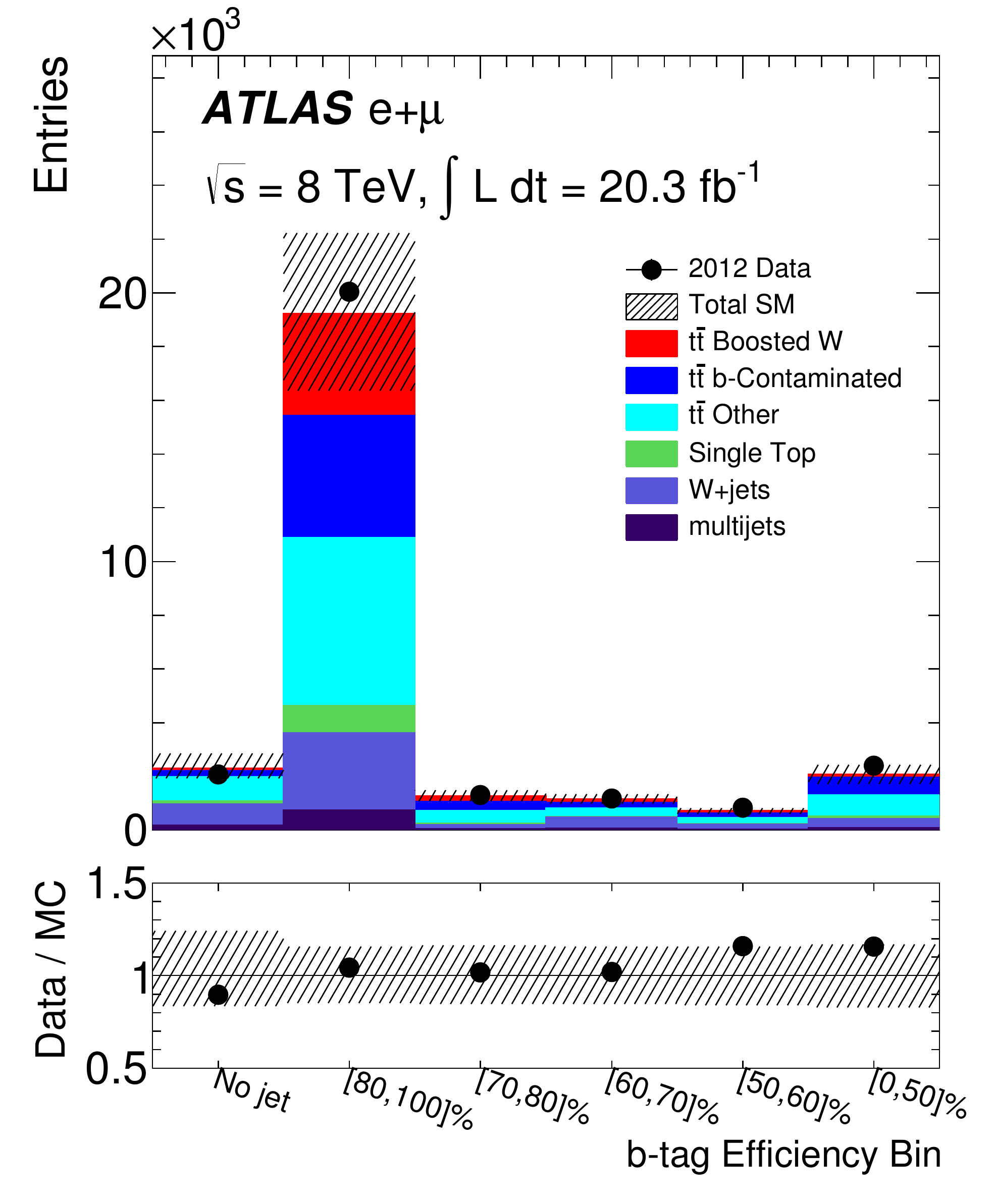}}
\caption{The efficiency-binned MV1 distribution for the (a) leading and (b) sub-leading matched small-radius in semi-leptonic $t\bar{t}$ events.  If there is no second small-radius jet with $p_\text{T}>25$~GeV and $\Delta R<1$ to the selected large-radius jet axis, the event is put in the `No jet' category in (b).  The uncertainty band includes all the experimental uncertainties on the jet $p_\text{T}$ and jet mass and those related to the $b$-tagging described in Sect.~\ref{sec:systs}.  Statistical uncertainty bars are included on the data points but are smaller than the markers in many bins.}
\label{fig:datamv1}
\end{figure}

The remaining input to the boson tagger is the $b$-tagging discriminant for the matched small-radius jets.  The efficiency-binned MV1 distributions are shown in Fig.~\ref{fig:datamv1}(a) and~\ref{fig:datamv1}(b) with the same selection criteria as for the previous figures, except that the $b$-jet veto is removed.  The contamination due to the $b$-jet from the top-quark decay complicates a direct study of the MV1 distribution for boosted $W$ jets; contamination from the $b$-quark decay products is seen clearly in the MV1 distribution at lower values of the efficiency.  Most of the boosted $W$ jets are in the highest efficiency bin because they have no real $b$-hadron decay.

Overall, the simulation models all three input variables well.

\newpage
\clearpage

\section{Outlook}
\label{sec:outlook}

The simulation studies of the boson-type tagger presented in Sect.~\ref{sec:perfm} show that for bosons with $200$~GeV $<p_\text{T}<$ 400~GeV, it is possible to achieve $Z$-boson efficiencies of $\epsilon_Z=90\%$, $50\%$, and $10\%$ with $W^+$ boson rejections of $1.7, 8.3$ and $1000$, respectively.   Putting this into context, with $R(\epsilon_Z)$ defined as the lowest possible $W$-boson tagging efficiency at a fixed $Z$-boson tagging efficiency: 

\begin{itemize}
\item The $WZ$/$WW$ cross-section ratio is $\sim 20\%$~\cite{Aad:2014mda}.  At the 50\% type-tagger working point, one can change the ratio of events to 

\begin{align}
\frac{50\%}{R(50\%)}\times \frac{  \sigma(WZ)}{ \sigma(WW)} = \frac{50\%}{12\%}\times \frac{ \sigma(WZ)}{ \sigma(WW)} = \frac{50}{12}\times 20\% \approx 83\%,
\end{align}

\noindent with the possibility for a high-purity extraction of the $WZ$ cross section in the semileptonic channel ($\ell\nu q\bar{q}$).

\item Diboson resonances are predicted by many models of physics beyond the Standard Model.  The all-hadronic channel provides a significantly higher yield than the leptonic channels.  At the $90\%$ type-tagger working point, one can distinguish $ZZ$ from $WZ$ with a likelihood ratio of $0.9^2/(0.9\times 0.6)\sim 1.5$.  

\item At the 10\% type-tagger working point, a leptophobic flavour-changing neutral current in $t\bar{t}$ production (with decays like in the SM) with a branching ratio of 1\% would have the same number of events as the $t\rightarrow bW$ decay:\footnote{Up to impurities due to the high-occupancy $t\bar{t}$ environment.}

\begin{align}
\frac{10\%}{R(10\%)}\times \frac{\Gamma(t\rightarrow Zc)}{ \Gamma(t\rightarrow Wb)}=\frac{10\%}{0.1\%}\times \frac{ \Gamma(t\rightarrow Zc)}{\Gamma(t\rightarrow Wb)} = 100 \times 1\% = 100\%.
\end{align}

\end{itemize}

\begin{figure}[h!]
\centering
\subfloat[]{\includegraphics[width=0.5\textwidth]{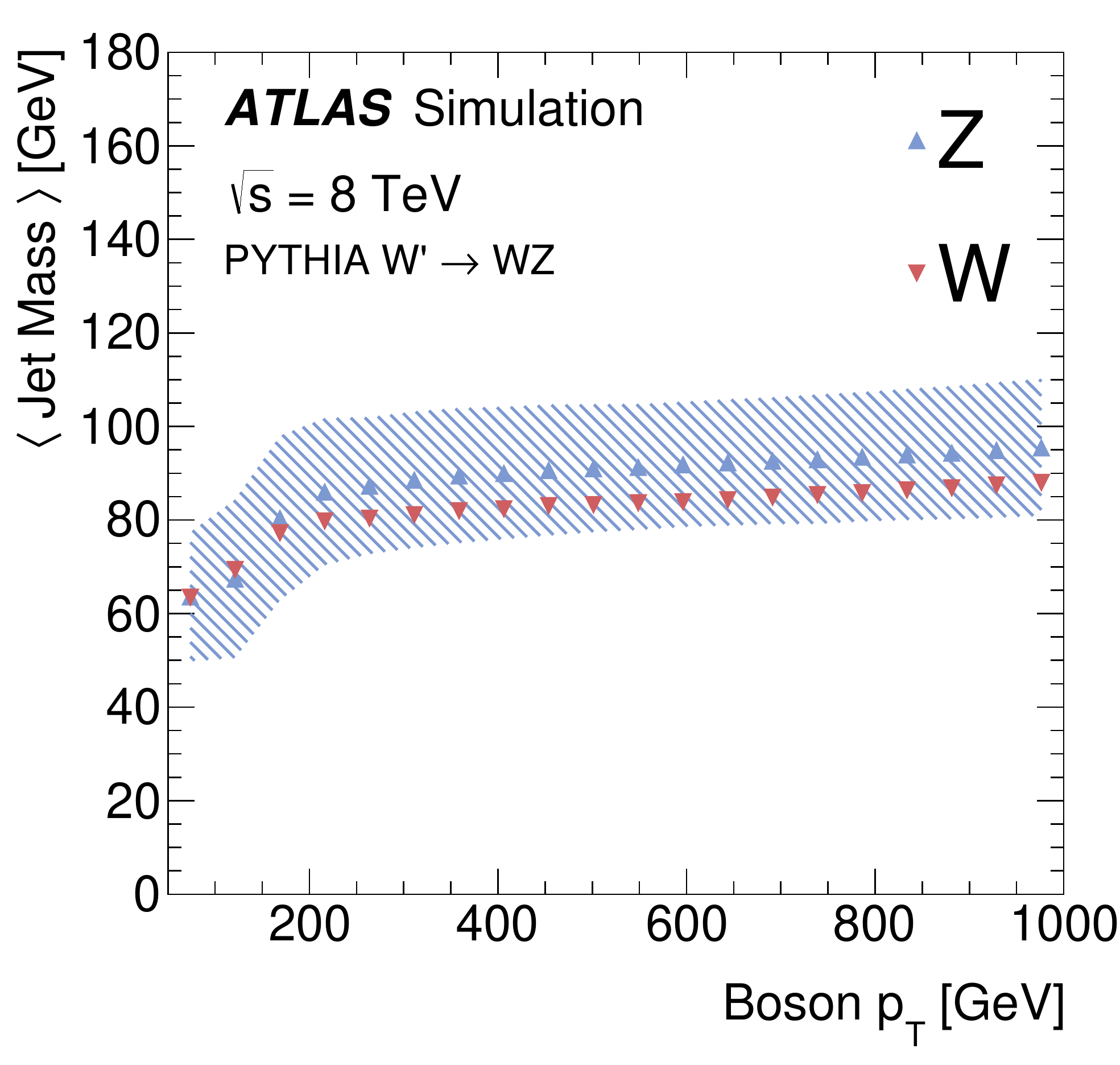}}\subfloat[]{\includegraphics[width=0.5\textwidth]{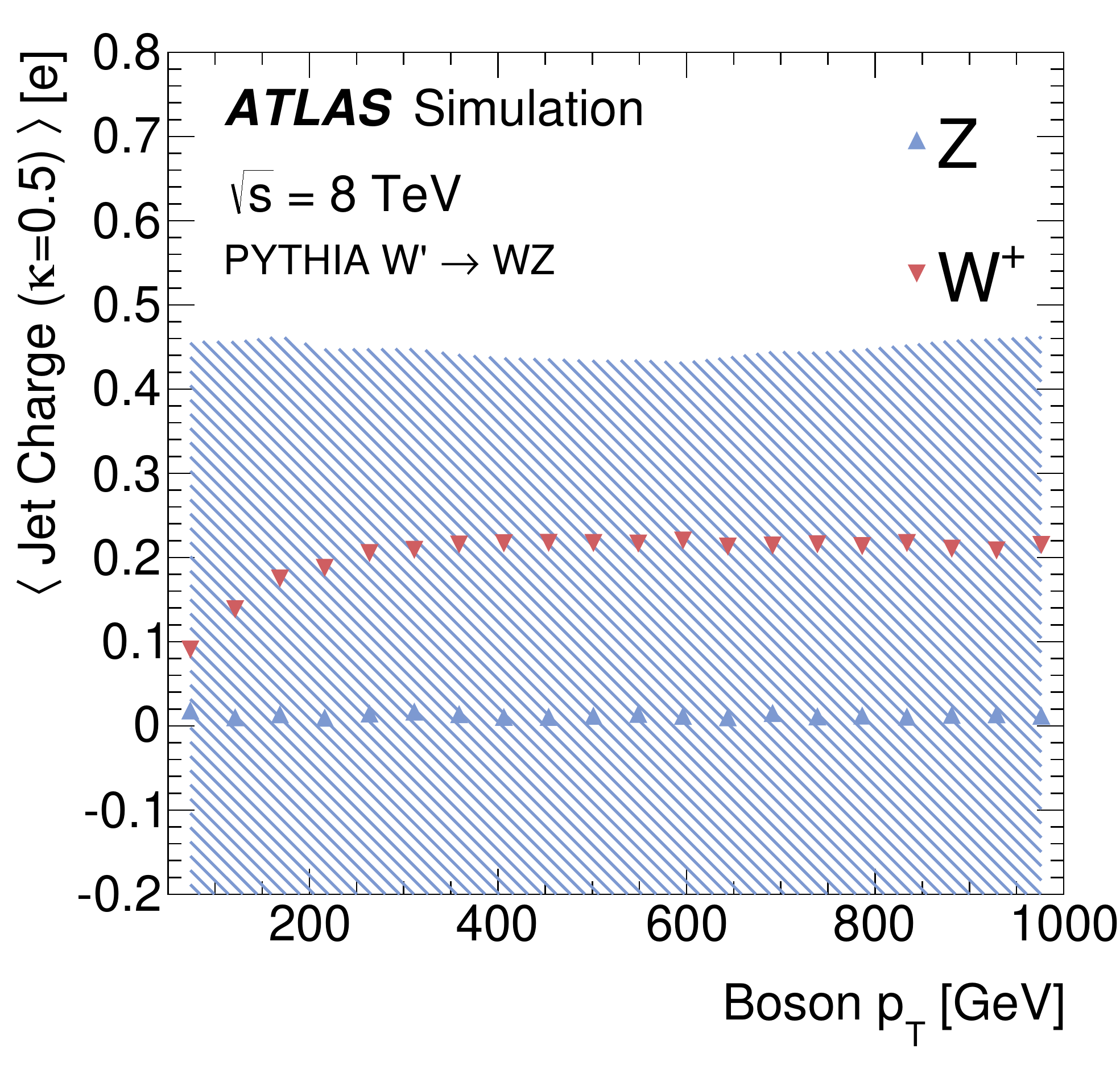}}\\
\subfloat[]{\includegraphics[width=0.5\textwidth]{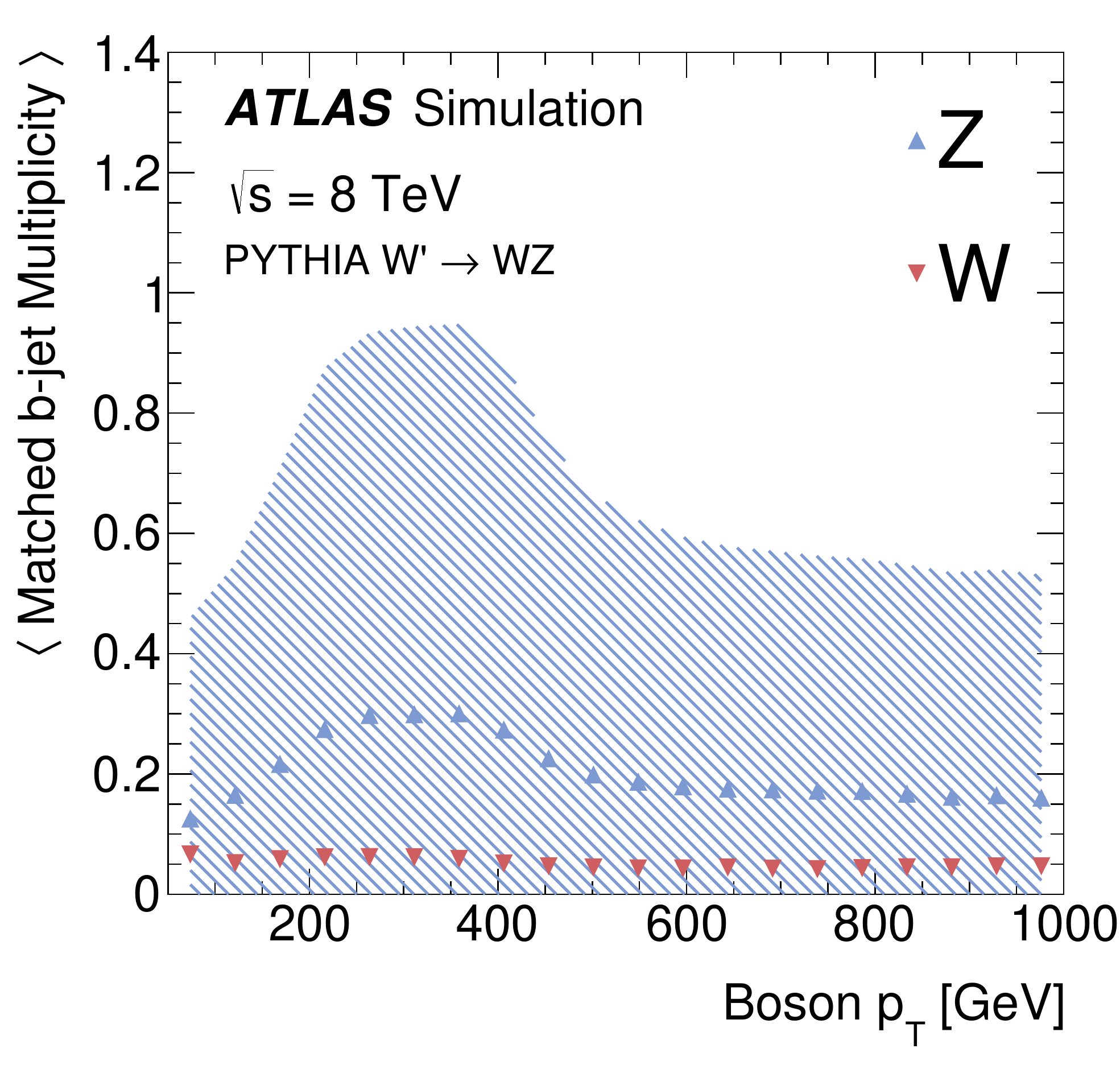}}
\caption{The boson $p_\text{T}$ dependence of the (a) jet mass, (b) jet charge, and (c) number of small-radius $b$-tagged jets matched to the large-radius jet.}
\label{fig:highpT2}
\end{figure}

Only the range $200$ GeV $<p_\text{T}<400$ GeV was studied thus far due to the availability of $W$ bosons in the data.  Figure~\ref{fig:highpT2} shows how the average and standard deviation of the jet mass, jet charge and multiplicity of the matched small-radius $b$-tagged jets distributions depend on jet $p_\text{T}$ in simulation up to 1 TeV.  As long as the jet $p_\text{T}$ is high enough so that a single jet captures all of the boson decay products, the jet mass and jet charge distributions are predicted to be largely independent of $p_\text{T}$.  The information from $b$-tagging degrades around 400 GeV as the two decay products from the boson become too close to resolve two separate jets.  

\section{Conclusions}
\label{sec:conc}

A tagger for distinguishing hadronically decaying boosted $Z$ bosons from $W$ bosons using the ATLAS detector has been presented.   It will most likely be used after a boson tagger has rejected most QCD multijet events.  Three discriminating variables are chosen which are sensitive to the differences in boson mass, charge, and branching ratios to specific quark flavours: large-radius jet mass, large-radius jet charge, and an associated small-radius jet $b$-tagging discriminant.  For moderate and high $Z$-boson tagging efficiencies, the jet mass is the most discriminating of the three variables, but there is significant improvement in discrimination when combining all three inputs into a single tagger.  At low $Z$-boson efficiencies, smaller than the $Z\rightarrow b\bar{b}$ branching ratio, the $b$-tagging discriminant is the most useful for rejecting $W$ bosons.  The full tagger is largely unaffected by many systematic uncertainties on the inputs, with the exception of the uncertainties on the jet-mass scale and resolution.  While it is not possible to measure the tagger efficiencies directly in data due to the lack of a pure sample of boosted, hadronically decaying $Z$ bosons, modelling of the likelihood function using hadronically decaying $W$ bosons has been studied in the data.  Overall, the simulation agrees well with the 20.3 fb${}^{-1}$ of $\sqrt{s}=8$ TeV $pp$ data collected at the LHC.  

\clearpage
\newpage

\appendix

\section{Correlations with 2-subjettiness}
\label{sec:tau21}

\begin{figure}[h!]
\centering
\subfloat[]{\includegraphics[width=0.95\textwidth]{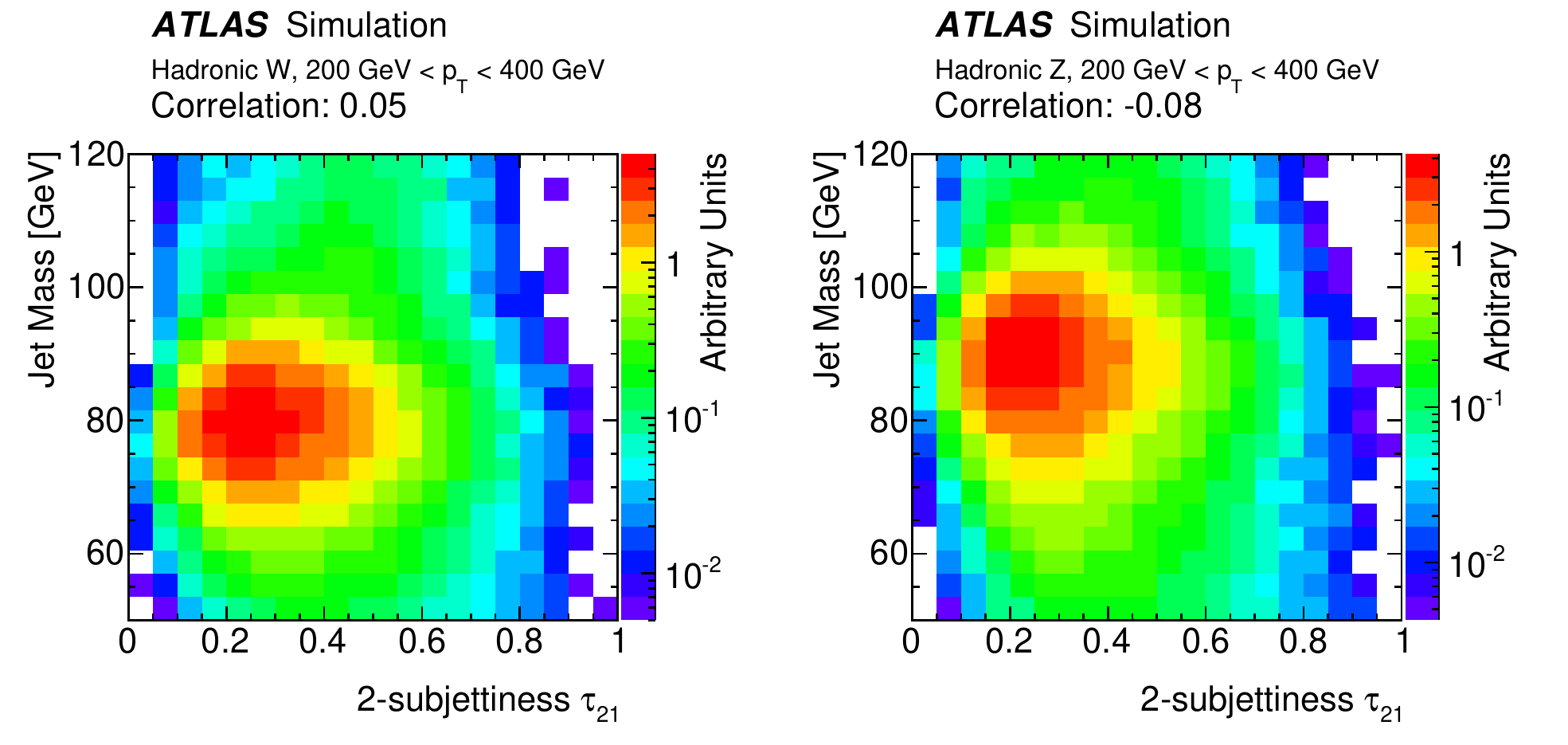}}\\
\subfloat[]{\includegraphics[width=0.95\textwidth]{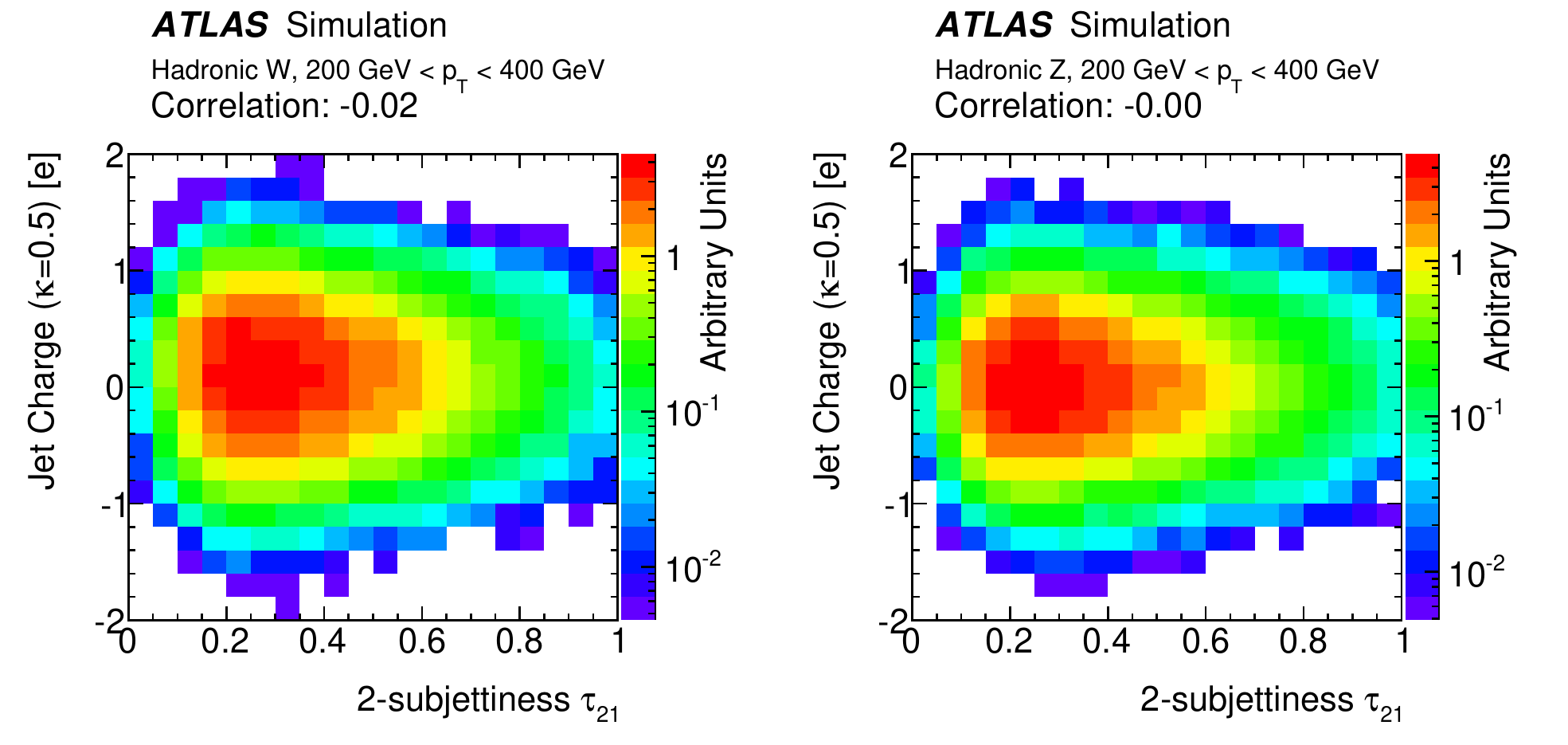}}\\
\caption{The joint distribution of (a) jet mass and 2-subjettiness and (b) jet charge and 2-subjettiness for (left) $W$ boson jets and (right) $Z$ boson jets.}
\label{fig:highpT}
\end{figure}

The tagger developed in this paper is designed to work in conjunction with a procedure for separating generic quark and gluon jets from boson jets. Figure~\ref{fig:highpT} shows the joint distribution of the jet mass and jet charge with a standard boson tagging variable 2-subjettiness, $\tau_{21}$.  The boson type tagger variables are nearly independent of $\tau_{21}$. 

\section*{Acknowledgments}



We thank CERN for the very successful operation of the LHC, as well as the
support staff from our institutions without whom ATLAS could not be
operated efficiently.

We acknowledge the support of ANPCyT, Argentina; YerPhI, Armenia; ARC, Australia; BMWFW and FWF, Austria; ANAS, Azerbaijan; SSTC, Belarus; CNPq and FAPESP, Brazil; NSERC, NRC and CFI, Canada; CERN; CONICYT, Chile; CAS, MOST and NSFC, China; COLCIENCIAS, Colombia; MSMT CR, MPO CR and VSC CR, Czech Republic; DNRF, DNSRC and Lundbeck Foundation, Denmark; IN2P3-CNRS, CEA-DSM/IRFU, France; GNSF, Georgia; BMBF, HGF, and MPG, Germany; GSRT, Greece; RGC, Hong Kong SAR, China; ISF, I-CORE and Benoziyo Center, Israel; INFN, Italy; MEXT and JSPS, Japan; CNRST, Morocco; FOM and NWO, Netherlands; RCN, Norway; MNiSW and NCN, Poland; FCT, Portugal; MNE/IFA, Romania; MES of Russia and NRC KI, Russian Federation; JINR; MESTD, Serbia; MSSR, Slovakia; ARRS and MIZ\v{S}, Slovenia; DST/NRF, South Africa; MINECO, Spain; SRC and Wallenberg Foundation, Sweden; SERI, SNSF and Cantons of Bern and Geneva, Switzerland; MOST, Taiwan; TAEK, Turkey; STFC, United Kingdom; DOE and NSF, United States of America. In addition, individual groups and members have received support from BCKDF, the Canada Council, CANARIE, CRC, Compute Canada, FQRNT, and the Ontario Innovation Trust, Canada; EPLANET, ERC, FP7, Horizon 2020 and Marie Sk?odowska-Curie Actions, European Union; Investissements d'Avenir Labex and Idex, ANR, Region Auvergne and Fondation Partager le Savoir, France; DFG and AvH Foundation, Germany; Herakleitos, Thales and Aristeia programmes co-financed by EU-ESF and the Greek NSRF; BSF, GIF and Minerva, Israel; BRF, Norway; the Royal Society and Leverhulme Trust, United Kingdom.

The crucial computing support from all WLCG partners is acknowledged
gratefully, in particular from CERN and the ATLAS Tier-1 facilities at
TRIUMF (Canada), NDGF (Denmark, Norway, Sweden), CC-IN2P3 (France),
KIT/GridKA (Germany), INFN-CNAF (Italy), NL-T1 (Netherlands), PIC (Spain),
ASGC (Taiwan), RAL (UK) and BNL (USA) and in the Tier-2 facilities
worldwide.

\printbibliography

\newpage \input{atlas_authlist}

\end{document}

%% file: WZTagger-meta-data.tex
\AtlasTitle{A new method to distinguish hadronically decaying boosted $Z$ bosons from $W$ bosons using the ATLAS detector}
\AtlasJournal{EPJC}
\AtlasAbstract{%
The distribution of particles inside hadronic jets produced in the decay of boosted $W$ and $Z$ bosons can be used to discriminate such jets from the continuum background.  Given that a jet has been identified as likely resulting from the hadronic decay of a boosted $W$ or $Z$ boson, this paper presents a technique for further differentiating $Z$ bosons from $W$ bosons.  The variables used are jet mass, jet charge, and a $b$-tagging discriminant.  A likelihood tagger is constructed from these variables and tested in the simulation of $W'\rightarrow WZ$ for bosons in the
transverse momentum range $200$~GeV~$<p_\text{T}<$~$400$~GeV in $\sqrt{s}=8$ TeV $pp$ collisions with the ATLAS detector at the LHC.  For $Z$-boson tagging efficiencies of  $\epsilon_Z=90\%$, $50\%$, and $10\%$, one can achieve $W^+$-boson tagging rejection factors ($1/\epsilon_{W^+}$) of $1.7$, $8.3$ and $1000$, respectively.  It is not possible to measure these efficiencies in the data due to the lack of a pure sample of high $p_\text{T}$, hadronically decaying $Z$ bosons.  However, the modelling of the tagger inputs for boosted $W$ bosons is studied in data using a $t\bar{t}$-enriched sample of events in $20.3$ fb${}^{-1}$ of data at $\sqrt{s}=8$ TeV.  The inputs are well modelled within uncertainties, which builds confidence in the expected tagger performance.
}
\PreprintIdNumber{CERN-PH-EP-2015-194}

%% file: atlas_authlist.tex
\begin{flushleft}
{\Large The ATLAS Collaboration}

\bigskip

G.~Aad$^{\rm 85}$,
B.~Abbott$^{\rm 113}$,
J.~Abdallah$^{\rm 151}$,
O.~Abdinov$^{\rm 11}$,
R.~Aben$^{\rm 107}$,
M.~Abolins$^{\rm 90}$,
O.S.~AbouZeid$^{\rm 158}$,
H.~Abramowicz$^{\rm 153}$,
H.~Abreu$^{\rm 152}$,
R.~Abreu$^{\rm 116}$
% ...
\bigskip
\\
% ...
$^{11}$ Institute of Physics, Azerbaijan Academy of Sciences, Baku, Azerbaijan\\
% ...
$^{85}$ CPPM, Aix-Marseille Universit{\'e} and CNRS/IN2P3, Marseille, France\\
% ...
$^{90}$ Department of Physics and Astronomy, Michigan State University, East Lansing MI, United States of America\\
% ...
$^{107}$ Nikhef National Institute for Subatomic Physics and University of Amsterdam, Amsterdam, Netherlands\\
% ...
$^{113}$ Homer L. Dodge Department of Physics and Astronomy, University of Oklahoma, Norman OK, United States of America\\
% ...
$^{116}$ Center for High Energy Physics, University of Oregon, Eugene OR, United States of America\\
% ...
$^{151}$ Institute of Physics, Academia Sinica, Taipei, Taiwan\\
$^{152}$ Department of Physics, Technion: Israel Institute of Technology, Haifa, Israel\\
$^{153}$ Raymond and Beverly Sackler School of Physics and Astronomy, Tel Aviv University, Tel Aviv, Israel\\
% ...
$^{158}$ Department of Physics, University of Toronto, Toronto ON, Canada\\
% ...
\end{flushleft}
